\documentclass[twocolumn,twocolappendix]{aastex63}

\usepackage{xspace}
\usepackage{hyperref}
\usepackage{multirow}
\usepackage{graphicx}
\usepackage{enumitem}
\usepackage{amssymb}
\usepackage{amsmath}
\usepackage{booktabs}
\usepackage{pifont}
\usepackage{array}
\newcommand{\PreserveBackslash}[1]{\let\temp=\\#1\let\\=\temp}
\newcolumntype{C}[1]{>{\PreserveBackslash\centering}p{#1}}
\newcolumntype{R}[1]{>{\PreserveBackslash\raggedleft}p{#1}}
\newcolumntype{L}[1]{>{\PreserveBackslash\raggedright}p{#1}}

\newcommand{\todo}{\ifmmode \text{\color{red}\Huge{\(\bullet\)}} \else {\color{red}{\Huge$\bullet$}}\fi}

\newcommand{\chandra}{\textit{Chandra}\xspace}
\newcommand{\xmm}{XMM-\textit{Newton}\xspace}
\newcommand{\nustar}{\textit{NuSTAR}\xspace}
\newcommand{\suzaku}{\textit{Suzaku}\xspace}
\newcommand{\Cahk}{\ifmmode \left[{\rm Ca\,\textsc{ii} H and K}\right\,\lambda3969,3934 \else Ca\,\textsc{ii} H and K$\,\lambda3969,3934$\fi}
\newcommand{\Catrip}{\ifmmode \left[{\rm Ca}\,\textsc{ii}\right\,\lambda8498, 8542, 8662 \else Ca\,\textsc{ii} $\,\lambda8498, 8542, 8662$\fi}
\newcommand{\mgb}{\ifmmode \left{\rm Mg\,\textsc{I}}\right\,$b$\,\lambda5183, 5172, 5167 \else Mg\,{\sc I} $b$$\,\lambda5183, 5172, 5167$\fi}
\newcommand{\NIIa}{\ifmmode \left[{\rm N}\,\textsc{ii}\right]\,\lambda6548 \else [N\,{\sc ii}]\,$\lambda6548$\fi}

\newcommand{\cmark}{\ding{51}}%
\newcommand{\xmark}{\ding{55}}%

\def\NIIHa{[\mbox{N\,{\sc ii}}]$\lambda 6583$/H$\alpha$}
\def\SIIHa{[\mbox{S\,{\sc ii}}]$\lambda\lambda 6716,6731$/H$\alpha$}
\def\OIHa{[\mbox{O\,{\sc i}}]$\lambda 6300$/H$\alpha$}
\def\OIIIHb{[\mbox{O\,{\sc iii}}]$\lambda 5007$/H$\beta$}

\def\OIII{[\mbox{O\,{\sc iii}}]$\lambda 5007$}

\def\civ{\mbox{C\,{\sc IV}}$\lambda 1549$}

\def\nev{[\mbox{Ne\,{\sc v}}]$\lambda 3426$}
\def\nevv{[\mbox{Ne\,{\sc v}}]$\lambda 3346$}

\defcitealias{mazzolari24}{M24}
\defcitealias{reiss25}{R25}

\shorttitle{BASS. XLIX. Characterization of highly luminous and obscured AGNs}
\shortauthors{Peca et al.}
\graphicspath{{./}{figures/}}

\usepackage{placeins}

\begin{document}

\title{BASS. XLIX. Characterization of highly luminous and obscured AGNs: local X-ray and \nev\ emission in comparison with the high-redshift Universe}


\author[0000-0003-2196-3298]{Alessandro Peca$^{\star}$}
\affiliation{Eureka Scientific, 2452 Delmer Street, Suite 100, Oakland, CA 94602-3017, USA; $^{\star}$peca.alessandro@gmail.com}
\affiliation{Department of Physics, Yale University, P.O. Box 208120, New Haven, CT 06520, USA}

\author[0000-0002-7998-9581]{Michael J. Koss}
\affiliation{Eureka Scientific, 2452 Delmer Street, Suite 100, Oakland, CA 94602-3017, USA; $^{\star}$peca.alessandro@gmail.com}
\affiliation{Space Science Institute, 4750 Walnut Street, Suite 205, Boulder, CO 80301, USA}

\author[0000-0002-5037-951X]{Kyuseok Oh}
\affiliation{Korea Astronomy \& Space Science institute, 776, Daedeokdae-ro, Yuseong-gu, Daejeon 34055, Republic of Korea}

\author[0000-0001-5231-2645]{Claudio Ricci}
\affiliation{N\'ucleo de Astronom\'ia de la Facultad de Ingenier\'ia, Universidad Diego Portales, Av. Ej\'ercito Libertador 441, Santiago 22, Chile}
\affiliation{Kavli Institute for Astronomy and Astrophysics, Peking University, Beijing 100871, People's Republic of China}

\author[0000-0002-3683-7297]{Benny Trakhtenbrot}
\affiliation{School of Physics and Astronomy, Tel Aviv University, Tel Aviv 69978, Israel}
\affiliation{Max-Planck-Institut f{\"u}r extraterrestrische Physik, Gie\ss{}enbachstra\ss{}e 1, 85748 Garching, Germany}
\affiliation{Excellence Cluster ORIGINS, Boltzmannsstra\ss{}e 2, 85748, Garching, Germany}

\author[0000-0002-7962-5446]{Richard Mushotzky}
\affiliation{Department of Astronomy, University of Maryland, College Park, MD 20742, USA}
\affiliation{Joint Space-Science Institute, University of Maryland, College Park, MD 20742, USA}

\author[0000-0001-7568-6412]{Ezequiel Treister}
\affiliation{Instituto de Alta Investigaci\'on, Universidad de Tarapac\'a, Casilla 7D, Arica, Chile}

\author[0000-0002-0745-9792]{C. Megan Urry}
\affiliation{Yale Center for Astronomy \& Astrophysics and Department of Physics, Yale University, P.O. Box 208120, New Haven, CT 06520-8120, USA}
\affiliation{Department of Physics, Yale University, P.O. Box 208120, New Haven, CT 06520, USA}

\author[0000-0001-6412-2312]{Andrealuna Pizzetti}
\affiliation{European Southern Observatory, Alonso de C{\'o}rdova 3107, Casilla 19, Santiago 19001, Chile}

\author[0000-0002-4377-903X]{Kohei Ichikawa}
\affil{Frontier Research Institute for Interdisciplinary Sciences, Tohoku University, Sendai 980-8578, Japan}
\affil{
Astronomical Institute, Tohoku University, Aramaki, Aoba-ku, Sendai, Miyagi 980-8578, Japan
}

\author[0000-0003-3450-6483]{Alessia Tortosa}
\affiliation{INAF - Osservatorio Astronomico di Roma, Via Frascati 33, 00078 Monte Porzio Catone (RM), Italy.}

\author[0000-0001-5742-5980]{Federica Ricci}
\affiliation{Instituto de Astrof{\'i}sica, Facultad de F{\'i}sica, Pontificia Universidad Cat{\'o}lica de Chile, Casilla 306, Santiago 22, Chile}
\affiliation{Dipartimento di Fisica e Astronomia, Università di Bologna, via Gobetti 93/2, 40129 Bologna, Italy}

\author[0000-0002-8177-6905]{Matilde Signorini}
\affiliation{Dipartimento di Matematica e Fisica, Università degli Studi di Roma Tre, via della Vasca Navale, 84, 00146 Rome, Italy}
\affiliation{INAF - Osservatorio Astrofisico di Arcetri, Largo Enrico Fermi 5, I-50125 Firenze, Italy}
\affiliation{European Space Agency (ESA), European Space Research and Technology Centre, Noordwijk, Netherlands }

\author[0000-0002-2603-2639]{Darshan Kakkad}
\affiliation{Centre for Astrophysics Research, University of Hertfordshire, Hatfield, AL10 9AB, UK}

\author[0000-0001-9910-3234]{Chin-Shin Chang}
\affiliation{Joint ALMA Observatory,  Avenida Alonso de Cordova 3107, Vitacura 7630355, Santiago, Chile}

\author[0009-0005-7383-6655]{Giovanni Mazzolari}
\affiliation{INAF – Osservatorio di Astrofisica e Scienza dello Spazio di Bologna, Via Gobetti 93/3, I-40129 Bologna, Italy}
\affiliation{Dipartimento di Fisica e Astronomia, Università di Bologna, Via Gobetti 93/2, I-40129 Bologna, Italy}

\author[0000-0002-9144-2255]{Turgay Caglar}
\affiliation{George P. and Cynthia Woods Mitchell Institute for Fundamental Physics and Astronomy, Texas A\&M University, College Station, TX, 77845, USA}
\affiliation{Leiden Observatory, PO Box 9513, 2300 RA Leiden, The Netherlands}

\author[0000-0002-1292-1451]{Macon Magno}
\affiliation{George P. and Cynthia Woods Mitchell Institute for Fundamental Physics and Astronomy, Texas A\&M University, College Station, TX, 77845, USA}
\affiliation{CSIRO Space and Astronomy, ATNF, PO Box 1130, Bentley WA 6102, Australia}

\author[0000-0002-4065-8304]{Ignacio del Moral-Castro}
\affiliation{Instituto de Astrof{\'i}sica, Facultad de F{\'i}sica, Pontificia Universidad Cat{\'o}lica de Chile, Casilla 306, Santiago 22, Chile}

\author[0000-0001-9379-4716]{Peter G. Boorman}
\affiliation{Cahill Center for Astronomy and Astrophysics, California Institute of Technology, Pasadena, CA 91125, USA}

\author[0000-0001-8211-3807]{Tonima T. Ananna}
\affiliation{Department of Physics and Astronomy, Wayne State University, Detroit, MI 48202, USA}

\author[0000-0002-4226-8959]{Fiona Harrison}
\affiliation{Cahill Center for Astronomy and Astrophysics, California Institute of Technology, Pasadena, CA 91125, USA}

\author[0000-0003-2686-9241]{Daniel Stern}
\affiliation{Jet Propulsion Laboratory, California Institute of Technology, 4800 Oak Grove Drive, MS 169-224, Pasadena, CA 91109, USA}

\author[0000-0002-1233-9998]{David Sanders}
\affiliation{Institute for Astronomy, University of Hawaii, 2680 Woodlawn Drive, Honolulu, HI 96822, USA}



\begin{abstract}

We present a detailed analysis of the most luminous and obscured Active Galactic Nuclei (AGNs) detected in the ultra-hard X-ray band (14--195\,keV) by \textit{Swift}/BAT. Our sample comprises 21 X-ray luminous ($\log L_{\rm X}/\mathrm{erg\,\, s^{-1}}>44.6$, 2--10\,keV) AGNs at $z<0.6$, optically classified as Seyfert 1.9 and 2. Using  \nustar, \xmm, \suzaku, and \chandra data, we constrain AGN properties such as absorption column density $N_{\rm H}$, photon index $\Gamma$, intrinsic $L_{\rm X}$, covering factor, and iron K$\alpha$ equivalent width.
We find median line-of-sight $\log N_{\rm H}/\mathrm{cm^{-2}}=23.5_{-1.2}^{+0.5}$ and 2--10\,keV rest-frame, de-absorbed $\log L_{\rm X}/\mathrm{erg\,\, s^{-1}}=44.7_{-0.6}^{+0.8}$, at the 5$^{th}$ and 95$^{th}$ percentiles. 
For sources with black hole mass estimates (12/20), we find a weak correlation between $\Gamma$ and Eddington ratio ($\lambda_{\rm Edd}$). Of these, six ($50\pm13\%$) lie in the $N_{\rm H}$-$\lambda_{\rm Edd}$ ``forbidden region'' and exhibit a combined higher prevalence of $N_{\rm H}$ variability and outflow signatures, suggesting a transitional phase where AGN feedback may be clearing the obscuring material.
For the 13/21 sources with multi-epoch X-ray spectroscopy, $82^{+6}_{-16}\%$ exhibit variability in either 2-10\,keV flux ($73^{+9}_{-16}\%$) or line-of-sight $N_{\rm H}$ ($33^{+15}_{-10}\%$).
For the 20/21 sources with available near-UV/optical spectroscopy, we detect \nev\ in 17 ($85^{+5}_{-11}$\%), confirming its reliability to probe AGN emission even in heavily obscured systems.
When renormalized to the same \OIII\ peak flux as $z = 2$–9 narrow-line AGNs identified with JWST, our sample exhibits significantly stronger \nev\ emission, suggesting that high-redshift obscured AGNs may be intrinsically weaker in \nev\, or that \nev\ is more challenging to detect in those environments.
The sources presented in this work serve as a benchmark for high-redshift analogs, showing the potential of \nev\ to reveal obscured AGNs and the need for future missions to expand X-ray studies into the high-redshift Universe.

\end{abstract}



\section{Introduction} \label{sec:intro}
Active Galactic Nuclei (AGNs) are among the most energetic objects in the Universe, powered by the accretion of matter onto supermassive black holes (SMBHs) at the centers of their host galaxies \citep{antonucci93,urry95}.
AGNs have been studied across multiple wavelengths, allowing for detailed investigations into how obscuration and extinction impact our understanding of their intrinsic properties \citep[e.g.,][]{padovani17,hickox18,caglar20}.
However, AGNs that are both heavily obscured (with equivalent hydrogen column densities $N_{\rm H} > 10^{23}$ cm$^2$) and highly luminous ($L_{\rm X} > 10^{44}$ erg s$^{-1}$, 2--10\,keV) remain underrepresented in population studies and therefore still not well understood \citep[e.g.,][]{ananna19,ni21,peca24,lamassa24}. This is due to the rarity of highly luminous sources and because of the challenges of detecting heavily obscured objects in the UV, optical, and soft X-ray ($<10$\,keV) bands \citep[e.g.,][]{hickox18,lamassa19_2,peca23,peca24}.
Identifying such AGNs is a critical step in understanding these objects, as they likely represent a crucial phase in galaxy evolution, often linked to rapid SMBH growth and feedback processes that can regulate or suppress star formation in their host galaxies \citep[e.g.,][]{hopkins06,hopkins08,treister12}. In this context, all-sky surveys in ultra-hard X-rays ($>$100\,keV), such as those carried out by the Swift/Burst Alert Telescope \citep{baumgartner13,oh18} and INTEGRAL \citep{bird16,krivonos22}, meet both requirements, providing a nearly unbiased census of AGNs by detecting even the most obscured sources across the entire sky \citep[e.g.,][]{ricci15,ricci17,ananna22}.

The local Universe ($z < 0.6$) provides a unique opportunity to study these AGNs with great detail. Due to their proximity and relative brightness, we can obtain high signal-to-noise ratio (SNR) spectra, including X-rays from telescopes like \nustar\ \citep{harrison13}, \xmm\ \citep{jansen01}, \suzaku\ \citep{mitsuda07}, and \chandra \citep{weisskopf00}. These instruments offer a powerful combination of broad energy coverage and high sensitivity, allowing us to penetrate the obscuring material and characterize the properties of these heavily obscured AGNs. 
High-quality (i.e., high-SNR) spectra with sufficient counts are critical for reliably constraining absorption, which, in turn, enables accurate estimates of intrinsic luminosity. When combined with the black hole mass ($M_{\rm BH}$), this allows the estimate of the Eddington ratio ($\lambda_{\rm Edd}=L_{\rm bol}/L_{\rm Edd}$), which represents a key quantity for understanding how these SMBHs evolve and interact with their host galaxy environment \citep[e.g.,][]{ananna22,ricci22b}.
Furthermore, detailed local studies provide essential benchmarks for interpreting the plethora of new high-redshift ($z>2$) AGN discoveries with the James Webb Space Telescope (JWST), which have proven challenging to characterize at other wavelengths due to limited SNR. Robust local observations thus enable the testing and refinement of theoretical models of AGN evolution and feedback, while also guiding predictions and observational strategies for current and future facilities \citep[e.g.,][]{ananna22, sobolewska11, fabian17, marchesi20, peca24axis, boorman_hexp}.

This work presents a detailed characterization of a sample of heavily obscured and highly luminous AGNs selected from the BAT AGN Spectroscopic Survey (BASS; \citealp{koss17,koss22_dr2overview}). BASS is a large-scale spectroscopic survey of AGNs detected by the \textit{Swift}/BAT telescope in the ultra-hard X-rays (14–-195\,keV), that leverages the vast multiband data available across the full electromagnetic spectrum, including X-ray \citep[e.g.,][]{ricci17,gupta21,marcotulli22,tortosa23}, optical-UV \citep[e.g.,][]{oh22,gupta24}, near-infrared \citep[e.g.,][]{lamperti17,denbrok22,riccif22}, mid-infrared \citep[e.g.,][]{ichikawa17,ichikawa19,pfifle23}, millimeter \citep[e.g.,][]{koss21,kawamuro23,ricci23}, and radio \citep[e.g.,][]{baek19,smith20a,smith20b}. 
By leveraging \nustar’s hard X-ray (3–-79\,keV) capabilities to study obscuration in AGNs \citep[e.g.,][]{marchesi18,lamassa19,kammoun20}, complemented by the softer X-ray coverage of \xmm, \suzaku, and \chandra down to $\sim$0.5\,keV, we investigate the physical properties of these sources, further supported by the multi-band datasets from BASS.
Additionally, using optical spectroscopy and focusing on the \nev\ emission line, we compare our results directly with those emerging from recent JWST-selected AGN samples at high redshift, showing potential differences between local obscured AGNs and distant populations. Finally, we utilize our best-fit spectral models to illustrate how detailed studies of obscured AGNs could be expanded with future X-ray observatories such as \textit{AXIS} and \textit{NewAthena}.

The paper is organized as follows. In Section \S \ref{sec:data_all}, we describe the sample selection and the multi-wavelength data used in this work. Section \S \ref{sec:Xdata} describes the X-ray data reduction process, while Section \S \ref{sec:analysis} outlines the spectral modeling strategy adopted. Results from the X-ray and optical analyses are presented in Sections \S \ref{sec:results_xrays} and \S \ref{sec:nev}, respectively. We discuss our findings in Section \S \ref{sec:discussion}, while a summary is provided in Section \S \ref{sec:summary}.
Throughout this paper, we assumed a $\Lambda$CDM cosmology with the fiducial parameters $H_0=70$ km s$^{-1}$ Mpc$^{-1}$, $\Omega_m=0.3$, and $\Omega_{\Lambda}=0.7$.
Errors are reported at the 90\% confidence level or as the 5$^{th}$ and 95$^{th}$ percentiles if not stated otherwise.
Uncertainties in proportions are computed using the binomial distribution \citep{cameron11}.

\section{Data and sample selection} \label{sec:data_all}

\subsection{Sample selection}
\begin{figure}[t]
    \centering
    \includegraphics[scale=0.55]{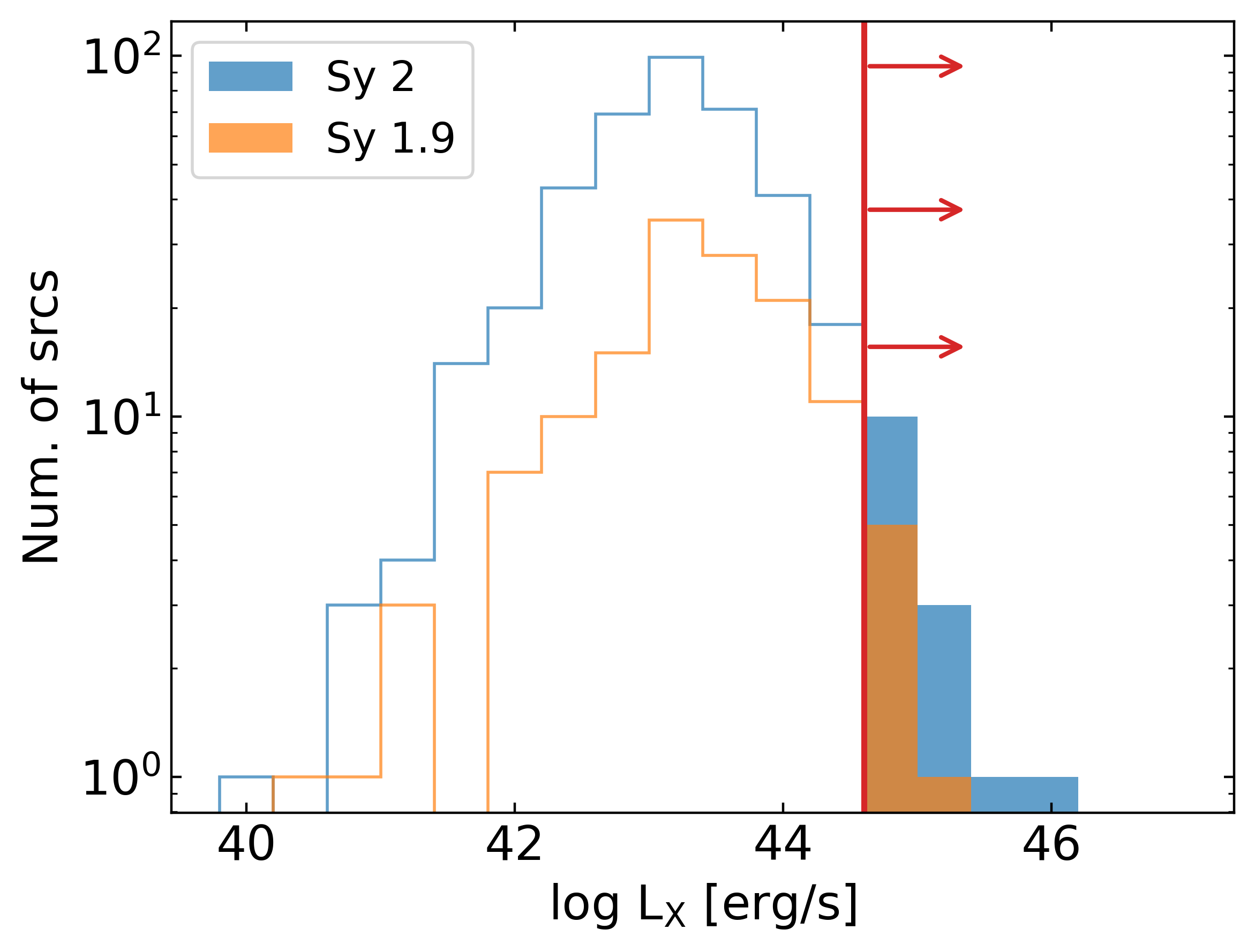}
    \caption{The selected sample consists of 21 AGNs with $\log L_{\rm X}/\mathrm{erg\,\, s^{-1}}>44.6$, indicated by the red line, classified as Seyfert 1.9 and 2 \citep{koss22_dr2catalog}. The filled histograms represent the selected sources, while the full histograms show the full sample of Seyfert 1.9 and 2 (orange and blue, respectively) from BASS DR3 (Koss et al. in prep.). Additional details on the sample selection are provided in the text.}
    \label{fig:sample_selection}
\end{figure}

We selected our sample from the BASS DR2 catalog \citep{koss22_dr2overview,koss22_dr2catalog}, which is based on the \textit{Swift}/BAT 70-month catalog \citep{baumgartner13}, and from the upcoming DR3 release (Koss et al, in prep.), which expands the BASS sample by including newly detected AGNs from the extended \textit{Swift}/BAT 105-month catalog \citep{oh18}.
We selected sources with absorption-corrected 2–10\,keV luminosity $\log L_{\rm X}/\mathrm{erg\,\, s^{-1}}>44.6$. The luminosity values for this selection were derived from the de-absorbed fluxes reported by \citet{ricci17}, who performed a detailed X-ray spectral analysis of sources from the \textit{Swift}/BAT 70-month catalog with redshifts obtained from BASS DR1 \citep{koss17}. For the additional sources in the 105-month catalog, we estimated luminosities based on the 14--195\,keV fluxes. To do so, we converted these to the 2–10\,keV band using a power-law model with a photon index of $\Gamma = 1.8$, which corresponds to the median value for non-beamed AGNs in \cite{ricci17}. This approach leverages the fact that the 14–-195\,keV flux is relatively unaffected by obscuration up to $N_{\rm H} \sim 10^{24}$ cm$^{-2}$ \citep{ricci15,ananna22}, ensuring consistency across catalogs while expanding the sample with the latest \textit{Swift}/BAT data.
To focus on obscured AGNs, we followed the optical classifications in \citet{koss22_dr2catalog} excluding Seyfert 1 objects. This resulted in selecting sources with a narrow H$\beta$ emission line and either a broad (FWHM$>$1000 km s$^{-1}$) or narrow H$\alpha$ emission line, corresponding to Seyfert 1.9 and Seyfert 2, classifications, respectively.
Additionally, we excluded beamed AGNs and removed one source (BAT ID 1240) with high Galactic extinction, E(B-V)=0.99, and Galactic $N_{\rm H}\sim7\times 10^{21}$ cm$^{-2}$ \citep{kalberla05}, as this level of foreground extinction introduces significant uncertainty in the intrinsic optical classification, emission line measurements, and X-ray column density analysis.
Our final sample consists of 21 sources at redshift $z < 0.6$, of which six are optically classified as Seyfert 1.9 and 15 as Seyfert 2 (see Figure \ref{fig:sample_selection}). Of these 21, 12 are from the BASS DR2 catalog, while nine are from the upcoming DR3, effectively almost doubling the number of high-luminosity and optically obscured AGNs.
We expect no additional sources meeting our criteria in the forthcoming BASS DR3 release, as optical spectroscopy has already been obtained for the relevant \textit{Swift}/BAT 105-month sources.

\subsection{X-ray data}
All sources have been observed with \nustar\ at least once, and five of them twice (BAT IDs 32, 80, 199, 505, 714). To provide additional coverage at softer X-ray energies, we incorporated \xmm\ data, where available, for 11 sources (BAT IDs 80, 119, 199, 476, 505, 714, 787, 1204, 1296, 1346, and 1515). Among these, two sources had multiple \xmm\ observations (BAT IDs 80 and 1204). Since BAT ID 1204 has been observed several times, we selected two observations, prioritizing longer exposure times and minimizing the temporal separation between observations to ensure high-SNR spectra while mitigating potential variability-induced biases in the soft X-ray band \citep[e.g.,][]{laha25}.
Given its large collecting area over a broad energy range (0.5–-10\,keV) and its greater availability for our sources compared to other soft X-ray telescopes, \xmm\ was prioritized for soft X-ray coverage when available.
Additionally, for two sources lacking \xmm\ coverage (BAT ID 1248 and BAT ID 32), we included \chandra\ and \suzaku\ observations, respectively, to supplement the dataset.
All the used X-ray observations are detailed in Table \ref{tab:sources}.

\begin{table*}[!ht]

\centering\footnotesize
\begin{tabular}{ccccccccc}
\toprule
\multirow{2}{*}{Source Name} & \multirow{2}{*}{\shortstack{BAT \\ ID}} & \multirow{2}{*}{Redshift} & \multicolumn{3}{c}{XMM} & \multicolumn{3}{c}{NuSTAR} \\
\cmidrule(lr){4-6} \cmidrule(lr){7-9}
               &       &       & Date       & PN/M1/M2                     & Exp.(ks) & Date       & FPMA/B         & Exp.(ks) \\
\midrule
2MASX J00343284-0424117          & 20   & 0.213 & -          & - \,\, - \,\, -                & -         & 2022-07-21 & \cmark \,\,\cmark & 20.7/20.5 \\
2MASX J01290761-6038423          & 80   & 0.203 & 2017-10-31 & \cmark \,\,\cmark \,\,\cmark   & 21.5/27.1 & 2016-12-03 & \cmark \,\,\cmark & 21.1/20.9 \\
                                 &      &       & 2021-06-17 & \cmark \,\,\xmark \,\,\xmark   & 10.3      & 2020-08-30 & \cmark \,\,\cmark & 25.0/24.9 \\
2MASS J02162672+5125251          & 119  & 0.422 & 2006-01-24 & \cmark \,\,\xmark \,\,\xmark   &  8.8      & 2024-12-18 & \cmark \,\,\cmark & 22.7/22.4 \\
LEDA 2816387                     & 199  & 0.108 & 2017-11-20 & \cmark \,\,\cmark \,\,\cmark   & 12.2/19.2 & 2016-05-06 & \cmark \,\,\cmark & 24.6/24.4 \\
                                 &      &       &            & - \,\, - \,\, -                &           & 2021-06-04 & \cmark \,\,\cmark & 26.6/26.5 \\
CXO J095220.1-623234             & 476  & 0.252 & 2009-01-20 & \cmark \,\,\cmark \,\,\cmark   & 12.2/20.3 & 2023-07-12 & \cmark \,\,\cmark & 27.7/27.5 \\
SDSS J102103.08-023642.6         & 494  & 0.294 & -          & - \,\, - \,\, -                & -         & 2023-05-09 & \cmark \,\,\cmark & 19.6/19.4 \\
SDSS J103315.71+525217.8         & 505  & 0.140 & 2018-06-02 & \cmark \,\,\cmark \,\,\cmark   & 23.6/28.0 & 2018-06-02 & \cmark \,\,\cmark & 30.6/30.4 \\
                                 &      &       & -          & - \,\, - \,\, -                & -         & 2020-06-24 & \cmark \,\,\cmark & 23.0/22.8 \\
SDSS J113915.13+253557.9         & 555  & 0.219 & -          & - \,\, - \,\, -                & -         & 2023-01-16 & \cmark \,\,\cmark & 21.1/20.8 \\
LEDA 511869                      & 714  & 0.076 & 2018-02-09 & \cmark \,\,\cmark \,\,\cmark   & 6.3/17.3  & 2016-05-25 & \cmark \,\,\cmark & 23.6/23.4 \\
                                 &      &       &            & - \,\, - \,\, -                &           & 2020-03-25 & \cmark \,\,\cmark & 21.5/21.5 \\
PKS 1549-79                      & 787  & 0.150 & 2008-09-21 & \cmark \,\,\xmark \,\,\xmark   & 45.7      & 2016-07-12 & \cmark \,\,\cmark & 17.3/17.3 \\
RBS 2043                         & 1204 & 0.597 & 2013-11-13 & \cmark \,\,\cmark \,\,\cmark   & 100.2/127.8 & 2021-01-07 & \cmark \,\,\cmark & 19.7/19.6 \\
                                 &      &       & 2013-11-22 & \cmark \,\,\cmark \,\,\cmark   & 80.5/92.1 &     -      &    - \,\, -    &     -     \\
SWIFT J0231.4+3605               & 1241 & 0.321 & -          & - \,\, - \,\, -                & -         & 2024-08-30 & \cmark \,\,\cmark & 20.7/20.5 \\
2MASX J06215493-5214343          & 1291 & 0.209 & -          & - \,\, - \,\, -                & -         & 2024-09-01 & \cmark \,\,\cmark & 19.6/19.4 \\
GALEXASC J063634.15+591319.6     & 1296 & 0.204 & 2025-03-14 & \cmark \,\,\cmark \,\,\cmark   & 15.7/19.7 & 2023-02-12 & \cmark \,\,\cmark & 22.7/22.4 \\
2MASX J09003684+2053402          & 1346 & 0.235 & 2007-04-13 & \cmark \,\,\cmark \,\,\cmark   & 9.9/15.7  & 2024-11-25 & \cmark \,\,\cmark & 22.7/22.5 \\
2MASS J17422050-5146223          & 1515 & 0.218 & 2025-02-27 & \cmark \,\,\cmark \,\,\cmark   & 12.1/17.2 & 2023-03-11 & \cmark \,\,\cmark & 16.1/16.0 \\
SWIFT J2036.0-0028               & 1586 & 0.203 & -          & - \,\, - \,\, -                & -         & 2022-10-13 & \cmark \,\,\cmark & 20.6/20.4 \\
SWIFT J2103.3-2144               & 1595 & 0.289 & -          & - \,\, - \,\, -                & -         & 2022-10-08 & \cmark \,\,\cmark & 19.1/18.9 \\
APMUKS(BJ) B233953.42-593202.5   & 1630 & 0.246 & -          & - \,\, - \,\, -                & -         & 2022-06-22 & \cmark \,\,\cmark & 21.9/21.6 \\

\hline \\

\multirow{2}{*}{Source name} & \multirow{2}{*}{\shortstack{BAT \\ ID}} & \multirow{2}{*}{Redshift}  & \multicolumn{3}{c}{Suzaku} & \multicolumn{3}{c}{NuSTAR} \\
\cmidrule(lr){4-6} \cmidrule(lr){7-9}
               &       &       & Date       & XIS-FI/BI                     & Exp.(ks) & Date       & FPMA/B         & Exp.(ks) \\

\midrule

ESP 39607                        & 32   & 0.201 & 2010-12-19  & \cmark\,\, \cmark                & 60.7/60.7        & 2023-05-08 & \cmark \,\,\cmark & 21.2/21.0 \\
                                 &      &       & -          & - \,\, - \,\, -                & -         & 2024-08-22 & \cmark \,\,\cmark & 21.6/21.4 \\

\hline \\

\multirow{2}{*}{Source name} & \multirow{2}{*}{\shortstack{BAT \\ ID}} & \multirow{2}{*}{Redshift}  & \multicolumn{3}{c}{Chandra} & \multicolumn{3}{c}{NuSTAR} \\
\cmidrule(lr){4-6} \cmidrule(lr){7-9}
               &       &       & Date       & ACIS-S                     & Exp.(ks) & Date       & FPMA/B         & Exp.(ks) \\

\midrule

GALEXMSC J025952.92+245410.8     & 1248  & 0.206 & 2024-11-16 & \cmark                     & 9.9       & 2023-03-26 & \cmark \,\,\cmark & 22.3/22.1 \\
\bottomrule
\end{tabular}
\caption{Data description for each source. From the left: Source name of the multi-band counterpart, BAT ID, redshift. The remaining columns provide the observation dates, camera availability, and exposure times (after data reduction) for \xmm and \nustar (top), \suzaku and \nustar (middle), and \chandra and \nustar (bottom). Additional details on single sources are summarized in Appendix \ref{app:srcs_notes}.}
\label{tab:sources}
\end{table*}

\subsection{Optical data} \label{sec:opt_data}
Existing optical spectra were available for all 21 sources from the BASS survey: 15 from the Very Large Telescope (VLT) X-shooter instrument \citep{xshooter_ref}, four from the Double Spectrograph (DBSP, \citealp{doublespec_ref}) on the 5m Hale telescope at Palomar Observatory, and two from the Sloan Digital Sky Survey (SDSS, \citealp{sdss_ref}). Of these, 9/21 were unpublished in BASS DR2 \cite{oh22}, and 1/21 (BAT ID 476) has been updated with a new VLT/X-shooter spectrum. For these 10 spectra, the fitting was performed following \cite{oh11,oh22} by de-redshifting the spectra and correcting them for Galactic foreground extinction, fitting the continuum emission using galaxy templates, and measuring the emission lines. 

Additionally, for four sources (BAT IDs 555, 1248, 1296, and 1586), new optical integral field unit (IFU) observations were conducted with the Keck Cosmic Web Imager (KCWI, \citealp{kcwi_ref}) on UT 30 March 2024 and UT 8 October 2024. KCWI was configured with the small slicer and the BL and RL gratings, centered at wavelengths of 4500\,\AA\ and 7150\,\AA, respectively, covering spectral ranges of 3420–5600\,\AA\ and 5600–8900\,\AA\ utilizing the 5600 Å dichroic with a total on-source observing time of 30 minutes.  In this setup, the slitlets were 0.35\farcs\ wide in the east-west direction, with a pixel scale of 0.147\farcs/pixel in the north-south direction and an 8.4\arcsec\ × 20.4\arcsec\ field of view. The average spectral resolution was R$\sim$3600 for both gratings. Flux calibration was performed using nightly standards. Data reduction was completed with the standard KCWI DRP (v1.01). For five additional sources (BAT IDs 32, 199, and 1346), observations from the Multi-Unit Spectroscopic Explorer (MUSE) instrument \citep{muse_ref} at the VLT were used. The fully reduced phase 3 ``MUSE-DEEP'' observations from the ESO archive were used\footnote{https://www.eso.org/rm/api/v1/public/releaseDescriptions/102}.

\subsubsection{Black hole masses} \label{sec:mbh}
In this work, 12/21 ($\sim$57\%) black hole masses were available for our sample, all computed using the velocity dispersion method of \cite{koss22_veldisp}.
Of these 12 masses, five (BAT IDs 20, 80, 714, 787, and 1204) were obtained from the BASS DR2 catalog, while the remaining seven (BAT IDs 32, 199, 555, 1248, 1296, 1346, and 1586) were newly estimated with the same procedure using the KCWI and MUSE data.
In brief, the continuum and the absorption features were fit using the penalized PiXel Fitting software (pPXF; \citealp{cappellari04}) to measure the central velocity dispersion for the galaxy, using templates from VLT/X-shooter, with the prominent emission lines masked. Measurements of the \Cahk\ \mgb\ region (3880--5550\,\AA) were performed, as the redshift of the sources causes the \Catrip\ triplet to shift into the near-infrared (NIR). 
We refer the reader to \citet{koss22_veldisp} for further details.

We excluded one $M_{\rm BH}$ measurement (BAT ID 476) from BASS DR2, which reported an unusually low black hole mass due to an apparent broad H$\alpha$ component. New VLT/X-shooter observations, however, show that the H$\alpha$ line is actually narrow, with a strong blue wing originating from \NIIa, rather than a true broad component. Unfortunately, the updated spectrum did not permit a revised $M_{\rm BH}$ estimate.
For this object, as well as the other sources in our sample lacking black hole masses, the available spectroscopic data did not meet the criteria outlined in \citet{koss22_veldisp}, which include adequate SNR and spectral resolution in the stellar absorption features. As a result, reliable $M_{\rm BH}$ estimates could not be obtained for nine sources in our sample.
The new spectra and black hole masses will be included in the forthcoming BASS DR3 release.

\section{Data Reduction} \label{sec:Xdata}

\subsection{\nustar}
\nustar data were reduced using NuSTARDAS v2.1.2. We calibrated and cleaned the downloaded datasets using the standard \texttt{nupipeline} routine.
The spectral extraction was done using the \texttt{nuproducts} command for the two FPMA and FPMB cameras.
We defined circular extraction regions for each source centered on the source position for spectral extraction. Starting from 60\arcsec, roughly corresponding to the \nustar\ half-power diameter (HPD), we manually adapted the extraction radius to optimize the SNR \citep[e.g.,][]{peca21}. Considering all the sources, we used extraction radii in the range [35-60]\arcsec.
The background was extracted from circular regions with a 90\arcsec\ radius near each source, positioned on the same chip while avoiding contamination from nearby sources and CCD gaps. Each background spectrum was inspected to ensure adequate sampling and to avoid empty channels.

\subsection{\xmm}
\xmm data were reduced using SAS v21.0.0, following the standard SAS threads for the EPIC PN, MOS1, and MOS2 cameras\footnote{\href{https://www.cosmos.esa.int/web/xmm-newton/sas-threads}{https://www.cosmos.esa.int/web/xmm-newton/sas-threads}}. In brief, after downloading the observation data files (ODFs), we produced the calibration index files and the summary files using the SAS commands \texttt{cifbuild} and \texttt{odfingest}. We reprocessed the ODFs using the commands \texttt{epproc} and \texttt{emproc} for PN and MOS cameras, respectively. Then, we filtered the events files for the flaring particle background, as described in the same SAS thread.

The spectral extraction procedure was similar to that of \nustar\ data, with source extraction radii starting at 30\arcsec (corresponding to $\sim$90\% of the XMM encircle energy fraction) and then adapted for each source by optimizing the SNR. The final range of extraction radii was [25-40]\arcsec. The background was extracted from circular regions with different radii in the range [65-85]\arcsec, avoiding CCD gaps and nearby sources. Each background spectrum was checked to ensure good sampling and avoid empty channels.
We ran \texttt{evselect} to extract the source and background spectra and calculated appropriate source and background scaling using the \texttt{backscal} command. Finally, the response matrices were obtained with \texttt{arfgen} and \texttt{rmfgen} SAS tasks.

\subsection{\suzaku}
In this work, we primarily used \xmm\ to obtain coverage at energies softer than those observed with \nustar. For one source, ESP 39607 (BAT ID 32), we instead used \suzaku\ due to the unavailability of \xmm\ observations.
\suzaku/XIS data \citep{koyama07} were used for this source. For most of its operational period, the XIS consisted of three cameras: the front-illuminated (FI) XIS 0 and XIS 3, and the back-illuminated (BI) XIS 1 (hereafter BI-XIS).
Following \cite{ricci17}, we reprocessed the data for each XIS camera and extracted spectra from the cleaned event files using a circular aperture with a 1.7\arcmin\ radius centered on the source. The background was from a source-free annulus centered on the source, with inner and outer radii of 3.5 and 5.7\arcmin, respectively. The extraction regions were selected following the same procedure described earlier.
Response matrices were generated using the \texttt{xisrmfgen} and \texttt{xissimarfgen} tasks \citep{ishisaki07}. The spectra from XIS 0 and XIS 3 were then merged using \texttt{mathpha}, \texttt{addrmf}, and \texttt{addarf} HEASoft\footnote{\href{https://heasarc.gsfc.nasa.gov/docs/software/heasoft/}{https://heasarc.gsfc.nasa.gov/docs/software/heasoft/}} tools.

\subsection{\chandra}
For one additional source, GALEXMSC 025952.92+245410.8 (BAT ID 1248), we used \chandra\ due to the unavailability of \xmm\ observations. For the data reduction we used CIAO v4.16.0 \citep{fruscione06} and followed the dedicated thread for point-like sources extraction\footnote{\href{https://cxc.cfa.harvard.edu/ciao/threads/pointlike/}{https://cxc.cfa.harvard.edu/ciao/threads/pointlike/}}. First, we reprocessed the data using the \texttt{chandra\_repro} tool. Then, we extracted the source and background spectra using the \texttt{specextract} command, where we specified the keywords weight=no and correctpsf=yes specific for point-like extraction. To determine source and background extraction regions, we followed the same procedure applied for \nustar\ and \xmm. The source region was defined as a circle with a radius of 2.2\arcsec\, while the background region was defined as an annulus with inner and outer radii of 8\arcsec and 50\arcsec, respectively, avoiding CCD gaps and ensuring good background sampling.

\subsection{Spectral binning}
All the spectra were binned to a minimum of 20 counts/bin, as the number of net counts (i.e., background subtracted) exceeded 200 for all the spectra \citep[e.g.,][]{ricci17}. For \nustar and \chandra the grouping was applied as part of the respective spectral extraction pipelines, for \xmm we used the SAS tool \texttt{specgroup}, and for \suzaku we used the HEASoft \texttt{grppha} tool.

\section{X-Ray Spectral Fitting} \label{sec:analysis}
We used PyXspec v2.1.3\footnote{Equivalent to XSPEC v12.14.0 \citep{xspec}.} \citep{pyxspec} to fit the extracted spectra. All available spectra were fitted simultaneously for each source, using the $\chi^2$ statistic.
We used Markov Chain Monte Carlo (MCMC) simulations for each fit, applying the Goodman-Weare algorithm with 50 walkers, a total chain length of 10$^6$, and a burn-in phase of $10\%$. Each chain started after a first fit to assess the model parameters.
The energy range used for the spectral fitting was 0.5--10\,keV for \xmm, 0.7--10\,keV for \suzaku, and 0.5--7\,keV for \chandra. For \nustar, the lower boundary was 3.0\,keV, while the upper boundary was set in the range [30-50]\,keV depending on where the background starts to dominate the source emission. 

We fitted the sample using the following baseline model:
\begin{multline} \label{eq:main_model}
    Model = phabs \times \left( apec \right. \\
    \left. + C_{AGN} \times \left( Torus\,Model + f_{scatt}\times zcutoffpl \right) \right) ,
\end{multline}
where \textit{phabs} represents the Galactic absorption at the source position \citep{kalberla05}; \textit{apec}\footnote{Solar abundances were assumed \citep{xspec_abundances}.} is a thermal emission component from collisionally-ionized diffuse gas, modeling the soft X-ray emission from the host galaxy; and $C_{AGN}$ allows the primary AGN emission to vary between spectra taken at different epochs. The \textit{Torus Model} encapsulates absorption and reflection components from the obscuring material, as well as the primary cut-off power-law describing the intrinsic AGN continuum (see details in Section~\ref{sec:torus}). We also included a secondary cut-off power-law component (\textit{zcutoffpw}) to account for emission scattered into the line of sight and leaking through the torus. The photon index $\Gamma$ of this secondary power-law was tied to that of the primary AGN emission, since it represents a fraction of the same intrinsic continuum, while its normalization was constrained to vary between 0 and 20\% of the primary component through a multiplicative constant $f_{scatt}$, in line with typical scattering fractions found in previous AGN studies (e.g., \citealp{marchesi16b, gupta21, peca23}).
The cut-off energy of the power-law components is set to 200\,keV, to match what is observed in the local Universe \citep[e.g.,][]{ricci17}.
In principle, an additional constant between the different instruments can be added to consider possible cross-calibration effects. However, we found this constant to be $\sim$1.1 in agreement with other studies (e.g., \citealp{madsen17,balokovic21}). Therefore, we did not include this additional parameter to avoid possible degeneracies (e.g., \citealp{marchesi22}). 

For sources with only \nustar spectra available, the \textit{apec} model and the secondary power-law component were turned off, as these components primarily contribute in the soft X-ray band not covered by \nustar, but accessible with \xmm, \chandra, or \suzaku.
A second \textit{apec} component was added when needed to improve the fit results, as detailed in Section \ref{sec:results_xrays_dic}. Similarly, additional adjustments for individual sources were applied to improve the fit, as detailed in Appendix \ref{app:srcs_notes}.

The free parameters in our baseline model are the temperature and normalization of the \textit{apec} component, the main power-law normalization, the photon-index $\Gamma$, the line-of-sight $N_{\rm H}$, and the relative secondary power-law normalization, also known as the scattering fraction, $f_{scatt}$. Additional free parameters from each torus model are included, as detailed in Section~\ref{sec:torus}.
Furthermore, for sources with multiple observations, we allowed $C_{AGN}$ and the line-of-sight $N_{\rm H}$ to vary between different epochs. This allows for accounting for possible intrinsic flux variations and potential occultation events caused by the obscuring material. To avoid degeneracies between parameters, the photon index $\Gamma$ was linked between the different epochs \citep[e.g.,][]{torres-alba23,pizzetti25}.

\subsection{Torus modelling} \label{sec:torus}
Several physically motivated torus models have been developed to study obscuration in AGNs, with many examples such as borus02 \citep{balokovic18}, BNsphere \citep{brightman11}, and XCLUMPY \citep{tanimoto19,tanimoto20}. The need of multiple models arises from differences in their geometrical assumptions and treatment of obscuration, as highlighted in several studies \citep[e.g.,][]{saha22, kallova24, boorman25}. For this work, we selected the widely used MYTorus \citep{murphy09}, RXTorusD \citep{reflex}, which includes the interaction between X-ray photons and dust grains, and UXCLUMPY \citep{buchner19}, a clumpy torus model able to reproduce eclipsing events in the obscurer. A detailed description of these models is given below.

\subsubsection{MYTorus}
The MYTorus model \citep{murphy09} features a cylindrical, azimuthally symmetric torus with a fixed half-opening angle (i.e., the angular extent of the region where the torus does not obscure the central AGN) of $60^\circ$, filled with a uniform, neutral, cold reprocessing material. This model integrates the main components of an obscured AGN X-ray spectrum, such as obscuration, reflection, and fluorescent emission line components, using three distinct tables to treat them consistently: the zero-order continuum, the scattered component, and the emission lines component (See Eq. \ref{eq:mytorus_dec} and Eq \ref{eq:mytorus_coup}).

First, we employed the model in its decoupled configuration \citep{yaqoob12,yaqoob15}, which separates the line-of-sight column density ($N_{\rm H, los}$) from the average column density ($N_{\rm H, ave}$). In this configuration, the zeroth-order continuum (i.e., the photons that escape the torus without scattering) is independent of the inclination angle, fixed at $\theta_i = 90^\circ$. This makes the zeroth-order continuum a line-of-sight quantity unaffected by the torus geometry. We consider both edge-on and face-on scenarios to account for potential patchiness and varied torus configurations, as well as resulting Compton-scattering and emission line features. In the edge-on case, with $\theta_{i,S,L} = 90^\circ$, forward scattering is simulated and weighted by $A_{S,L90}$, representing a more uniform torus where photons are primarily reprocessed by the material between the AGN and the observer. In the face-on case, with $\theta_{i,S,L} = 0^\circ$, backward scattering is modeled and weighted by $A_{S,L0}$, indicating a patchier torus structure where photons scattered from the back of the torus have fewer interactions before reaching the observer. When $A_{S,L90}$ and $A_{S,L0}$ can vary freely, the configuration is called ``decoupled''. 
In XSPEC syntax:
\begin{multline} \label{eq:mytorus_dec}
Torus\,Model = mytorus Ezero v00.fits \times zcutoffpw + \\
             A_{S,0} \times mytorus scatteredH200 v00.fits + \\
             A_{L,0} \times mytl V000010nEp000H200 v00.fits + \\
             A_{S,90} \times mytorus scatteredH200 v00.fits + \\
             A_{L,90} \times mytl V000010nEp000H200 v00.fits.
\end{multline}
where the Compton-scattered and emission line components are weighted differently using the multiplicative constants $A_S$ and $A_L$, respectively.
The additional free parameters when using this torus model are $A_{S,L,90}$, $A_{S,L,0}$, and the average $N_{\rm H, ave}$. These parameters are linked between the spectra when multi-epoch spectra are fitted simultaneously.

Second, we utilized the MYTorus model in its ``coupled'' configuration, where the inclination angle, $\theta_i$ is allowed to vary freely: 
\begin{multline} \label{eq:mytorus_coup}
Torus\,Model = mytorus Ezero v00.fits \times zcutoffpw + \\
             A_{S,\theta} \times mytorus scatteredH200 v00.fits + \\
             A_{L,\theta} \times mytl V000010nEp000H200 v00.fits.
\end{multline}
In this configuration, the scattered and emission line components represent the same inclination angle, and the only absorption parameter is the $N_{\rm H, los}$, therefore reducing the number of model components.
The additional free parameters of this torus model are $\theta_i$, $A_{S,\theta}$, and $A_{L,\theta}$. As with the decoupled configuration, these additional parameters were tied across epochs when fitting multi-epoch spectra simultaneously.

\subsubsection{RXTorusD}
Next, we used the RXTorusD model \citep{reflex}, which is based on the ray-tracing code for X-ray reprocessing, RefleX \citep{paltani17}. This model includes absorption and reflection from the torus with varying torus covering factors. In particular, the covering factor is defined as the ratio of the minor to the major axis of the torus ($r/R$). In this model, the $N_{\rm H, los}$ is directly linked to the the equatorial column density ($N_{H,eq}$) via:
\begin{equation}
\begin{cases}
N_{\rm H, los}(\theta_i) = N_{H,eq} \left(1 - \frac{R^2}{r^2} \cos^2 \theta_i \right)^{1/2}, & \cos \theta_i < \frac{r}{R}, \\
N_{\rm H, los}(\theta_i) = 0, & \cos \theta_i \geq \frac{r}{R},
\end{cases}
\end{equation}

In its new version, RXTorusD considers the different effects of the interaction between X-ray photons and dust grains, such as dust scattering, near-edge X-ray absorption fine structures, and shielding. This model accounts for changes in the cross-sections of photon-gas interactions based on the fraction of metals in dust grains (the dust depletion factor). It also includes other physical processes, such as Rayleigh scattering and molecular gas scattering, which lead to significant differences in the predicted X-ray spectra for the same set of geometrical and physical parameters \citep{reflex}.

We adopted the following torus configuration\footnote{From \href{https://www.astro.unige.ch/reflex/xspec-models}{https://www.astro.unige.ch/reflex/xspec-models}}, which accounts for reprocessed and continuum emission, respectively:
\begin{multline} \label{eq:reflex}
Torus\,Model = RXTorus\_rprc.mod + \\
             RXTorus\_cont.mod \times zcutoffpw.
\end{multline}

The additional free parameters for this torus model component are the inclination angle 
$\theta_i$ and $r/R$, which are linked between the spectra when multi-epoch observations are fitted simultaneously.

\subsubsection{UXCLUMPY}
The last model we used was UXCLUMPY \citep{buchner19}. This model is designed to reproduce and model the column density and cloud eclipsing events in AGN tori, considering their angular sizes and frequency. UXCLUMPY differs from the previously mentioned models by including the torus clumpiness and cloud dispersion. The model aims to reproduce a cloud distribution with various hydrogen column densities based on observed eclipse event rates, assuming the clouds follow circular Keplerian orbits on random planes for simplicity. The distribution's dispersion is controlled by the parameter TOR$\sigma$ ($\sigma \in [0-84]$), where a higher value indicates a larger dispersion and covering factor of the clouds.
To model strong reflection features, UXCLUMPY includes an additional inner ring of Compton-thick ($\log N_{\rm H}/\mathrm{cm^{-2}}>24$) material, whose covering factor is measured by the parameter CTKcover ($\in [0-0.6]$). TOR$\sigma$ and CTKcover together provide a robust framework to probe the torus geometry by modeling the cloud distribution and the extent of the reflecting material. 
In the geometries discussed previously, $N_{\mathrm{H, ave}}$ and $N_{H,eq}$ are closely related to $N_{\mathrm{H, los}}$. In contrast, UXCLUMPY adopts a unified obscurer model, where a single clumpy torus geometry defined by TOR$\sigma$ and CTKcover can be observed under a wide range of $N_{\mathrm{H, los}}$ ($10^{20} - 10^{26}$ cm$^{-2}$). The used configuration\footnote{\url{https://github.com/JohannesBuchner/xars/blob/master/doc/uxclumpy.rst}} is: 
\begin{multline} \label{eq:uxclumpy}
Torus\,Model = uxclumpy{\rm-}cutoff.fits + \\
             f_{scatt} \times uxclumpy{\rm-}cutoff{\rm-}omni.fits.
\end{multline}
the first table considers all the emissions from the primary component, and the second table accounts for the secondary power-law emission, therefore replacing the last term of Eq. \ref{eq:main_model}.
The additional free parameters of this torus component are the inclination angle $\theta_i$, TOR$\sigma$, and CTKcover, which are linked between the spectra when multi-epoch spectra are fitted simultaneously.

\section{Results I: X-ray characterizzation} \label{sec:results_xrays}

\subsection{Determine the best-fit model} \label{sec:results_xrays_dic}
When fitting data with different models, selecting the best-fit model is a critical step, and several methods can be employed for this purpose \citep[e.g.,][]{buchner14,boorman25}. For instance, several works \citep[e.g.,][]{arcodia18,sicilian22,peca23} emphasized the importance of using information criteria such as the Akaike Information Criterion (AIC; \citealp{akaike1974}) and the Bayesian Information Criterion (BIC; \citealp{schwarz1978}), which are applicable to both nested and non-nested models. These tests are likelihood-based, relying on the maximum likelihood obtained through chi-squared minimization, and they include a penalty term based on the number of free parameters to mitigate overfitting. While these criteria provide a reliable method for model selection, they do not consider the full parameter space of the likelihood distribution.
In this work, we adopted the Deviance Information Criterion (DIC; \citealp{Spiegelhalter_2002}) as our model selection metric. DIC is a Bayesian model selection approach that, like AIC and BIC, balances goodness of fit (from the likelihood function) and model complexity (the number of free parameters). However, DIC generalizes these criteria by incorporating the full posterior probability distribution, rather than relying on point estimates, to evaluate the best-fit model \citep{wilkins22}. 
To fully explore the posterior distributions required for DIC, we utilize the MCMC chains from the fitting procedure. 
The DIC is defined as:

\begin{equation}
\text{DIC} = \overline{D(\theta)} + p_{DIC} ,
\end{equation}
\noindent
where \( D(\theta) = -2\log L(\theta) \) is the deviance, \( \overline{D(\theta)} \) is the expectation of the deviance over the posterior distribution of the parameters, and \( p_{DIC} \) is the effective number of parameters. We used PyXspec to calculate the DIC, where \( p_{DIC} \) is determined according to the definition provided by \citet{gelman04}, \begin{equation}
p_{DIC} = \frac{1}{2} \overline{\mathrm{var}(D(\theta))} ; 
\end{equation}
i.e., half the variance of \( D(\theta) \) over the chain. 
The model with the lowest DIC is preferred \citep{Spiegelhalter_2002,gelman04}. 
The DIC criterion favors the UXCLYMPY model for 9/21 sources, the MYTorus (coupled) model for 5/21 sources, the RXTorusD model for 4/21 sources, and the MYTorus (decoupled) model for 3/21 sources. The best-fit spectra are shown in Appendix \ref{app:srcs_notes}. 

In addition to this approach, we also tested simplified modeling where some of the geometrical parameters of the torus models were fixed. Specifically, we compared our baseline setup to a configuration where the inclination angle was fixed at 75$^{\circ}$ for MYTorus, with default values of r/R=0.5 for RXTorusD, and TOR$\sigma$=28 and CTKcover=28 for UXCLUMPY. For each pair of models (e.g., RXTorusD with and without free geometrical parameters), we evaluated the differences in DIC.
A threshold of $\Delta$DIC$>$2 was used, as it is considered to indicate ``substantial'' support for one model over another \citep[e.g.,][]{burnham2002model}. In all cases, the models with fixed parameters were not preferred ($\Delta$DIC$<$2), supporting our choice to allow these geometrical parameters to vary, at least from a statistical perspective.

\begin{table*}[!ht]
\centering
\footnotesize 

\begin{tabular}{cccccccccc}
\hline
\multirow{2}{*}{\shortstack{BAT \\ ID}} & Redshift &  Type  & Flux                    & $\mathrm{\Gamma}$  & log $\mathrm{N_H}$ & log $\mathrm{L_X}$ & log $\mathrm{\lambda_{Edd}}$\\
       &          &        & ($\mathrm{erg/s/cm^{2}}$)   &                      & ($\mathrm{cm^{-2}}$)& ($\mathrm{erg/s}$)  &  \\
\hline
20 & 0.213 & Sy2     & $27.2_{-0.5}^{+0.8}$   & $1.71_{-0.19}^{+0.21}$ & $23.7_{-0.3}^{+0.5}$ & $45.0_{-0.3}^{+0.3}$ & $-0.8_{-0.2}^{+0.2}$ &  \\
32 & 0.201 & Sy2     & $14.7_{-0.6}^{+0.5}$   & $1.72_{-0.10}^{+0.11}$ & $23.6_{-0.1}^{+0.2}$ & $44.7_{-0.1}^{+0.1}$ & $-0.3_{-0.2}^{+0.2}$ &  \\
80 & 0.203 & Sy2     & $4.9_{-0.3}^{+0.2}$    & $1.64_{-0.19}^{+0.23}$ & $24.0_{-0.1}^{+0.2}$ & $44.4_{-0.2}^{+0.3}$ & $-1.3_{-0.2}^{+0.2}$ &  \\
119 & 0.422 & Sy2    & $56.9_{-4.3}^{+3.3}$   & $1.72_{-0.05}^{+0.06}$ & $22.5_{-0.2}^{+0.2}$ & $45.5_{-0.1}^{+0.1}$ & ... &  \\
199 & 0.108 & Sy1.9  & $8.5_{-0.4}^{+0.3}$    & $1.71_{-0.12}^{+0.14}$ & $23.9_{-0.1}^{+0.1}$ & $44.4_{-0.2}^{+0.2}$ & $-0.4_{-0.2}^{+0.2}$ &  \\
476 & 0.252 & Sy1.9$^{*}$    & $41.5_{-1.1}^{+0.8}$   & $1.85_{-0.07}^{+0.07}$ & $23.0_{-0.1}^{+0.1}$ & $45.0_{-0.1}^{+0.1}$ & ... &  \\
494 & 0.294 & Sy2    & $10.2_{-5.4}^{+2.5}$   & $2.29_{-0.40}^{+0.46}$ & $23.4_{-0.4}^{+0.3}$ & $45.2_{-0.6}^{+0.7}$ & ... &  \\
505 & 0.14  & Sy2    & $3.4_{-0.3}^{+0.2}$    & $1.85_{-0.15}^{+0.13}$ & $24.1_{-0.2}^{+0.6}$ & $44.1_{-0.1}^{+0.1}$ & ... &  \\
555 & 0.219 & Sy2    & $31.1_{-1.7}^{+1.1}$   & $1.88_{-0.20}^{+0.31}$ & $22.8_{-0.5}^{+0.4}$ & $44.7_{-0.3}^{+0.3}$ & $-0.9_{-0.2}^{+0.2}$ &  \\
714 & 0.076 & Sy2    & $10.3_{-0.5}^{+0.3}$   & $1.60_{-0.10}^{+0.11}$ & $23.9_{-0.1}^{+0.1}$ & $44.0_{-0.1}^{+0.1}$ & $-1.5_{-0.2}^{+0.2}$ &  \\
787 & 0.15  & Sy2    & $65.5_{-28.8}^{+23.4}$ & $1.77_{-0.03}^{+0.05}$ & $22.5_{-0.1}^{+0.1}$ & $44.7_{-0.1}^{+0.1}$ & $-0.4_{-0.2}^{+0.2}$ &  \\
1204 & 0.597 & Sy2   & $58.4_{-0.7}^{+0.5}$   & $1.72_{-0.07}^{+0.08}$ & $23.5_{-0.1}^{+0.1}$ & $45.7_{-0.1}^{+0.1}$ & $-0.3_{-0.2}^{+0.2}$ &  \\
1241 & 0.321 & Sy2   & $5.0_{-2.5}^{+1.1}$    & $2.14_{-0.41}^{+0.46}$ & $23.5_{-0.5}^{+0.3}$ & $44.9_{-0.7}^{+0.6}$ & ... &  \\
1248 & 0.206 & Sy2   & $10.9_{-5.3}^{+4.0}$   & $1.95_{-0.26}^{+0.36}$ & $23.4_{-0.1}^{+0.2}$ & $44.7_{-0.4}^{+0.5}$ & $-0.2_{-0.2}^{+0.2}$ &  \\
1291 & 0.209 & Sy2   & $5.3_{-2.6}^{+1.1}$    & $1.81_{-0.27}^{+0.55}$ & $23.5_{-0.4}^{+0.3}$ & $44.5_{-0.5}^{+0.8}$ & ... &  \\
1296 & 0.204 & Sy1.9 & $22.0_{-1.1}^{+0.5}$   & $1.60_{-0.08}^{+0.08}$ & $22.6_{-0.2}^{+0.2}$ & $44.6_{-0.1}^{+0.1}$ & $-0.9_{-0.3}^{+0.2}$ &  \\
1346 & 0.235 & Sy2   & $24.7_{-0.5}^{+0.6}$   & $1.74_{-0.14}^{+0.14}$ & $23.6_{-0.1}^{+0.3}$ & $44.7_{-0.2}^{+0.2}$ & $-1.1_{-0.2}^{+0.2}$ &  \\
1515 & 0.218 & Sy1.9 & $27.2_{-2.6}^{+1.6}$   & $1.72_{-0.03}^{+0.03}$ & $21.6_{-0.8}^{+0.5}$ & $44.6_{-0.6}^{+0.1}$ & ... &  \\
1586 & 0.203 & Sy1.9 & $23.0_{-1.2}^{+0.4}$   & $1.86_{-0.13}^{+0.15}$ & $23.2_{-0.2}^{+0.5}$ & $44.7_{-0.2}^{+0.2}$ & $-0.3_{-0.2}^{+0.2}$ &  \\
1595 & 0.289 & Sy1.9 & $19.7_{-2.0}^{+1.9}$   & $1.87_{-0.15}^{+0.19}$ & $22.3_{-0.8}^{+0.4}$ & $44.9_{-0.2}^{+0.8}$ & ... &  \\
1630 & 0.246 & Sy2   & $4.3_{-1.8}^{+0.7}$    & $2.16_{-0.36}^{+0.55}$ & $23.7_{-0.3}^{+0.3}$ & $44.8_{-0.5}^{+0.9}$ & ... &  \\
\hline
\end{tabular}

\vspace{5mm}

\begin{tabular}{ccccccccc}
\hline
\multirow{2}{*}{\shortstack{BAT \\ ID}} & $\mathrm{\theta_{i}}$ & $\mathrm{f_{scatt}\%}$ & C.F. &  kT    &  kT   & Best Model & $\mathrm{\chi^2/dof}$ \\
       & ($\deg$)                &                         &      & (keV)  & (keV) &            &                         \\
\hline
20 & $75_{-18}^{+10}$   & ...                  & $0.5_{-0.2}^{+0.4}$ & ... & ... & RXTorusD & 149.8/172 &  \\
32 & $70_{-8}^{+16}$    & $0.9_{-0.2}^{+0.3}$  & ...                 & $0.4_{-0.1}^{+0.1}$ & ... & MYTorus-coup. & 479.9/473 &  \\
80 & $69_{-7}^{+17}$    & $0.2_{-0.1}^{+0.2}$  & ...                 & $0.8_{-0.5}^{+0.7}$ & ... & MYTorus-coup. & 195.2/191 &  \\
119 & $70_{-19}^{+14}$  & $0.7_{-0.4}^{+0.4}$  & $0.8_{-0.2}^{+0.2}$ & ... & ... & RXTorusD & 536.9/565 &  \\
199 & $69_{-7}^{+8}$    & $0.7_{-0.3}^{+0.4}$  & $0.4_{-0.2}^{+0.3}$ & $0.2_{-0.1}^{+0.1}$ & $0.9_{-0.1}^{+0.2}$ & RXTorusD & 343.5/314 &  \\
476 & ...               & $0.6_{-0.4}^{+0.3}$  & ...                 & $0.6_{-0.6}^{+2.2}$ & ... & MYTorus-dec. & 613.8/639 &  \\
494 & $35_{-26}^{+39}$  & ...                  & $>0.21$             & ... & ... & UXCLUMPY & 36.9/31 &  \\
505 & ...               & $0.5_{-0.2}^{+0.3}$  & ...                 & $0.6_{-0.5}^{+1.0}$ & ... & MYTorus-dec. & 107.9/118 &  \\
555 & ... & ... & ... & ... & ... & MYTorus-dec. & 133.4/140 &  \\
714 & $>72$             & $0.3_{-0.1}^{+0.1}$  & ... & ... & ... & MYTorus-coup. & 251.1/233 &  \\
787 & $>77$             & $2.3_{-0.7}^{+2.6}$  & $0.4_{-0.1}^{+0.1}$ & ... & ... & UXCLUMPY & 1187.2/1209 &  \\
1204 & $>83$            & $14.4_{-6.8}^{+2.5}$ & ...                 & $1.9_{-0.9}^{+5.0}$ & $7.1_{-2.6}^{+2.4}$ & MYTorus-coup. & 7433.6/7019 &  \\
1241 & $42_{-31}^{+34}$ & ...                  & $0.6_{-0.2}^{+0.4}$ & ... & ... & UXCLUMPY & 21.2/18 &  \\
1248 & $42_{-30}^{+32}$ & $<17.6$              & $0.7_{-0.4}^{+0.3}$ & ... & ... & UXCLUMPY & 62.0/70 &  \\
1291 & $43_{-31}^{+33}$ & ...                  & $0.6_{-0.4}^{+0.2}$ & ... & ... & UXCLUMPY & 32.7/28 &  \\
1296 & $53_{-31}^{+23}$ & $7.9_{-5.0}^{+5.5}$  & $0.5_{-0.2}^{+0.2}$ & ... & ... & UXCLUMPY & 352.9/327 &  \\
1346 & $77_{-12}^{+9}$  & $3.7_{-3.7}^{+8.4}$  & ...                 & $1.3_{-0.1}^{+6.1}$ & $3.1_{-0.5}^{+0.1}$ & MYTorus-coup. & 958.3/952 &  \\
1515 & $64_{-36}^{+12}$ & ...                  & $0.5_{-0.1}^{+0.1}$ & ... & ... & UXCLUMPY & 1054.3/1009 &  \\
1586 & $76_{-36}^{+10}$ & ...                  & $0.4_{-0.3}^{+0.3}$ & ... & ... & RXTorusD & 111.2/118 &  \\
1595 & $58_{-23}^{+30}$ & ...                  & $0.5_{-0.3}^{+0.2}$ & ... & ... & UXCLUMPY & 74.3/81 &  \\
1630 & $48_{-34}^{+31}$ & ...                  & $0.6_{-0.4}^{+0.1}$ & ... & ... & UXCLUMPY & 18.3/20 &  \\
\hline
\end{tabular}

\caption{Results from the spectral analysis and derived quantities. \textit{Top table}, from the left: BAT ID, redshift and optical classification from \cite{koss22_dr2catalog} ($^*$: for BAT ID 476 see Section \ref{sec:mbh}), observed 2--10\,keV flux in units of 10$^{-13}$ erg/s/cm$^2$, photon index $\Gamma$, logarithm of the line-of-sight $N_{\rm H}$, logarithm of the intrinsic (rest-frame and absorption corrected) 2--10\,keV luminosity, and logarithm of the Eddington ratio $\lambda_{\rm Edd}$.
\textit{Bottom table}, from the left: Inclination angle $\mathrm{\theta_i}$, percentage of scattering fraction $\mathrm{f_{scatt}\%}$, covering factor C.F., temperature of the 
\textit{apec} components, chosen best-fit model, and best-fit chi-squared over degrees of freedom. Note that the inclination angle for the MYTorus model in decoupled configuration is not reported, as it is not a parameter of the model, nor is it the covering factor for both MYTorus models (see text for details).} \label{tab:best_results}
\end{table*}

Table \ref{tab:best_results} summarizes the model selection and best-fit results. Parameters and associated uncertainties are derived from the MCMC chains and reported as the 50$^{\rm th}$ percentile along with the 5$^{\rm th}$ and 95$^{\rm th}$ percentiles. All luminosities presented hereafter are intrinsic, meaning they are corrected for absorption and k-corrected to the rest-frame 2--10\,keV band.
For sources with multi-epoch observations, we report the average values of $N_{\rm H, los}$ (hereafter referred to simply as $N_{\rm H}$) and $L_{\rm X}$, as our primary goal is to characterize the global properties of the sample. We verified that the $N_{\rm H}$ values from the different torus models are consistent within the uncertainties (see Appendix~\ref{app:srcs_notes}).
Single-epoch measurements are also provided and discussed in Section \ref{sec:variability}.
The median values for the full sample are $\log N_{\rm H}/\mathrm{cm}^{-2} = 23.5_{-1.2}^{+0.5}$, $\log L_{\rm X}/\mathrm{erg\,s}^{-1} = 44.7_{-0.6}^{+0.8}$, and $\Gamma = 1.78_{-0.17}^{+0.39}$. 
For sources overlapping with the analysis of \citet{ricci17}, we find good agreement, as detailed in Appendix~\ref{app:ricci_comp}.

Out of the initial selection, we report that five sources are slightly below the luminosity threshold used for the sample selection, but still in the high luminosity regime ($L_{\rm X} > 10^{44}$ erg/s). This minor discrepancy is likely due to using different datasets and models for the spectral analysis, as well as possible variability, given that the BAT data represent an average flux since \textit{Swift}’s launch in 2004.
We also find that one source (BAT ID 1515) is not classified as X-ray obscured ($\log N_{\rm H}/\mathrm{cm^{-2}} < 22$); however, this is consistent with the fact that sources classified as Seyfert 1.9 might indeed exhibit low levels of X-ray obscuration \citep[e.g.,][]{panessa02,oh22}.
This corresponds to an X-ray obscured fraction for high-luminosity, Seyfert 1.9 and 2 sources of $95^{+2}_{-9}\%$ in our sample, consistent with the $85^{+4}_{-7}\%$ ($91^{+3}_{-16}\%$) from \citet{ricci17} for the same optical selection and $\log L_{\rm X}/\mathrm{erg\,s}^{-1} > 44$ ($>44.6$).

\subsection{Photon index versus Eddington ratio}
We investigate the relationship between the Eddington ratio ($\lambda_{\rm Edd}$) and the X-ray photon index ($\Gamma$), which has been extensively studied by several works \citep[e.g.,][]{shemmer06,shemmer08,risaliti09,brightman13,brightman16,trakhtenbrot17,liu21_lambda,huang20,kamraj22}, using samples selected in different ways. For instance, \citet{risaliti09} analyzed a sample of optically selected SDSS AGNs up to redshift 4.5, while \citet{brightman13} focused on unobscured AGNs in the COSMOS field up to redshift 2.5. This was later extended by \citet{brightman16}, who studied megamaser Compton-thick AGNs in the local Universe using \nustar, and by \citet{trakhtenbrot17},  who analyzed the full BASS DR1 sample leveraging the hard X-ray selection of \textit{Swift}/BAT. More recent studies have further explored this relation for low- and high-Eddington regimes, with \citet{huang20} and \citet{liu21_lambda} finding steeper slopes or unobscured AGNs in the high-Eddington range at relatively low redshifts ($\lesssim 0.5$). In contrast, \citet{kamraj22} analyzed unobscured AGNs including \nustar\ data, and found no strong evidence of a significant correlation.
These mixed results suggest that the $\lambda_{\rm Edd}$-$\Gamma$ relation may not be universal and could depend on factors such as sample selection, luminosity range, X-ray energy coverage, spectral modeling techniques, and the methods used to determine $\lambda_{\rm Edd}$. 

Two primary explanations have been proposed to account for this correlation. The first links the observed trend to the impact of accretion disk radiation on the AGN corona \citep[e.g.,][]{shemmer08,trakhtenbrot17}. At higher $\lambda_{\rm Edd}$, intense UV/optical emission from the accretion disk enhances the cooling of the corona via Comptonization, lowering its temperature and optical depth, and producing steeper X-ray spectra. Conversely, at lower $\lambda_{\rm Edd}$, the reduced photon supply results in a hotter and more compact corona, leading to flatter X-ray spectra.
The second, alternative explanation, was proposed by \citet{ricci18}, who showed through simulations that runaway pair production could drive this trend. They suggested that AGNs avoid the region in the temperature-compactness parameter space where pair production becomes significant, resulting in the temperature of the Comptonized plasma decreasing with the Eddington ratio. This effect can therefore explain the $\lambda_{\rm Edd}$-$\Gamma$ trend, as the photon index is related to the temperature of the plasma \citep[e.g.,][]{ricci18,laha25}.

We computed the Eddington ratio, $\lambda_{\rm Edd} = L_{\rm bol}/L_{\rm Edd}$, where $L_{\rm bol}$ is the bolometric luminosity and $L_{\rm Edd}$ is the Eddington luminosity, given by $L_{\rm Edd} = 1.26 \times 10^{38} \times M_{\rm BH}/M_{\odot}$ erg/s. 
$L_{\rm bol}$ was computed from the X-ray luminosities derived in this work using the \cite{duras20} bolometric correction. 
The median bolometric luminosity is $\log L_{\rm bol}/{\rm erg}\,\,{\rm s}^{-1}\sim 46.1$.
12/21 sources in our sample have estimated $M_{\rm BH}$, and therefore $\lambda_{\rm Edd}$. The uncertainties in  $\lambda_{\rm Edd}$  were estimated by combining in quadrature the error on  $L_{\rm X}$ and the typical scatter in $M_{\rm BH}$, which is dominated by systematics rather than uncertainties in emission line fitting \citep{koss22_veldisp}. For BASS DR2, the estimated uncertainty on $M_{\rm BH}$ is $\sim$0.5 dex \citep{riccif22, caglar23}, reflecting the intrinsic scatter in virial and velocity dispersion-based methods.
The median value across our sample is $\log \lambda_{\rm Edd} = -0.59$.

We found a weak correlation between $\lambda_{\rm Edd}$ and $\Gamma$.  Using the \texttt{scipy.stats} module, we computed Spearman, Pearson, and Kendall’s tau correlation coefficients of 0.59, 0.60, and 0.55, with corresponding p-values of 0.043, 0.036, and 0.014, respectively. These values correspond to a significance level between 2 and $3\sigma$, which we consider indicative of a weak correlation.
We performed a linear fit using the MCMC tool \texttt{emcee} \citep{emcee}, explicitly accounting for asymmetric uncertainties in both x and y. To do so, we carried out 1000 Monte Carlo realizations, in which the data points were randomly drawn from two half-Gaussian distributions centered on the best-fit value, with widths defined by the respective upper and lower uncertainties\footnote{\href{https://github.com/alessandropeca/BALinFit}{https://github.com/alessandropeca/BALinFit}}. This approach ensures proper propagation of asymmetric measurement errors into the regression analysis.
The best-fit slope and intercept are $0.14_{-0.11}^{+0.13}$ and $1.85_{-0.09}^{+0.11}$, respectively, where the uncertainties are reported at 1$\sigma$. The standard deviation is 0.04 dex. These results are consistent when fitting with \texttt{linmix} \citep{kelly07}, a hierarchical Bayesian model that accounts for error bars in both x and y directions. Since \texttt{linmix} requires symmetric uncertainties, we approximated them by averaging the asymmetric errors, yielding $0.14\pm0.11$ and $1.81 _{-0.08}^{+0.09}$, for slope and intercept, respectively.
Due to the limited number of data points available, we did not attempt to separate our sample into sub- and super-Eddington regimes.

\begin{figure}[!tp]
    \centering
    \includegraphics[scale=0.55]{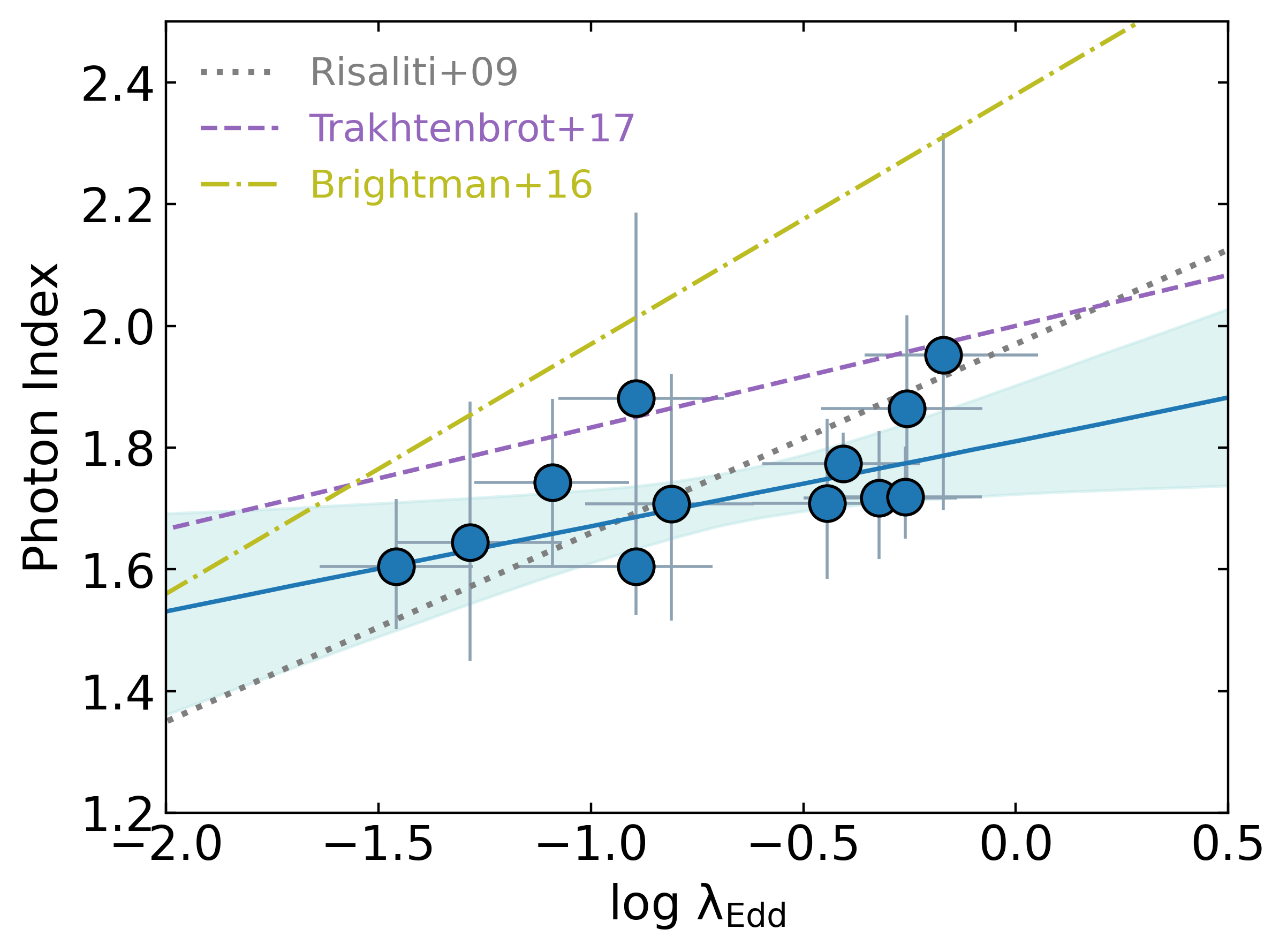}
    \caption{Eddington ratio ($\lambda_{\rm Edd}$) versus photon index ($\Gamma$) from our sample (blue points), with the best-fit linear relation ($y = 0.14x - 1.85$) shown as a blue line. The shaded region represents the 1$\sigma$ uncertainty. For comparison, the best-fit relations from \citealp{risaliti09} (grey dotted line), \citealp{trakhtenbrot17} (purple dashed line), and \citealp{brightman16} (olive dotted-dashed line) are shown. Note that \citealp{trakhtenbrot17} used three methods to compute the best fit; we report results for the two methods that agree with each other (see \citealp{trakhtenbrot17} for details). Comparisons with other works are discussed in the text and omitted here for clarity.}
    \label{fig:gamma_edd}
\end{figure}

Our results are shown in Figure \ref{fig:gamma_edd}. Within the errors, we found agreement with the results of \citet{trakhtenbrot17}, who analyzed the full BASS DR1 sample, suggesting a similar behavior for both obscured AGNs (our sample) and the broader AGN population. Similarly, \citet{brightman16} analyzed a sample of obscured AGNs and found consistency with the relation reported by \citet{brightman13} for unobscured sources. The slopes reported by \citet{brightman13,brightman16} are higher but consistent with our results, while we observed a discrepancy in the normalizations. 
For unobscured AGNs, studies such as \citet{risaliti09}, \citet{liu21_lambda}, and \citet{huang20} reported steeper slopes that remain consistent with our results, albeit with slightly higher normalizations. Interestingly, our findings also align with the best-fit normalization reported by \citet{kamraj22}, who used \nustar\ data but did not find a significant correlation.
Nonetheless, the scatter of data points, uncertainties in the best-fit relations, different methods for determining $\Gamma$ and $\lambda_{\rm Edd}$, and variations in sample selection hinder a direct comparison of the results,  with a deeper analysis of these effects lying beyond the scope of this study.

\subsection{The obscuration and Eddington ratio forbidden region}\label{sec:forbidden_region}

\begin{figure}[!tp]
    \centering
    \includegraphics[scale=0.55]{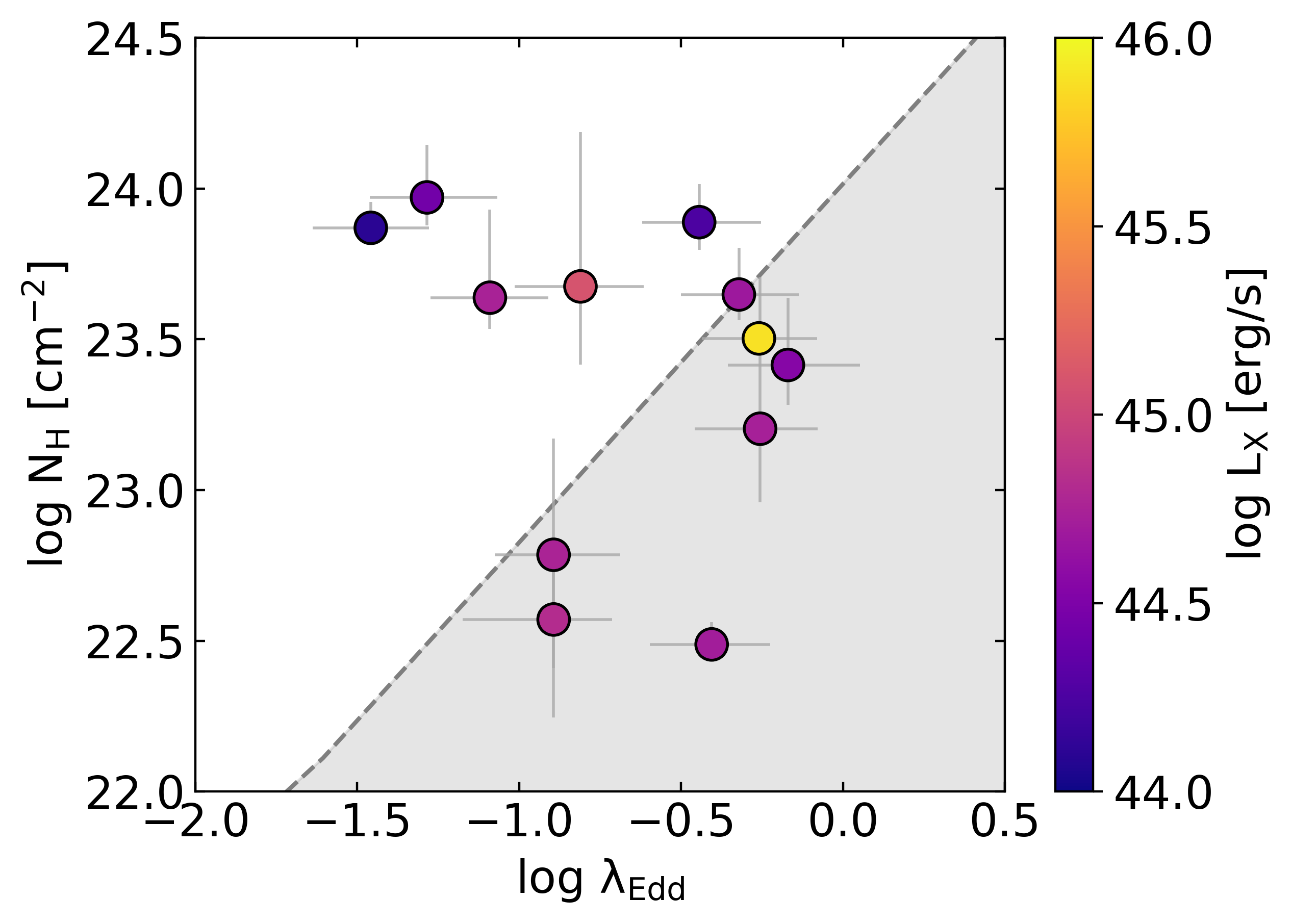}
    \caption{Eddington ratio ($\lambda_{\rm Edd}$) versus absorption ($N_{\rm H}$) for the 12 sources for which $\lambda_{\rm Edd}$ was computed, color-coded by X-ray luminosity ($L_{\rm X}$). 6/12 sources are inside the so-called ``forbidden'' region (shaded grey), where AGNs are expected to be caught in a transitional phase where outflows/winds and/or changes in $N_{\rm H}$ are expected.}
    \label{fig:nh_forb}
\end{figure}

Figure \ref{fig:nh_forb} shows the Eddington ratio $\lambda_{\rm Edd}$ versus the derived absorption $N_{\rm H}$. The grey shaded region marks the so-called ``forbidden region'' where AGNs are not expected to persist due to the effects of radiation pressure on dusty gas from the central AGN engine \citep[e.g.,][]{fabian08,ricci22nat}. When the Eddington ratio exceeds the effective Eddington limit for a given column density, radiation pressure overcomes gravitational forces, expelling the obscuring material and creating a ‘blowout’ region where long-lived dusty clouds cannot survive \citep[e.g.,][]{fabian09}.
AGNs found within this region are likely caught in a short-lived transitional phase, during which radiation pressure actively reduces the column density \citep[e.g.,][]{kammoun20, ricci22b, ricci23}. This phase may involve variability in both accretion rate and obscuration, and can be associated with outflows/winds \citep[e.g.,][]{kakkad16, musimenta23}.

In our sample, 6 out of 12 ($50\pm13$\%) sources (BAT IDs 555, 787, 1204, 1248, 1296, and 1586) are clearly in this region, with one additional source (BAT ID 32) lying very close to the threshold. 
This suggests that high-luminosity, obscured AGN are frequently observed in this transitional phase. Among the sources in the forbidden region, two objects (BAT IDs 787 and 1248) show an outflow component in \OIII\ \citep{oh22}, with BAT ID 787 also showing winds in the X-rays \citep{tombesi14,mestici24}. Additionally, X-ray flux variability driven by changes in $N_{\rm H}$ is observed in BAT IDs 787, 1204, and 1296 (see Section \ref{sec:variability}), with two sources lacking multi-epoch data.
Among the sources outside the forbidden region, outflows in \OIII\ are found in two sources (BAT IDs 80 and 199), and none show flux changes due to $N_{\rm H}$ varibility, with the exception of one source for which we do not have multi-epoch data available.

While \OIII\ outflows are observed both inside and outside the forbidden region (two sources in each case), it is noteworthy that one of the outflow sources outside this area (BAT ID 199) lies not far away from the boundary, suggesting a potential connection between outflows and both the region itself and its immediate vicinity.  Interestingly, another source near the boundary (BAT ID 32) has signs of activity, showing an ultrafast inflow in the X-rays \citep{peca_ufi}. 
On the other hand, $N_{\rm H}$ variability is more prevalent for sources in the forbidden region. Indeed, among those with multi-epoch data, all but one within the region show variability, whereas none outside the region do.
To summarize, the concentration of sources with $N_{\rm H}$ variability and the presence of inflows and outflows within or near the forbidden region support the interpretation that these AGNs are undergoing a transitional phase in their evolution, marked by increased and variable activity.
However, we acknowledge that the relatively small number of objects in our sample prevents us from drawing more detailed statistical conclusions about these phenomena. Additional investigation of, for example, X-ray winds and outflows would be worthwhile to understand their role in the transitional phase further, but such analysis is beyond the scope of this paper.

These findings can be placed in the broader context of AGN obscuration models. \citet{ananna22b} presented a scenario where the simple geometric unified model applies to AGNs at low-$\lambda_{\rm Edd}$ regime, but at a transitional value of $\log \lambda_{\rm Edd} \gtrsim -1.7$ \citep{fabian08}, the ratio of obscured to all AGNs reflects temporal evolution rather than just geometric evolution, i.e., it reflects how much time AGNs spend in the obscured state. According to this model, at our median  $\log \lambda_{\rm Edd} = -0.59$, this ratio is about 20\% for AGNs with $\log N_{\rm H}/\mathrm{cm}^{-2} \geq 22$. As the median $\log N_{\rm H}$ of our sample is higher ($\simeq 23.5$), this may indicate the persistence of higher $\log N_{\rm H}$ obscuring matter for a given $\lambda_{\rm Edd}$. This is also in agreement with the \citet{fabian08} and \citet{ricci17nature} scenario where the Compton-thick level of obscuration is expected to be persistent even at very high Eddington ratios ($\lambda_{\rm Edd} \simeq 1$).
Finally, we report no trend with luminosity, although we note that the three sources with $\log L_{\rm X}<44.5$ erg/s lie well outside this transitional region.

\subsubsection{Comparison with \citealp{baer19}}
Our work can be compared to \citet{baer19}, who selected a sample of 28 luminous ($\log L_{\rm bol}/\mathrm{erg\,s}^{-1} \gtrsim 45.3$) and optical Seyfert 2 AGNs from the BASS DR1 survey. Six sources in our sample overlap with theirs. The differences between the samples arise from various factors, including the exclusion of 22 out of 28 sources from \citet{baer19} in our study due to their X-ray 2--10\,keV luminosity falling below our chosen threshold. On the other hand, our sample includes 6 BASS DR1 sources not present in \citet{baer19}, either because they were not yet classified in BASS DR1, were reclassified in BASS DR2, or were not included in the X-ray analysis of \citet{ricci17}.

\citet{baer19} showed that high-luminosity AGNs in generally avoid the forbidden region, with a few exceptions. This is not in disagreement with our findings, as among the sources in our sample that lie in this region, only two overlap with the \citet{baer19} sample. These two sources (BAT IDs 555 and 1204) are located near the demarcation lines in our work, and the minor differences with \citet{baer19} likely arise from different fitting methods, datasets, and variability.
Comparing the two full samples, we find consistent results with \citet{baer19}, confirming that highly X-ray luminous and obscured AGN follow standard optical diagnostic diagrams and infrared classification schemes, as we further discuss in Appendix \ref{app:opt_ir_class}.

\subsection{Covering factor}
The covering factor plays a crucial role in understanding the modeling of the AGN structure. 
Recent studies \citep[e.g.,][]{ricci17nature,ananna22,ananna22b,ricci22b,ricci23b} show that the covering factor varies significantly with the Eddington ratio, reflecting the dynamics between accretion processes and obscuring material. At low $\lambda_{\rm Edd}$, AGNs tend to have larger covering factors, indicative of more extensive obscuration which may arise from dense, gravitationally bound clouds. As $\lambda_{\rm Edd}$ increases, radiation pressure from the central source begins to push and expel the obscuring material, reducing the covering factor at higher accretion rates \citep[e.g.,][]{ricci22b,ananna22b}.

In this work, not all torus models allow for an estimate of the covering factor. Specifically, the MYTorus model does not provide such an estimate, as it assumes a fixed geometry for the torus, as described in Section \ref{sec:torus}. Due to this limitation, we could not estimate the covering factor for the eight sources where this model was preferred. 
For the remaining 13 sources, we obtained covering factor estimates using model-specific approaches. For the four sources modeled with RXTorusD, the covering factor was taken directly from the $r/R$ ratio, a model parameter that directly represents the covering fraction \citep{ricci23}. In UXCLUMPY, instead, the covering factor is not an explicit model parameter. However, it can be derived by interpolating the tabulated values provided by \citet{boorman_hexp}, which relate the TOR$\sigma$ and CTKcover parameters to the covering factor. For the nine sources in our sample modeled with this model, we applied this method and used the MCMC-derived posterior distributions of these parameters to estimate the associated uncertainties.

\begin{figure}[!tp]
    \centering
    \includegraphics[scale=0.55]{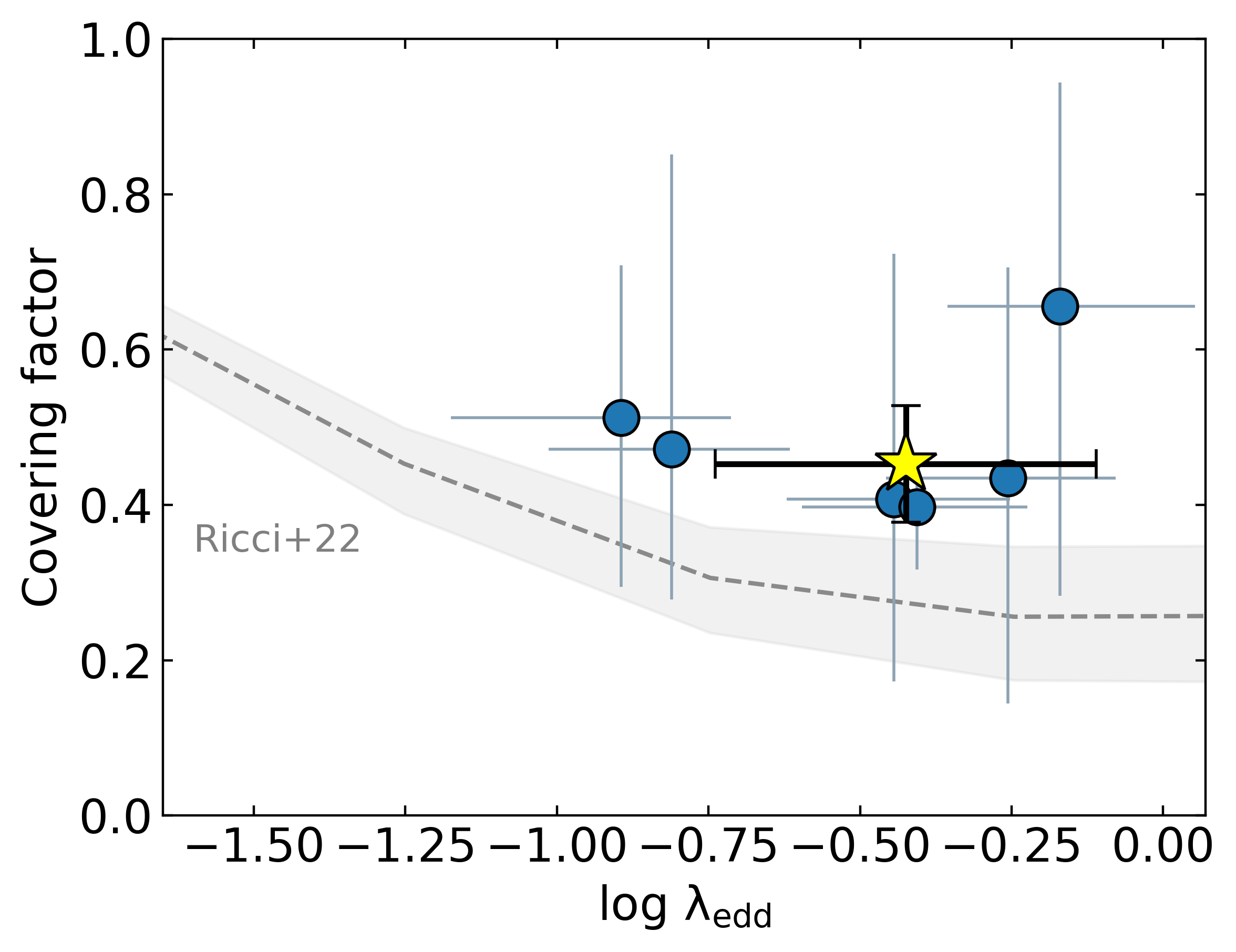}
    \caption{Eddington ratio ($\lambda_{\rm Edd}$) versus covering factor for the sources where both measurements could be obtained (six sources). The yellow star represents the median value, and its uncertainties are computed as median absolute deviations. The grey dashed line is the relation by \citep{ricci22b} with associated 1$\sigma$ uncertainties.}
    \label{fig:cov_edd}
\end{figure}

The median covering factor for the 13 sources where this measurement could be computed is $0.53 \pm 0.09$. The uncertainty is estimated as the median absolute deviation, excluding one source for which only a lower limit was available.
In Figure \ref{fig:cov_edd}, we show our results against $\lambda_{\rm Edd}$. This figure includes only six points, as $\lambda_{\rm Edd}$ was available only for these sources. We compare these results with the relation revised by \citet{ricci22b}. Interestingly, our median value (yellow star) lies above the curve. This effect can be attributed to the fact that optically obscured AGNs tend to have higher covering factors than the broader AGN population \citep{ricci11,elitzur12,ichikawa15}. However, we note that within the uncertainties all points are consistent with the expected relation.

\begin{figure}[!tp]
    \centering
    \includegraphics[scale=0.55]{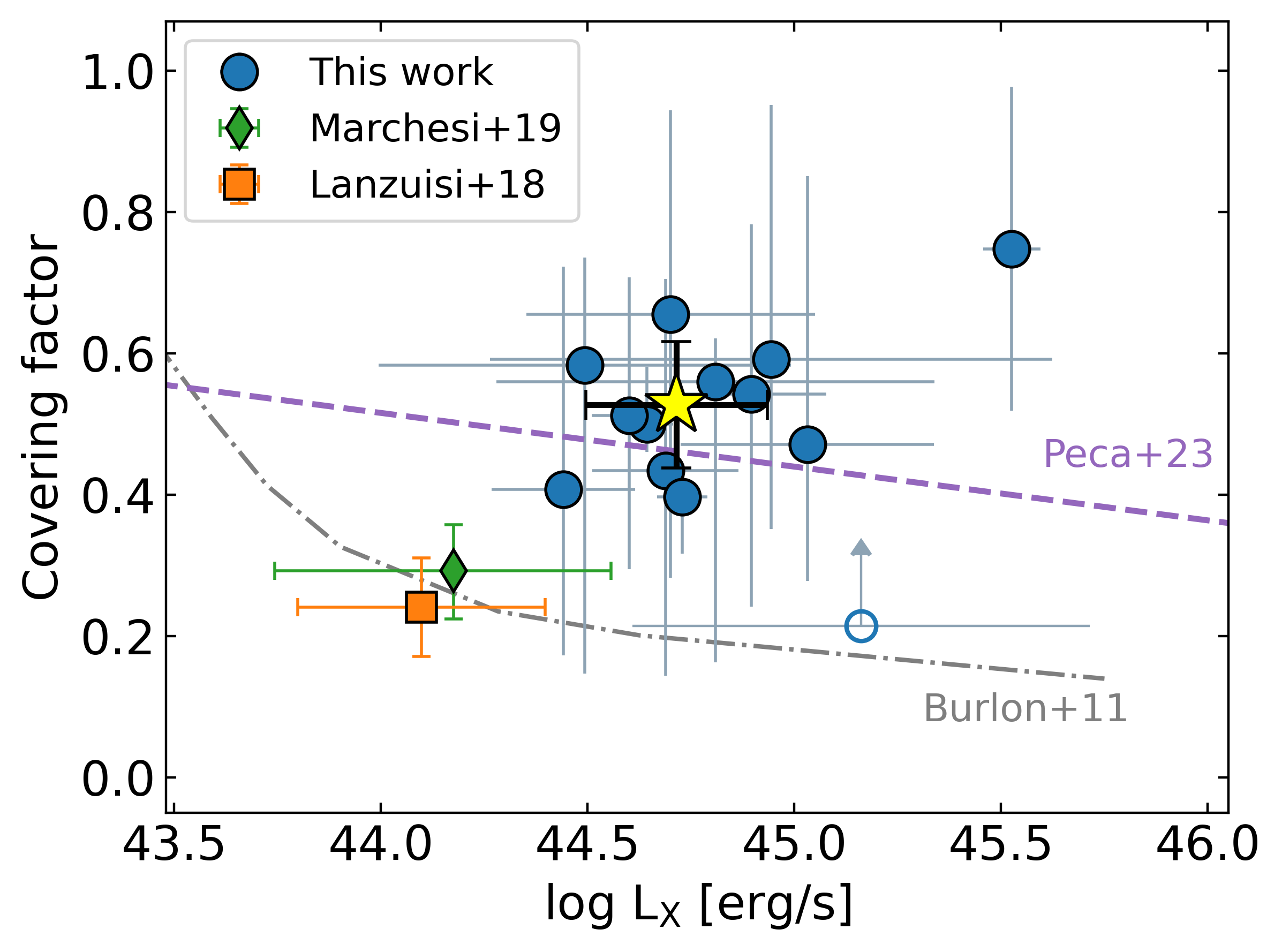}
    \caption{Intrisic 2--10\,keV X-ray luminosity ($L_{\rm X}$) versus covering factor for the 13 sources in our sample where the latter could be estimated. The yellow star indicates the median value, with uncertainties are computed as median absolute deviations. The single lower limit (blue empty point) was excluded from the median calculation. For comparison, we also show the covering factor result from \cite{marchesi19} (green diamond), obscured Compton-thin fraction by \cite{peca23} (purple dashed line), and Compton-thick fractions by \cite{lanzuisi18} (orange square) and \cite{burlon11} (grey dotted-dashed line).}
    \label{fig:cov_lum}
\end{figure}

Another well-studied relation is the one between the covering factor and X-ray luminosity \citep[e.g.,][]{brightman15,marchesi19}. The covering factor, being a direct geometrical proxy for the fraction of obscured AGNs, can be used to investigate the observed trend in which the obscured AGN fraction decreases with increasing luminosity \citep{treister05, hasinger08,aird15,buchner15,peca23}. This trend has been attributed to the combination of different mechanisms, including increasing radiation pressure, dust sublimation, geometrical effects, and the clearing of gas clouds \citep[e.g.,][]{honig07,akylas08,hasinger08,mateos17,matt19}. 

In Figure \ref{fig:cov_lum} we show the derived covering factor versus X-ray luminosity. Notably, we extend earlier analyses \citep[e.g.,][]{marchesi19, zhao21, tanimoto20,tanimoto22} to higher intrinsic luminosities ($\log L_{\rm X}/\mathrm{erg\, s^{-1}}>44.5$). For example, \cite{marchesi19} studied a sample of local Compton-thick AGNs at lower luminosities  ($\log L_{\rm X}/\mathrm{erg\, s^{-1}} \sim 44$) and found lower covering factors compared to ours. 
When compared with other works on the fraction of obscured AGNs, our results show higher values than the Compton-thick fractions found by \citet{lanzuisi18} at $z < 1$ and \citet{burlon11} at $z < 0.3$, while aligning well with the Compton-thin ($22<\log N_{\rm H}/\mathrm{cm^{-2}}<24$) fractions from \citet{peca23} at $z<1$ and other works as well \citep[e.g.,][]{aird15,buchner15,ananna19}. This agrees with the nature of our sample, which contains heavily obscured AGNs, though only one source exceeds the Compton-thick threshold. 
These comparisons support the scenario in which the covering factor of the Compton-thin material is larger than that of the Compton-thick gas \citep[e.g.,][]{tanimoto20, zhao21}.
Finally, we do not report any trend with luminosity, likely due to the limited luminosity range of our sample.

\subsection{Iron K$\alpha$ Emission Lines}
We proceeded with the characterization of the iron K$\mathrm{\alpha}$ emission lines as follows. The torus models selected for the spectral analysis include the iron line in a self-consistent manner, meaning it is built into the model and not treated as an independent, free parameter.  To address this limitation, we performed a new fit using the MYTorus model in its simpler, coupled configuration, disabling the emission lines module. A Gaussian line at a fixed rest-frame energy of 6.4\,keV was then manually added. We assumed a fixed width of $\sigma$=10 eV, consistent with an unresolved component \citep[e.g.,][]{vignali15,nanni18}.
We verified that the fitted power-law continuum in this configuration was consistent, within uncertainties, with the best-fit model preferred by the DIC criterion for the main analysis. This ensured that the measurements of the emission line were performed on a model consistent with the one used for the spectral analysis. 
We considered the emission line significant if the difference in the fit statistic between the model with and without the Gaussian line was $\Delta \chi^2> 2.7$, corresponding to a 90\% confidence level for the additional line normalization free parameter.
We computed the rest-frame equivalent width for the 9/19 ($47\pm11$\%) sources where this criterion was met. For the remaining sources, we computed the 3$\sigma$ upper limits. We do not report any values for the two sources located in galaxy clusters (BAT ID 1204 and 1346), as their spectra are dominated by thermal emission from the \textit{apec} component, which includes multiple emission lines, thereby making the measurement of the iron K$\alpha$ line unreliable. The results are shown in Table \ref{tab:ka_results}.
Note that for GALEXMSC J025952.92+245410.8 (BAT ID 1248), the values reported for the iron K$\alpha$ line are based solely on the \nustar\ spectra, as the \chandra\ spectrum has a limited number of counts, causing the line significance to fall below the detection threshold.

\begin{table}[!ht]
\centering\hspace{-1cm}
\begin{tabular}{cccc}
\hline
\multirow{2}{*}{\shortstack{BAT \\ ID}}  &  K$\mathrm{\alpha}$ EW & $\mathrm{\Delta \chi^2}$\\
       & (eV)                    &                  \\
\hline
20 & $<0.18$ &  $<2.7$  &  \\
32 & $0.10_{-0.05}^{+0.16}$ & $22.2$ &  \\
80 & $0.28_{-0.10}^{+0.17}$ & $24.3$ &  \\
119 & $<0.19$ &  $<2.7$  &  \\
199 & $0.11_{-0.07}^{+0.07}$ & $6.8$ &  \\
476 & $0.08_{-0.04}^{+0.05}$ & $15.1$ &  \\
494 & $<0.46$ &  $<2.7$  &  \\
505 & $0.32_{-0.24}^{+0.05}$ & $15.5$ &  \\
555 & $<0.17$ &  $<2.7$  &  \\
714 & $0.19_{-0.05}^{+0.09}$ & $22.5$ &  \\
787 & $<0.04$ &  $<2.7$  &  \\
1204 &  ... &  ... &  \\
1241 & $<0.56$ &  $<2.7$  &  \\
1248 & $0.18_{-0.16}^{+0.18}$ & $4.0$ &  \\
1291 & $<0.89$ &  $<2.7$  &  \\
1296 & $0.10_{-0.06}^{+0.03}$ & $9.0$ &  \\
1346 &  ... &  ... &  \\
1515 & $0.10_{-0.05}^{+0.09}$ & $11.8$ &  \\
1586 & $<0.23$ &  $<2.7$  &  \\
1595 & $<0.24$ &  $<2.7$  &  \\
1630 & $<1.08$ &  $<2.7$  &  \\
\hline
\end{tabular}
\caption{Rest-frame Fe K$\alpha$ equivalent width (EW) measurements using the MYTorus model (see text for details). For sources where the inclusion of the emission line led to a significant fit improvement at the 90\% confidence level ($\Delta \chi^2 > 2.7$) we report the measured EW, while for the remaining sources we provide 3$\sigma$ upper limits. No values are reported for BAT ID 1204 and 1346, as their spectra are dominated by thermal emission, preventing any line measurement. Columns, from left to right: source name, rest-frame EW, and $\Delta \chi^2$ between models with and without the emission line.}
\label{tab:ka_results}
\end{table}

\begin{figure*}[!tp]
    \centering
    \includegraphics[scale=0.8]{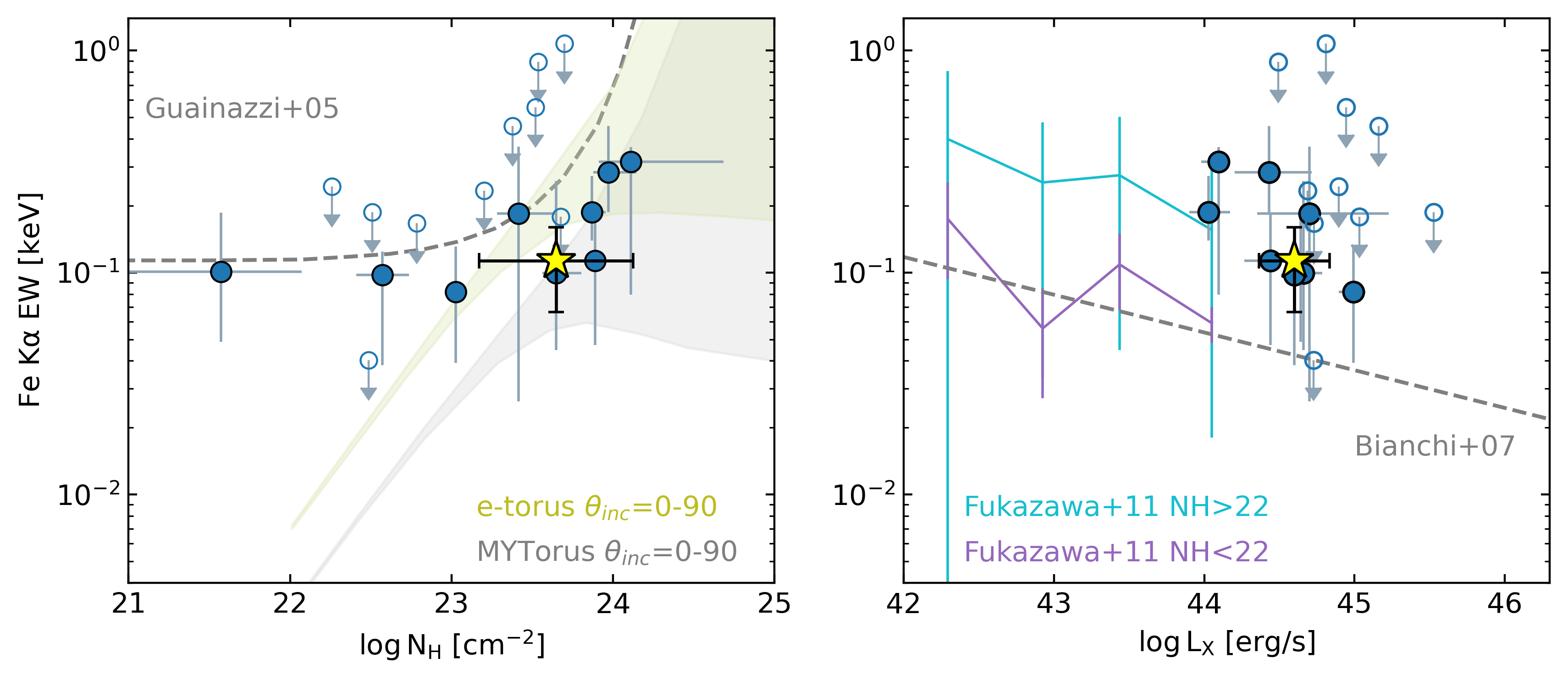}
    \caption{\textit{Left:} Fe K$\alpha$ rest-frame equivalent width as a function of the derived $N_{\rm H}$. The shaded areas represent model predictions from e-torus \citep{ikeda09} (olive) and MYTorus \citep{mytorus} (grey), showing the range of all possible inclination angles. The torus opening angles assumed in the models are 30$^\circ$ and 60$^\circ$, respectively. The empirical relation from \citet{guainazzi05} is shown as a grey dashed line.
    \textit{Right:} Fe K$\alpha$ rest-frame equivalent width as a function of X-ray luminosity. Cyan and purple points show binned results from \citet{fukazawa11} for obscured and unobscured AGNs, respectively. Each point represents the average in a given $L_{\rm X}$ bin, with error bars corresponding to one standard deviation. The dashed grey line indicates the relation found by \citet{bianchi07} for unobscured AGNs. 
    In both panels, empty points indicate 3$\sigma$ upper limits, where error bars on the x-axis are omitted for clarity but correspond to $\sim$0.5 dex. No values are reported for BAT ID 1204 and 1346, as their spectra are dominated by thermal emission from the \textit{apec}, preventing any measurements of the emission line. The yellow stars represent the median values of the constrained values, with uncertainties computed as median absolute deviations.}
    \label{fig:ironka}
\end{figure*}

Figure \ref{fig:ironka} (left panel) shows that the EWs from our work are consistent with predictions made by different models when compared against $N_{\rm H}$. Specifically, we compare our measurements with the e-torus model from \cite{ikeda09} and with the MYTorus model \cite{mytorus}, considering all possible inclination angles. Additionally, we compare our results to the empirical best-fit relation derived by \cite{guainazzi05} from a sample of Seyfert 2 galaxies at $z<0.1$. Our data confirm their findings, showing consistency with a correlation between EW and $N_{\rm H}$ for column densities above $\log N_{\rm H}/\mathrm{cm^{-2}}\sim 23$, and a flattening at lower values around EWs of about 100 eV. This flattening is also consistent with the results of \cite{fukazawa11}, who measured iron K$\alpha$ EWs within the range of [50-120] eV for sources with $\log N_{\rm H}/\mathrm{cm^{-2}}<23$. Such behavior supports an origin of the iron line emission from the inner walls of the torus or the broad line region \citep[e.g.,][]{ghisellini94,guainazzi05,fukazawa11}. Our median EW lies slightly below the \cite{guainazzi05} relation, which can be explained by the lower luminosity of their sample ($\log L_{\rm X}/\mathrm{erg\,\, s^{-1}}\sim 42$), as discussed in what follows.

In Figure \ref{fig:ironka} (right panel), we compare our EW results against $L_{\rm X}$. Many established relations in the literature connect X-ray luminosity with the equivalent width of the iron K$\alpha$ line, for both unobscured \citep[e.g.,][]{iwasawa93,bianchi07} and obscured \citep{ricci14,ricci14b,boorman18} AGNs. 
This trend, known as the X-ray Baldwin Effect \citep{iwasawa93}, takes its name from the discovery of its analog in the UV for the \civ\ line \citep{baldwin77}. In X-rays, this effect is attributed to factors such as the ionization of circumnuclear material, which weakens line emission, continuum dilution with increasing luminosity, and a luminosity-dependent decrease in the torus covering factor and line-of-sight $N_{\rm H}$, which diminishes reprocessed emission \citep[e.g.,][]{ricci14,ricci14b,boorman18}.
We compare our measurements with the relation from \citet{bianchi07}, which was derived for unobscured AGNs, and the values reported by \citet{fukazawa11}, who analyzed both obscured and unobscured AGNs using \textit{Suzaku} data. Although the limited number of data points prevents us from looking for possible correlations, our results are more consistent with those of obscured AGNs from \citet{fukazawa11}. This aligns with expectations, as obscured AGNs typically exhibit higher equivalent widths due to the suppression of the continuum by absorption and the stronger contribution of reprocessed emission from dense circumnuclear material \citep[e.g.,][]{ricci14, ricci14b,lanzuisi15,peca21}. As discussed in \citet{ricci14}, this enhancement in EW for obscured AGNs is primarily an observational effect, potentially biasing its use as a diagnostic in studies of the Baldwin Effect.

\begin{table*}[!ht] 
\centering\hspace{-1.2cm}
\footnotesize
\begin{tabular}{cccccccccccc}
\hline
\multirow{2}{*}{\shortstack{BAT \\ ID}} & Flux       & $\mathrm{C_{AGN}}$ & Log $\mathrm{N_H}$  & Date & Instr. & p-value \\
            & (erg/s/cm$^2$) &                     & (cm$^{-2}$)   &  &   & (Flux/$\mathrm{N_H}$) \\
\hline

32 & $18.9_{-0.9}^{+0.5}$    & 1                   & $23.7_{-0.1}^{+0.2}$ & 2010-12-19 & Suzaku & 2e-13/0.93\\
   & $15.6_{-1.0}^{+0.9}$    & $0.8_{-0.7}^{+0.8}$ & $23.6_{-0.2}^{+0.2}$ & 2023-05-08 & NuSTAR & \\
   & $9.6_{-0.7}^{+0.9}$     & $0.6_{-0.5}^{+0.7}$ & $23.7_{-0.2}^{+0.2}$ & 2024-08-22 & NuSTAR & \\[4pt] 
80 & $5.3_{-0.3}^{+0.3}$     & 1                   & $23.9_{-0.2}^{+0.3}$ & 2016-12-03 & NuSTAR & 3e-10/0.95\\
   & $3.6_{-0.5}^{+0.2}$     & $1.1_{-0.3}^{+0.4}$ & $24.1_{-0.2}^{+0.3}$ & 2017-10-31 & XMM & \\
   & $4.9_{-0.6}^{+0.2}$     & $1.0_{-0.3}^{+0.4}$ & $23.9_{-0.2}^{+0.3}$ & 2020-08-30 & NuSTAR & \\
   & $5.9_{-0.9}^{+0.4}$     & $1.5_{-0.4}^{+0.6}$ & $24.0_{-0.2}^{+0.3}$ & 2021-06-17 & XMM & \\[4pt] 
119 & $86.0_{-3.3}^{+3.5}$   & 1                   & $22.5_{-0.1}^{+0.2}$ & 2006-01-23 & XMM & $<$1e-16/0.95\\
    & $27.9_{-7.9}^{+5.5}$   & $0.3_{-0.1}^{+0.2}$ & $22.5_{-0.4}^{+0.2}$ & 2024-12-18 & NuSTAR & \\[4pt] 
199 & $10.3_{-0.6}^{+0.5}$   & 1                   & $23.9_{-0.2}^{+0.2}$ & 2016-05-06 & NuSTAR & 3e-7/0.99\\
    & $6.5_{-0.7}^{+0.4}$    & $0.6_{-0.3}^{+0.5}$ & $23.9_{-0.2}^{+0.2}$ & 2017-11-20 & XMM & \\
    & $8.6_{-0.5}^{+0.5}$    & $0.9_{-0.3}^{+0.4}$ & $23.9_{-0.2}^{+0.2}$ & 2021-06-04 & NuSTAR & \\[4pt] 
476 & $38.0_{-1.0}^{+0.6}$   & 1                   & $23.1_{-0.1}^{+0.1}$ & 2009-01-20 & XMM & 2e-10/4e-2\\
    & $44.9_{-1.9}^{+1.5}$   & $1.1_{-0.1}^{+0.1}$ & $22.9_{-0.2}^{+0.1}$ & 2023-07-12 & NuSTAR & \\[4pt] 
505 & $3.7_{-0.5}^{+0.3}$    & 1                   & $23.9_{-0.2}^{+1.0}$ & 2018-06-02 & NuSTAR & 0.43/0.82\\
    & $3.2_{-0.6}^{+0.2}$    & $0.8_{-0.4}^{+1.0}$ & $24.0_{-0.1}^{+0.5}$ & 2018-06-02 & XMM & \\
    & $3.4_{-0.2}^{+0.7}$    & $1.3_{-0.5}^{+0.9}$ & $24.3_{-0.5}^{+0.6}$ & 2020-06-24 & NuSTAR & \\[4pt] 
714 & $7.0_{-0.7}^{+0.7}$    & 1                   & $23.9_{-0.1}^{+0.2}$ & 2016-05-25 & NuSTAR & 4e-7/0.80\\
    & $11.8_{-1.1}^{+0.4}$   & $1.4_{-0.2}^{+0.3}$ & $23.9_{-0.1}^{+0.1}$ & 2018-02-09 & XMM & \\
    & $12.0_{-0.6}^{+0.5}$   & $1.3_{-0.3}^{+0.3}$ & $23.8_{-0.1}^{+0.1}$ & 2020-03-25 & NuSTAR & \\[4pt] 
787 & $57.0_{-45.9}^{+45.5}$ & 1                   & $22.7_{-0.1}^{+0.1}$ & 2008-09-21 & XMM & 0.76/4e-13\\
    & $74.0_{-34.8}^{+10.9}$ & $1.0_{-0.2}^{+0.1}$ & $22.2_{-0.2}^{+0.3}$ & 2016-07-12 & NuSTAR & \\[4pt] 
1204 & $54.8_{-0.1}^{+0.6}$  & 1                   & $23.6_{-0.1}^{+0.1}$ & 2013-11-13 & XMM & $<$1e-16/3e-2\\
     & $56.2_{-0.5}^{+0.2}$  & $1.1_{-0.1}^{+0.1}$ & $23.6_{-0.1}^{+0.1}$ & 2013-11-22 & XMM & \\
     & $64.0_{-1.9}^{+1.2}$  & $1.1_{-0.1}^{+0.1}$ & $23.3_{-0.2}^{+0.2}$ & 2021-01-07 & NuSTAR & \\[4pt] 
1248 & $10.7_{-4.0}^{+1.2}$  & 1                   & $23.5_{-0.2}^{+0.3}$ & 2023-03-26 & NuSTAR & 0.82/0.68\\
     & $11.2_{-9.8}^{+8.0}$  & $0.8_{-0.3}^{+0.5}$ & $23.3_{-0.2}^{+0.3}$ & 2024-11-26 & Chandra & \\[4pt] 
1296 & $28.4_{-2.0}^{+0.8}$  & 1                   & $21.8_{-0.5}^{+0.4}$ & 2023-02-12 & NuSTAR & $<$1e-16/7e-7\\
     & $14.8_{-0.7}^{+0.4}$  & $0.7_{-0.6}^{+0.8}$ & $22.8_{-0.1}^{+0.1}$ & 2025-03-14 & XMM & \\[4pt] 
1346 & $22.8_{-0.5}^{+0.7}$  & 1                   & $23.7_{-0.1}^{+0.4}$ & 2007-04-13 & XMM & 3e-4/0.57\\
     & $26.6_{-0.8}^{+1.0}$  & $1.1_{-0.1}^{+0.2}$ & $23.6_{-0.2}^{+0.4}$ & 2024-11-25 & NuSTAR & \\[4pt] 
1515 & $22.2_{-5.1}^{+-1.9}$ & 1                   & $21.9_{-1.0}^{+0.5}$ & 2023-03-11 & NuSTAR & 4e-9/ - \\
     & $32.1_{-1.0}^{+2.6}$  & $1.4_{-0.3}^{+0.4}$ & $<21.1$              & 2025-02-27 & XMM & \\
\hline
\end{tabular}

\caption{Summary of parameters left free to vary in simultaneous fits for sources with multi-epoch data. From the \textit{left}: BAT ID, observed 2--10\,keV best-fit flux in units of 10$^{-13}$ erg/s/cm$^2$, $C_{AGN}$ (fixed to 1 for the first observation of each source), line-of-sight column density $N_{\rm H}$, observation date, telescope, and p-values on flux and $N_{\rm H}$ obtained from Equation \ref{eq:var}. The value $<$1e-16 indicates a p-value lower than the limits from our Python scripts.
For BAT ID 1515, we computed the p-values only for the flux, as $N_{\rm H}$ is constrained only as an upper limit in the second observation (see text for details).}
\label{tab:var}
\end{table*}

\subsection{Variability analysis}\label{sec:variability}
In the previous sections, we characterized our sample using the average values of the spectral parameters $L_{\rm X}$  and $N_{\rm H}$. Among the 21 sources, 13 have multi-epoch spectroscopy: 11 observed with \xmm\ and \nustar, one with \chandra\ and \nustar, and one observed twice with \nustar and one with \suzaku.
Here, we focus on these 13 sources and present single-epoch results. Specifically, we examine the two primary parameters left free to vary in the spectral fitting: $C_{\rm AGN}$, which accounts for intrinsic flux variability, and the line-of-sight $N_{\rm H}$. We also include the observed fluxes in the 2–10\,keV band.
The temporal evolution of these parameters is summarized in Table \ref{tab:var}, with a visual characterization shown in Figures \ref{fig:variability1}, \ref{fig:variability2}, and \ref{fig:variability3} in Appendix \ref{app:srcs_notes}.

While a detailed study of the variability of these parameters is beyond the scope of this paper and it is the subject of ongoing efforts, we quantified the variaiblity using the method proposed by \cite{torres-alba23}. This approach, which has been used to evaluate variability in  $N_{\rm H}$ and flux \citep{torres-alba23,sengupta25,pizzetti25}, employs a chi-squared-like procedure:
\begin{equation}\label{eq:var}
\chi^2 = \sum_{i=1}^{n} \frac{\left(N_{H,\mathrm{los},i} - \langle N_{H,\mathrm{los}} \rangle \right)^2}{\delta \left(N_{H,\mathrm{los},i}\right)^2}
\end{equation}
where the ``true'' value is the average value from the available $i$-observations, and the uncertainty $\delta$ is defined as the lower uncertainty when the $i$-value is less than the average or the upper uncertainty otherwise. The p-value is then computed by assuming chi-squared statistics. A source is considered variable if the p-value is less than 0.05 \citep{torres-alba23,pizzetti25}.
We applied this method to the observed flux and $N_{\rm H}$. We did not perform this computation for $C_{AGN}$, since its value is fixed to unity for the first observation in the simultaneous fitting procedure, limiting the method's applicability. Our results are presented in Table \ref{tab:var}, where we also report the derived $C_{AGN}$ values, as they still provide qualitative support for possible intrinsic variability.  For one source only (BAT ID 1515), we applied the method only on the flux, as the $N_{\rm H}$ value for the second observation is an upper limit. This is due to a combination of the source being unobscured ($\log N_{\rm H}/\mathrm{cm^{-2}}<22$) and lying behind a high Galactic column density ($N_{\rm H}=1.2\cdot10^{21}$ cm$^{-2}$, \citealp{kalberla05}), preventing our ability to constrain $N_{\rm H}$.

Overall, 11/13 sources ($85^{+5}_{-15}\%$) exhibit variability in either flux (10/13, $77^{+8}_{-15}\%$) or $N_{\rm H}$ (4/12, $33^{+15}_{-10}\%$). 
Additionally, 3/12 ($25^{+16}_{-8}\%$) are variable in both flux and $N_{\rm H}$, 6/12 ($50^{+13}_{-13}\%$) show flux variability without changes in $N_{\rm H}$, and 1/12 ($8^{+15}_{-3}\%$) displays variability in $N_{\rm H}$ only. These distinctions suggest that AGN variability may be driven by changes in $N_{\rm H}$, intrinsic flux, or a combination of both processes. 
It is worth mentioning that our results on the fraction of sources with $N_{\rm H}$ variability are in excellent agreement to what was found by \cite{torres-alba23} and \cite{pizzetti25} in a sample of lower luminosity ($\log L_{\rm X}/\mathrm{erg\,\, s^{-1}}\sim43$) local Compton-thin AGNs \citep{zhao21}.

\section{Results II: \nev\ emission line} \label{sec:nev}
\begin{figure*}[!tp]
    \centering
    \includegraphics[scale=0.52]{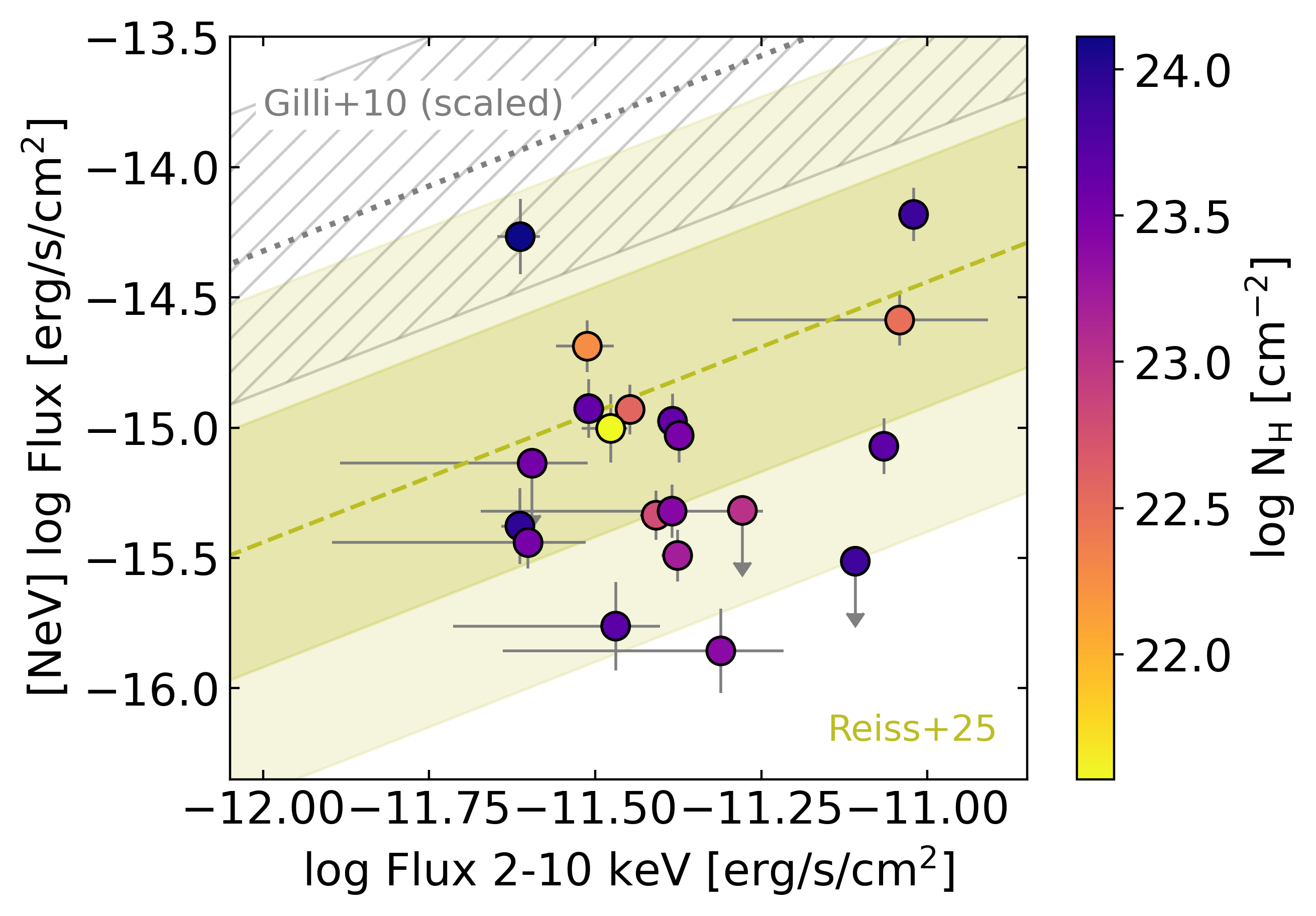} \hspace{2mm}
     \includegraphics[scale=0.52]{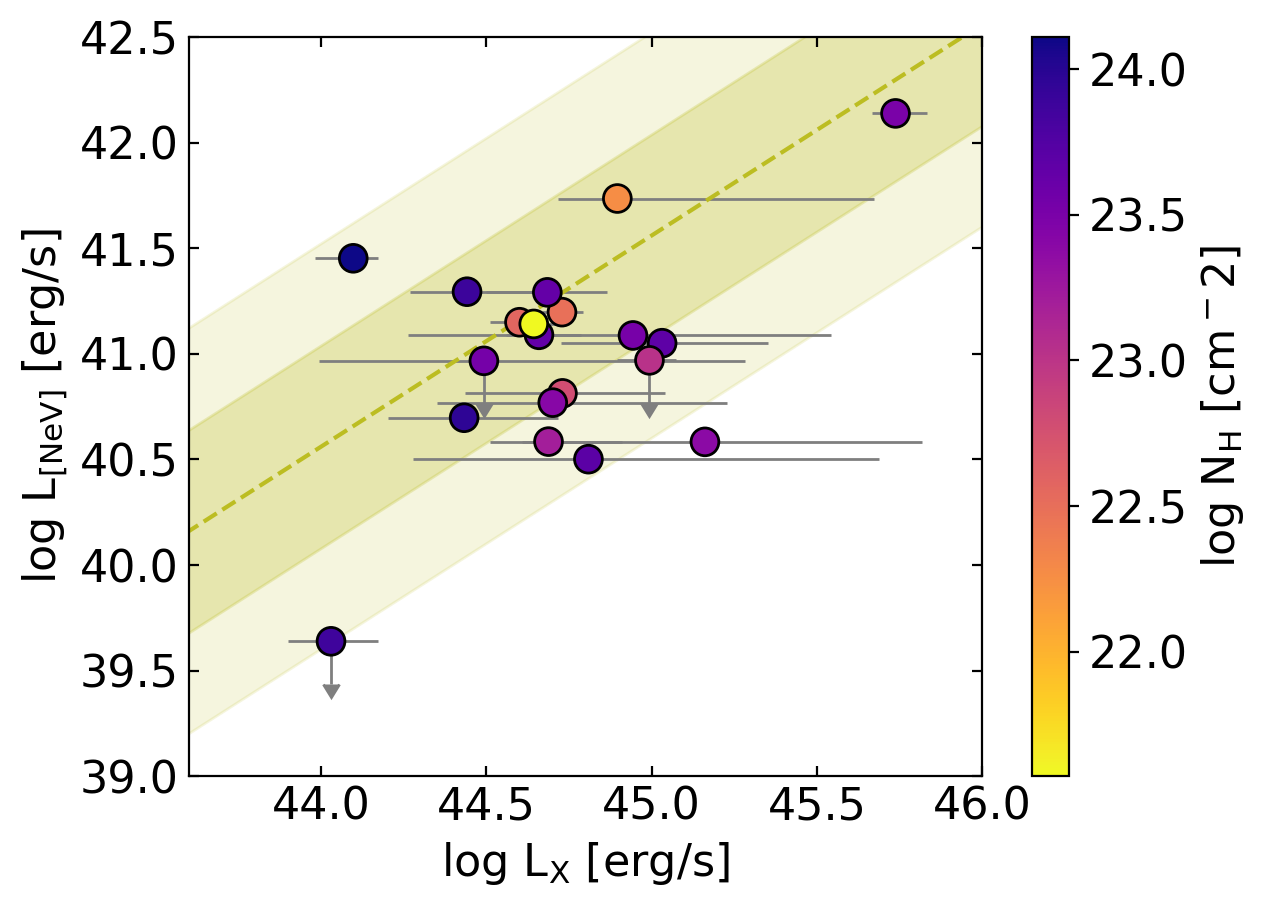}
    \caption{\textit{Left panel:} \nev\ flux versus absorption-corrected 2--10\,keV flux for our sample. For comparison, we show the \citetalias{reiss25} median relation for the BASS DR2 sample (olive dashed line), along with its 1$\sigma$ and 2$\sigma$ uncertainties (olive shaded areas), and the Compton-thick AGN threshold from \citet{gilli10} with its 1$\sigma$ uncertainty (grey dotted line and hatched area). Since \citet{gilli10} used observed fluxes (i.e., uncorrected for absorption), we applied a correction factor of 14 (see details in the text) to align with our intrinsic fluxes.
    \textit{Right panel:} Intrinsic (i.e., de-absorbed 2–10\,keV rest-frame) luminosity versus \nev\ luminosity. We compare our results with the \citetalias{reiss25} median value relation with its 1$\sigma$ and 2$\sigma$ uncertainties (olive dashed line and shaded areas). In both panels, data points are color-coded by the line-of-sight column density ($N_{\rm H}$). The \nev\ flux is not corrected for extinction or reddening. \vspace{2mm}}
    \label{fig:nev}
\end{figure*}

\subsection{\nev\ and X-ray obscuration}
The \nev\ emission line is a forbidden transition with a high-ionization potential of $h\nu>97$ eV. As such energetic photons are rarely produced by stellar processes, \nev\ is considered a reliable tracer of AGN activity \citep[e.g.,][]{zakamska03,gilli10,yuan16,negus23, cleri23a}.  
Unlike X-ray emission, which is produced near the central SMBH and may be heavily attenuated by the obscuring torus, the \nev\ line originates in the narrow-line region, located at larger scales \citep[e.g.,][]{hickox18,negus23}. As a result, \nev\ emission can remain observable even in heavily obscured AGNs where X-rays are suppressed, making it a valuable tracer for uncovering such systems \citep[e.g.,][]{mignoli13,vergani17,mignoli19}. In this context, combining X-ray and \nev\ emissions offers a powerful tool for probing obscuration, as the X-ray-to-\nev\ flux ratio can be used as a direct indicator of the absorbing column density \citep[e.g.,][]{gilli10, vignali15b, lanzuisi15b, barchiesi24,cavicchi25}.
Although powerful, it is worth mentioning that this method can be influenced by dust, gas attenuation, extreme star formation in the host galaxy, and geometrical effects, potentially affecting both emission components \citep[e.g.,]{buchner17, gilli22, cleri23a}.
With this in mind, in this section we investigate the \nev\ emission line in our sample of highly luminous, obscured AGNs and explore its relation with the X-ray emission.

For this study, we obtained emission line flux measurements from a compilation of optical observations performed with the VLT/X-shooter and KCWI instruments, as described in Section \ref{sec:opt_data}. Optical data were available for all but one source (BAT ID 119), which was observed under poor conditions, making it unsuitable for detailed analysis. The uncertainties on line measurements are less than 0.1 dex \citep{oh22, koss22_dr2catalog}, with overall uncertainties primarily driven by systematic calibration errors. To account for this, we added a 20\% systematic calibration error to the uncertainties provided by the pipeline (see Section \ref{sec:opt_data}) on the line measurements.
We detected \nev\ emission in 17/20 sources, corresponding to a detection rate of $85^{+5}_{-11}\%$. For the three additional sources, we derived 3$\sigma$ upper limits. Our detection rate is slightly higher but consistent with \cite{reiss25}, hereafter \citetalias{reiss25}, which analyzed narrow-line AGNs in the full BASS DR2 sample to study the \nev\ emission and found a detection rate as high as $\sim 70\%$ for high SNR spectra.

Our results can be compared with previous studies at slightly higher redshifts. For instance, \citet{barchiesi24}, analyzing a sample of \nev\ emitters at $0.6 < z < 1.2$, found that two-thirds are heavily X-ray obscured ($\log N_{\rm H}/{\rm cm}^{-2} > 23$) and that half are undergoing intense accretion ($\log \lambda_{\rm Edd} > -1$). These findings are consistent with our results, as we detect a high fraction of \nev\ emission among sources that are predominantly obscured and rapidly accreting, with median values of $\log N_{\rm H}/{\rm cm}^{-2} \simeq 23.5$ and $\log \lambda_{\rm Edd} \simeq -0.6$, respectively. This agreement is particularly noteworthy given that half of our sample lies within the forbidden region of the $N_{\rm H}$–$\lambda_{\rm Edd}$ parameter space, where radiation pressure is expected to clear out the obscuring material (see Section~\ref{sec:forbidden_region}), therefore demonstrating that \nev\ can be detected in sources in this evolutionary phase.
Furthermore, both \citet{barchiesi24} and \citet{vergani17}, who studied another sample of \nev\ emitters at similar redshifts, reported enhanced star formation rates compared to non-\nev\ emitters and interpreted this as a signature of the star formation quenching phase. Our sample is consistent with these studies, as we find comparable stellar masses, with a median $\log M_{\star}/M_\odot = 10.9$ based on the nine sources with available measurements \citep{koss22_dr2overview}, and star formation rates, with a median $\sim 30$ $M_\odot\,{\rm yr}^{-1}$ based on the seven sources with values reported by \citet{ichikawa17}. Additionally, our \nev\ luminosities are consistent with the \nev\ luminosity–stellar mass relations found by \citet{vergani17}.

Figure \ref{fig:nev} presents the \nev\ flux versus the observed 2--10\,keV X-ray flux (left panel) and the de-absorbed, 2--10\,keV rest-frame X-ray luminosity versus the \nev\ luminosity (right panel). Both panels are color-coded by the $N_{\rm H}$ derived from our results presented in Section \ref{sec:results_xrays}. For \nev\, fluxes, we accounted for Galactic extinction following the prescriptions by \cite{oh22} but did not correct for intrinsic extinction, as our goal is to examine the observed properties of \nev\ and ensure a meaningful comparison with literature results.
We first compare our results with \citetalias{reiss25}, which found $\log(L_{\rm [NeV]}/L_{\rm X}) \simeq -3.36$ in the BASS DR2 sample. Our luminosity plot shows less scatter than the flux plot, which is in line with their findings. Both distributions appear slightly skewed toward higher X-ray and lower \nev\ flux and luminosity. However, all our points are consistent when considering the 2$\sigma$ uncertainty on the \citetalias{reiss25} relation, indicating overall agreement. Furthermore, \citetalias{reiss25} have several upper limits for the \nev\ flux range covered in this work, which could potentially influence their relation at these fluxes.
We also confirm the absence of any correlation with $N_{\rm H}$.
Next, we compare our results with the Compton-thick AGN selection criterion from \citet{gilli10}, which identifies Compton-thick candidates as those with an X-ray-to-\nev\ flux ratio below 15. To align with our analysis, we rescaled their observed fluxes to intrinsic fluxes by applying a correction factor of 14, determined by assuming a power-law with $\Gamma=1.8$ and $N_{\rm H} = 10^{24}$ cm$^{-2}$.
Our sample appears skewed toward lower \nev\ fluxes and higher X-ray fluxes, even when accounting for the fact that our median $N_{\rm H}$ is slightly lower than the Compton-thick level. A key difference is that our sample comprises higher-luminosity AGNs, particularly at low redshifts, resulting in higher fluxes for the same level of $N_{\rm H}$. Moreover, the \citet{gilli10} sample may tend toward higher \nev\ fluxes, as the authors pointed out. Consequently, we conclude that the apparent discrepancy is likely attributable to differences in sample properties and selection criteria. Despite these differences, we note that the only object in our sample that is nominally Compton-thick (BAT ID 505) falls within the 1$\sigma$ region of the \citet{gilli10} relation.

\subsection{\nev\ stacking and comparison with JWST}\label{sec:nevJWST}
\begin{figure*}[!tp]
    \centering
    \includegraphics[scale=0.9]{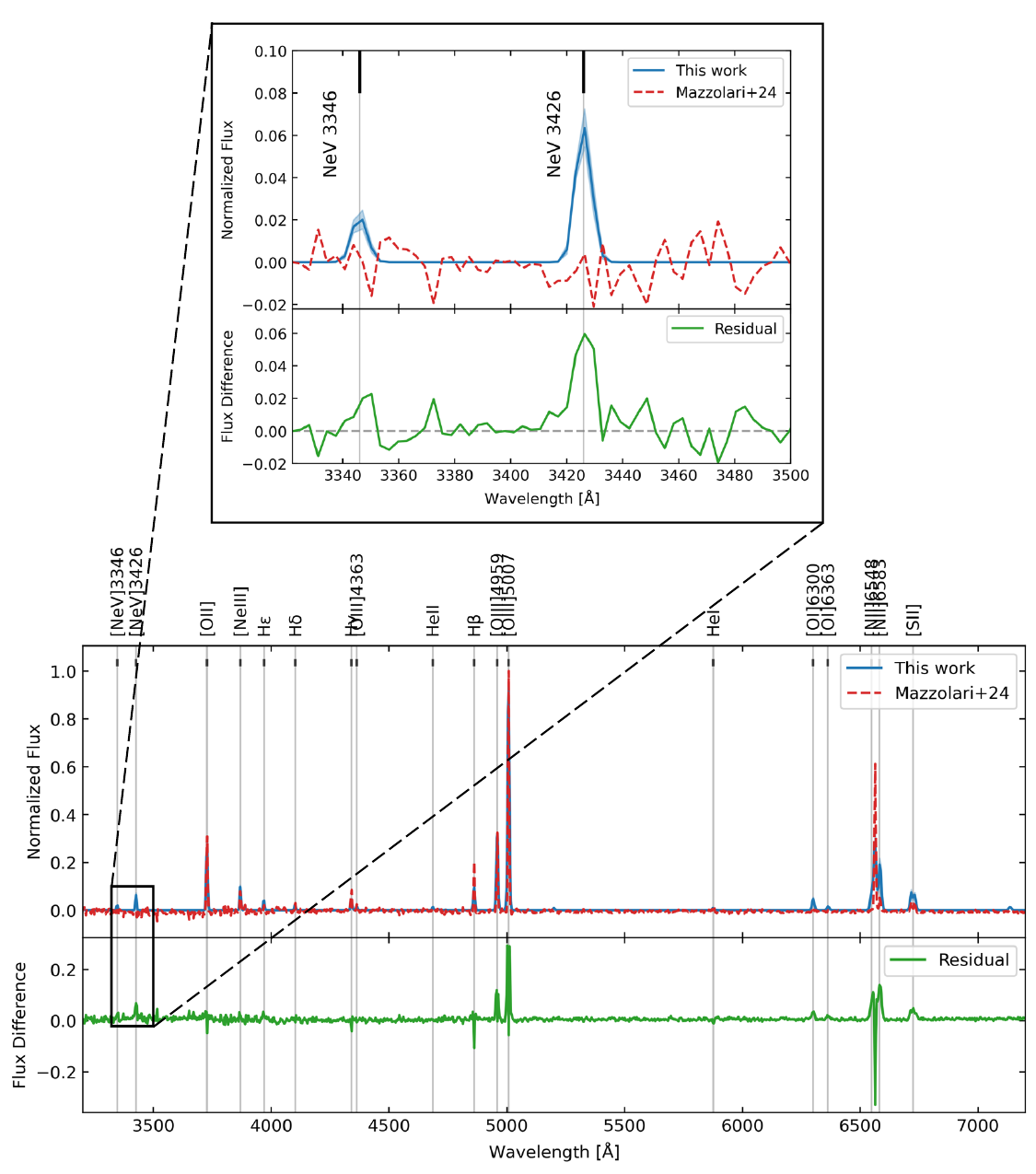}
    \caption{Comparison between the stacked high-redshift JWST spectrum from \citetalias{mazzolari24} (red dashed lines) and our stacked spectrum of low-redshift, highly luminous BASS AGNs (blue solid lines). The shaded blue region shows the 1$\sigma$ uncertainty derived via bootstrap resampling. The green lines represent the residual spectrum, computed as the difference between the two stacks.
    \textit{Top panel}: A zoom-in on the [NeV] emission lines complex from the full stacks shown in the bottom panel. While \nevv\ in our stack is consistent with the noise level in the JWST stacked spectrum, \nev\ stands out at approximately six times that level. In this region, we applied a 0.08 offset to the \citetalias{mazzolari24} stack, as their flux is systematically below zero, likely due to their stacking method. \textit{Bottom panel}: Full stacked spectra, highlighting the broader comparison between the two datasets. Major emission lines are labeled across the top for reference.}
    \label{fig:nev_stack}
\end{figure*}

This section aims to connect our sample with the high-redshift ($z>2$) Universe, particularly with JWST results. To this end, we used the emission-line best-fit spectra from \citet{oh22} for the sources available in BASS DR2, and derived the remaining using the same fitting procedure for consistency across all 21 sources in our sample.
The template spectra were stacked using a bootstrap procedure with 1000 iterations.
We then matched our stacked spectrum with the stacking analysis of \cite{mazzolari24}, hereafter \citetalias{mazzolari24}, who studied 52 narrow-line AGNs in the CEERS survey \citep{ceers} at $2 \lesssim z \lesssim 9$. As in \citetalias{mazzolari24}, we renormalized our stack using the \OIII\ emission line peak flux. Because our spectra have higher resolution ($R > 1000$; see \citealp{koss22_dr2catalog}), compared to the lower resolution of the JWST stacked spectrum ($R \sim 1000$), we degraded our data to match the resolution of \citetalias{mazzolari24} before stacking. To do this, we used the \texttt{SpectRes} code \citep{spectres}, which resamples spectra and their associated uncertainties onto an arbitrary wavelength grid. This function works with any grid of wavelength values, including non-uniform sampling, and preserves the integrated flux. 
The stacking results are shown in Figure\,\ref{fig:nev_stack}.
We note that, while the continuum was fitted and subtracted in \citetalias{mazzolari24}, here we used the emission line best-fit templates from \cite{oh22}. 

Interestingly, the only emission lines completely missing in the CEERS stacked spectrum are those from [NeV] doublet (\nev\ and \nevv). While other differences are visible in the residual spectrum (Figure \ref{fig:nev_stack} bottom panel), they primarily arise from variations in line widths or peak fluxes rather than the absence of specific emission lines. 
As shown in the zoom-in of Figure \ref{fig:nev_stack}, our stack clearly exhibits a strong \nev\ emission line, while the CEERS stack does not. Additionally, the \nevv\, emission line is absent. However, the lack of \nevv\, is not unexpected, as it is too faint to be detected and consistent with the noise level. In contrast, the \nev\ emission line strength from stacking our sample is up to six times higher than the noise level in the JWST stacked spectrum, where the noise was estimated as rms $\sim 0.01$ over the 3250–3700$\text{\AA}$ range, a region free of emission lines.
These results are consistent when performing the stacking using the best spectroscopic resolution of the spectra in our sample.

This discrepancy becomes particularly intriguing when we consider the quantitative expectations based on established scaling relations. The average expected ratio $\log(L_{\rm X}/L_{\rm [OIII]})$ for type 2 AGNs is $\simeq$2.1 \citep[e.g.,][]{georgantopoulos10,berney15,ueda15,malkan17,oh22}. When combined with the relation from \citetalias{reiss25}, $\log(L_{\rm [NeV]}/L_{\rm X}) \simeq -3.36$, we obtain an expected $L_{\rm [NeV]}/L_{\rm [OIII]}$ ratio of $\sim 0.06$. Such a ratio corresponds to a \nev\ flux well above the JWST noise level and matches what we observe in our stacked spectrum.
We note that these scaling relations span a wide range of bolometric luminosities, including values below those probed in our study and consistent with those found in \citetalias{mazzolari24} and other JWST studies of obscured AGNs at high redshift \citep[e.g.,][]{yang23,lyu24,maiolino24}, typically around $\log L_{\rm bol}/\mathrm{erg\,\, s^{-1}} \approx 44$–45. This supports the expectation that the observed \OIII\ emission in JWST spectra should be accompanied by detectable \nev\ emission, despite the relations being calibrated at low redshift and subject to possible evolutionary effects.

The presence of detectable \nev\ emission is particularly relevant given the debate surrounding high-redshift AGNs identified by JWST, many of which exhibit very weak or entirely undetected X-ray emission \citep[e.g.,][]{ananna24, maiolino24, mazzolari24}. As we know from prior studies that X-ray obscuration increases with redshift \citep[e.g.,][]{aird15,buchner15,vito18,vijarnwannaluk22,peca23,signorini23,pouliasis24}, we can invoke high levels of obscuration to explain their observed X-ray weakness. In this context, the \nev\ emission line can be exceptionally useful as an obscured AGN indicator, given its high detection rate in such systems, as shown both in this work and by \citetalias{reiss25}. We explore the implications of these findings in further detail in Section \ref{sec:discussion}.

\section{Discussion} \label{sec:discussion}
This work provides a comprehensive X-ray study of high luminosity ($L_{\rm X} \gtrsim 10^{44}$ erg/s), optically obscured (Seyfert 1.9 and 2) AGNs at redshifts $z < 0.6$. In addition to detailed X-ray spectral modeling, we analyzed the UV/optical \nev\ emission line associated with the X-ray sources, enabling a multi-wavelength perspective on this obscuration indicator. In this section, we discuss the broader implications of our findings.

\subsection{The high luminosity and high obscuration regime}
\begin{figure}[!tp]
    \centering
    \includegraphics[scale=0.56]{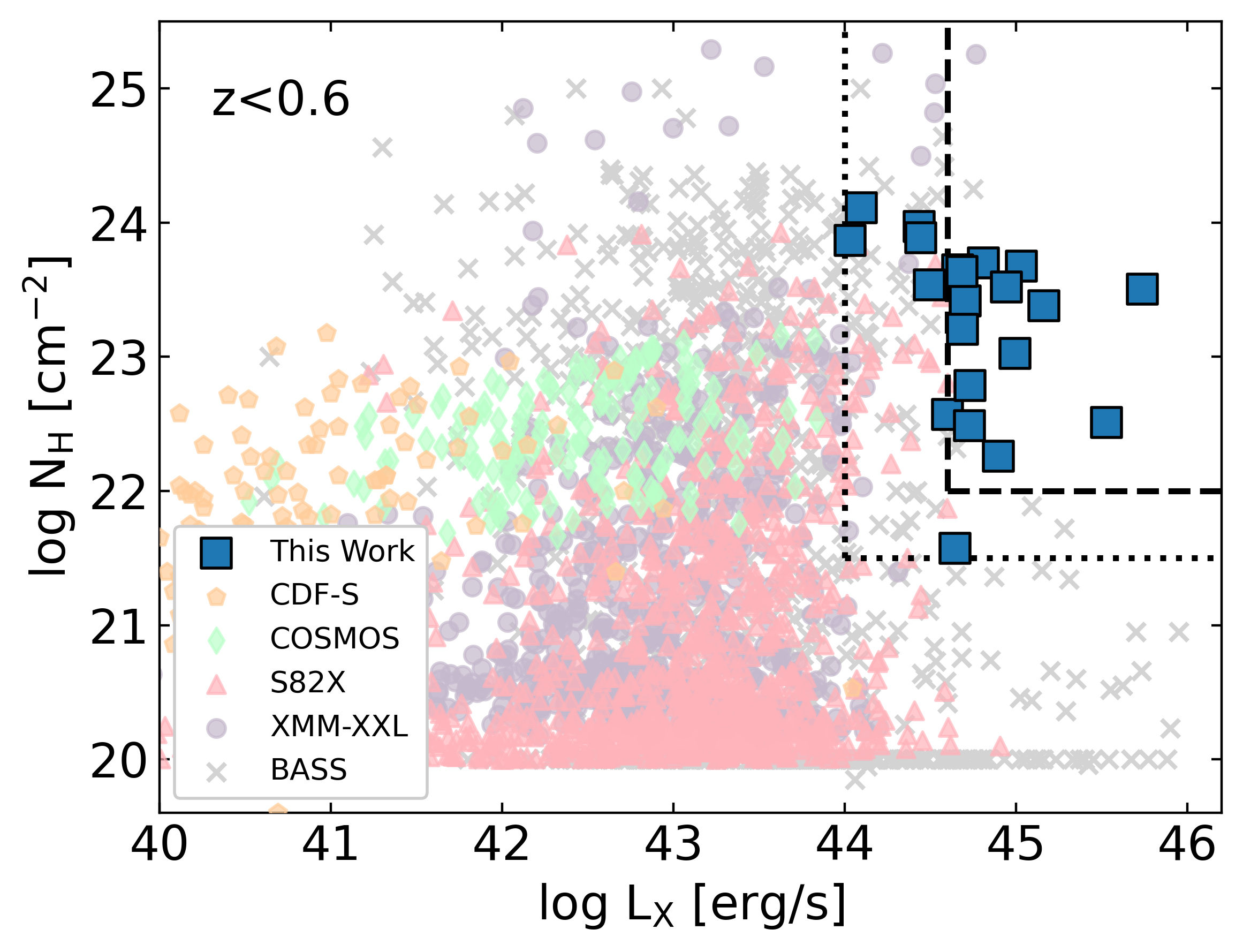}
    \caption{Intrinsic 2--10\,keV luminosity ($L_{\rm X}$) versus absorbing column density  ($N_{\rm H}$) for our sample (blue squares), compared with sources from other X-ray surveys with available spectral analyses. For consistency, only sources at $z < 0.6$ are shown to match the redshift range of our sample. Data form other surveys include: the Chandra Deep Field South (\citealp{liu17}; orange pentagons), COSMOS (\citealp{marchesi16b}; green diamonds), Stripe 82X (\citealp{peca23}; pink triangles), XMM-XXL (\citealp{liu16}; purple circles), and BASS 70-month catalog (\citealp{ricci17}; grey crosses). The region covered by our full sample is shown with black dotted lines, defined by $\log L_{\rm X}/\mathrm{erg\,s^{-1}} > 44$ and $\log N_{\rm H}/\mathrm{cm}^{-2} > 21.5$. Within this area, we highlight in black dashed lines a more extreme subregion, defined by $\log L_{\rm X} > 44.6$ and $\log N_{\rm H}/\mathrm{cm}^{-2} > 22$, which is uniquely sampled by our sources but only poorly populated by previous surveys (see text for details).}
    \label{fig:comparison}
\end{figure}

In Figure \ref{fig:comparison}, we compare our sample with other X-ray surveys, showing only sources at $z < 0.6$ to match the redshift range of our sample. The broad region covered by our full sample is outlined with black dotted lines and corresponds to $\log L_{\rm X}/\mathrm{erg\,s^{-1}} > 44$ and $\log N_{\rm H}/\mathrm{cm}^{-2} > 21.5$. Within this area, we highlight a more extreme subregion defined by $\log L_{\rm X}/\mathrm{erg\,s^{-1}} > 44.6$ and $\log N_{\rm H}/\mathrm{cm}^{-2} > 22$ with black dashed lines. This subregion is uniquely sampled by our dataset, whereas only three sources from other surveys fall within it.
Two of these are from \cite{ricci17}: one (BAT ID 800) is only mildly obscured in the X-rays ($\log N_{\rm H}/\mathrm{cm}^{-2}\sim 22.3$) and it is not included in our sample because it is classified as a Seyfert 1; the other, NGC 6240 (BAT ID 841), consists of two nuclei \citep{komossa03}. In \cite{ricci17}, NGC 6240 was fitted as a single source due to the limited resolution of \xmm\ data. However, when modeled separately, both nuclei have luminosities $\log L_{\rm X}/\mathrm{erg\,\, s^{-1}}<44.6$ \citep{puccetti16}, placing them outside our defined luminosity selection threshold. 
The third source, from XMM-XXL \citep{liu16}, exhibits a broad H$\beta$ line in the optical spectrum \citep{rakshit17}, and it also has a relatively low SNR in the X-ray band (only $\sim$60 photons detected in the 2--10\,keV \xmm\ PN spectrum). \citet{liu16} fitted the source with a double power-law model, reporting an unusually strong secondary component (scattering fraction $>$60\%), which likely led to an overestimation of $N_{\rm H}$ \citep{tokayer25}. In any case, these are only a few sources located near the edge of the defined region, leaving this extreme parameter space uniquely well-sampled by our dataset.

\subsection{On the \nev\ emission line}
We showed that our sample has an $85^{+5}_{-11}\%$ detection fraction for \nev, highlighting its utility as a tracer of AGN emission even in cases of heavy obscuration. This is particularly valuable for identifying AGNs where extreme absorption can completely suppress the X-ray emission. As such, \nev\ becomes particularly relevant in light of recent high-redshift AGN candidates discovered with JWST, many of which exhibit weaker-than-expected or entirely undetected X-ray emission, even in stacked analyses (e.g., \citealt{ananna24}; \citetalias{mazzolari24}). These include the so-called Little Red Dots \citep[LRDs; e.g.,][]{greene23,harikane23,kocevski23,kocevski24,matthee24,labbe25}, whose nature remains uncertain \citep[e.g.,][]{baggen24,kokubo24}. 
Proposed explanations for this X-ray weakness include heavy obscuration \citep[e.g.,][]{maiolino23,maiolino24}, super-Eddington accretion \citep[e.g.,][]{pacucci24,lambrides24,madau24}, or a weak or undeveloped X-ray corona \citep{juod24}. Still, the exact scenario remains unclear (e.g., \citealt{yue24}; \citetalias{mazzolari24}; \citealt{degraaf25}).
In this context, it is particularly puzzling that \nev\ emission is often undetected in JWST-selected samples, including LRDs. 
\citetalias{reiss25} extend our findings by showing that roughly half of all BASS AGNs exhibit \nev\ emission, and that this fraction can be as high as $\sim$70\% in high-SNR spectra. Importantly, they find that the detection fraction remains $\gtrsim$40-50\% across a wide range of luminosities, spanning from the high luminosities of our sample down to $\log L_{\rm X}/\mathrm{erg\,s^{-1}}\sim42$, with no clear dependence on X-ray obscuration or Eddington ratio.  
This suggests that, even though the high-redshift, obscured AGNs identified by JWST seem to have lower bolometric luminosities and accretion rates (e.g., \citealt{yang23}; \citealt{lyu24}; \citetalias{mazzolari24}; \citealt{maiolino24}) than those in our sample, we would still expect \nev\ emission to be detectable in a larger fraction of those sources.
However, only a few detections of \nev\ have been reported so far \citep[e.g.,][]{chisholm24,labbe24}, in particular for narrow-line AGNs in deep fields such as CEERS (\citetalias{mazzolari24}) and JADES \citep{eisenstein23,scholtz25}, challenging our current understanding of these objects.

Radiative transfer models by \citet{mckaig24} further emphasize this puzzle. Their calculations indicate that \nev\ is among the least dust-sensitive coronal lines, with its luminosity reduced by less than a factor of ten even under extreme conditions of super-solar metallicity and dusty gas, compared to a dust-free scenario. Thus, the general lack of detectable \nev\ emission in JWST-selected AGNs suggests potentially different physical conditions compared to those typically found in local sources.
An alternative explanation involves substantial, host-galaxy-scale obscuration by the interstellar medium (ISM). \citet{gilli22} showed that the observed increase in obscuration with redshift seen in X-ray surveys \citep[e.g.,][]{aird15,vijarnwannaluk22,peca23,pouliasis24} can be closely linked to the rising contribution of ISM-scale gas, which may reach column densities as high as $\log N_{\rm H}/\mathrm{cm^{-2}} \sim 24$ at $z \gtrsim 6$ (see also e.g., \citealt{circosta19,damato20}). While this supports the obscured AGN scenario as a possible explanation for the observed X-ray weakness, such high column densities can also correspond to a significant amount of dust, which could attenuate emission lines like \nev.

Extinction measurements\footnote{Computed as in \citetalias{mazzolari24}, using the Balmer decrement, assuming Case B recombination, and adopting the \cite{calzetti00} extinction law with $R_V=3.1$.} in our local sample and \citetalias{mazzolari24} are comparable, with median values of $A_V \approx 0.5$–$0.6$ (consistent also with \citealp{harikane23, maiolino23}), suggesting similar total dust effects. At the same time, our local galaxies have higher stellar masses (median log $M_\star/M_\odot = 10.9$) and thus larger gas and dust reservoirs \citep[e.g.,][]{santini14,salvestrini22}, compared to \citetalias{mazzolari24} and other JWST-selected hosts (log $M_\star/M_\odot < 10$; \citealp[e.g.,][]{harikane23,maiolino23,maiolino24}). The similarity in extinction despite these mass differences suggests complex variations in dust and gas geometry, distributions, or dust grain properties, making direct comparisons challenging and beyond the scope of this work. 
However, since we matched our sample to \citetalias{mazzolari24} in \OIII\ (Section~\ref{sec:nevJWST}), we expect at least comparable ionization conditions between the samples. Indeed, although JWST hosts have typically lower stellar masses and metallicities \citep[e.g.,][]{maiolino23,curti24}, these effects are mitigated by the \OIII\ matching, as stellar mass correlates with metallicity \citep[e.g.,][]{tremonti04,curti20} and higher metallicity weakens \OIII\ emission efficiency \citep[e.g.,][]{kewley02,maiolino08}. 
Therefore, the local and JWST stacks compared in this work are expected to have similar ionization conditions, making the observed differences in \nev\ detectability surprising.

Future and upcoming efforts using JWST data (e.g., with NIRSpec) will play a crucial role in resolving these open questions, providing deeper insights into the intrinsic properties and nature of these intriguing high-redshift AGNs.
For example, \citet{cleri23b} demonstrated that \nev\ can serve as a valuable diagnostic tool for identifying AGN activity in high-redshift line-emitting galaxies, particularly where traditional line-ratio diagnostics may be less effective. Indeed, recent JWST-based studies \citep[e.g.,][]{chisholm24, scholtz25} have successfully relied on \nev, along with other high-ionization emission lines, to argue for an AGN-dominated nature in several galaxies at $z > 5$.
In this regard, an upcoming publication by Trakhtenbrot et al. (submitted) further explores the utility of \nev\ to infer AGN luminosities, and thus constraints on early black hole growth, by leveraging available JWST data for high-redshift galaxies.

\subsection{Future X-ray studies}
This work underscores the importance of local AGNs for understanding the nature of X-ray emission and its connection to multi-wavelength properties. It also highlights the need for future facilities, such as the Advanced X-ray Imaging Satellite (\textit{AXIS}; \citealp{axis}) and \textit{NewAthena} \citep{nandra13,barret19}, to replicate such detailed analyses at higher redshifts. 
In what follows, we discuss the capabilities of the two instruments based on the most up-to-date planned exposures, survey coverages, and data products available from both teams at the time of writing\footnote{\href{https://blog.umd.edu/axis/}{https://blog.umd.edu/axis/}; \\ \href{https://www.mpe.mpg.de/ATHENA-WFI/}{https://www.mpe.mpg.de/ATHENA-WFI/}}. For \textit{NewAthena}, we specifically focus on the Wide Field Imager (WFI; \citealp{meidinger17}), which is expected to play a central role in future X-ray surveys. 

First, we quantify how many sources similar to those in our sample will be detected. To do so, we used the mock catalog of \citet{marchesi20}, which is based on the luminosity functions of \citet{gilli07} for $z<3$ and \citet{vito14,vito18} at higher redshifts. Assuming the \textit{AXIS} limiting fluxes \citep{axis} of $2.5 \times 10^{-17}$ erg s$^{-1}$ cm$^{-2}$ for the Intermediate survey (7000 arcmin$^2$) and $4 \times 10^{-18}$ erg s$^{-1}$ cm$^{-2}$ for the Deep survey (450 arcmin$^2$), we predict the detection of a total of $\sim150$ luminous ($\log L_{\rm X}/\mathrm{erg,s^{-1}}>44.6$) and obscured ($\log N_{\rm H}/\mathrm{cm}^{-2}>22$) AGNs. Of these, about 100 sources will be at $z<3$, and approximately 50 at $z>3$. 
For \textit{NewAthena}, we used the three planned surveys described in \citet{athena_prediction}, with limiting fluxes of $\sim 5\times10^{-17}$ erg s$^{-1}$ cm$^{-2}$ for the deeper $\sim$31 deg$^2$ survey, and $4\times10^{-17}$ and $7\times10^{-17}$ erg s$^{-1}$ cm$^{-2}$ for the wider surveys of approximately 180 and 370 deg$^2$, respectively. Using these sensitivities and areas, we predict the detection of $\sim$ 22700 luminous and obscured AGNs, with about 14800 sources at $z<3$ and 7900 at $z>3$. This significant increase reflects the spatial rarity of these objects \citep[e.g.,][]{ueda14,aird15,peca23}, which can only be captured in large numbers through wide-area surveys.

\begin{figure*}[!tp]
    \centering
    \includegraphics[scale=0.45]{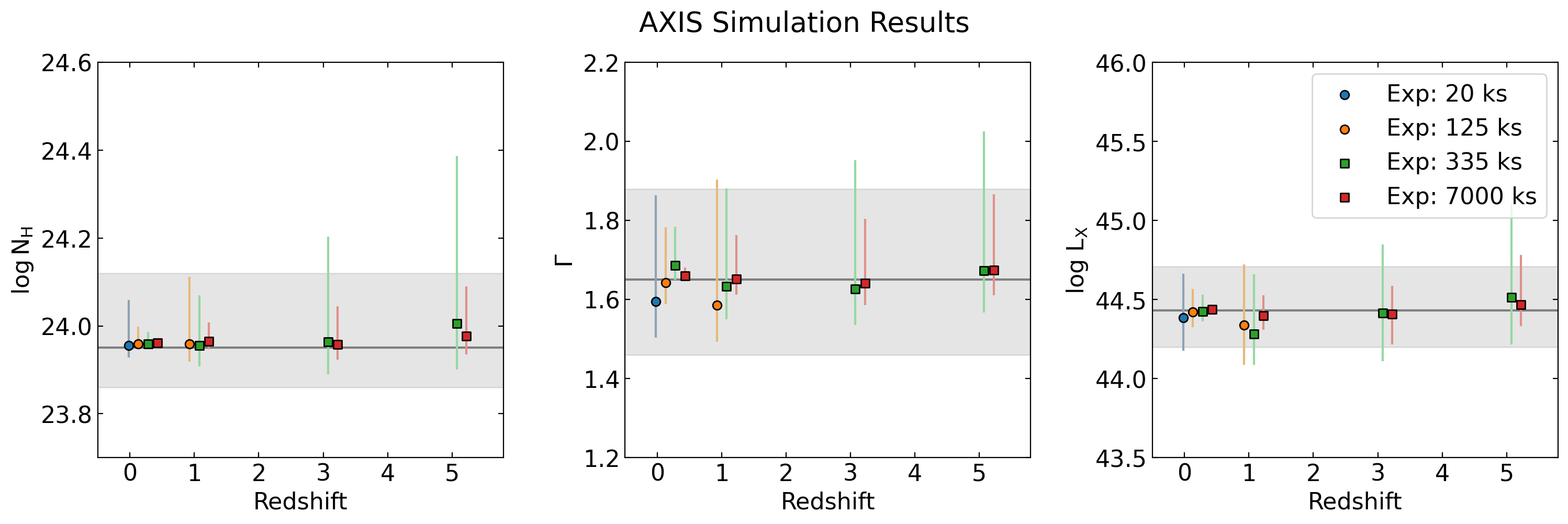}
    \includegraphics[scale=0.45]{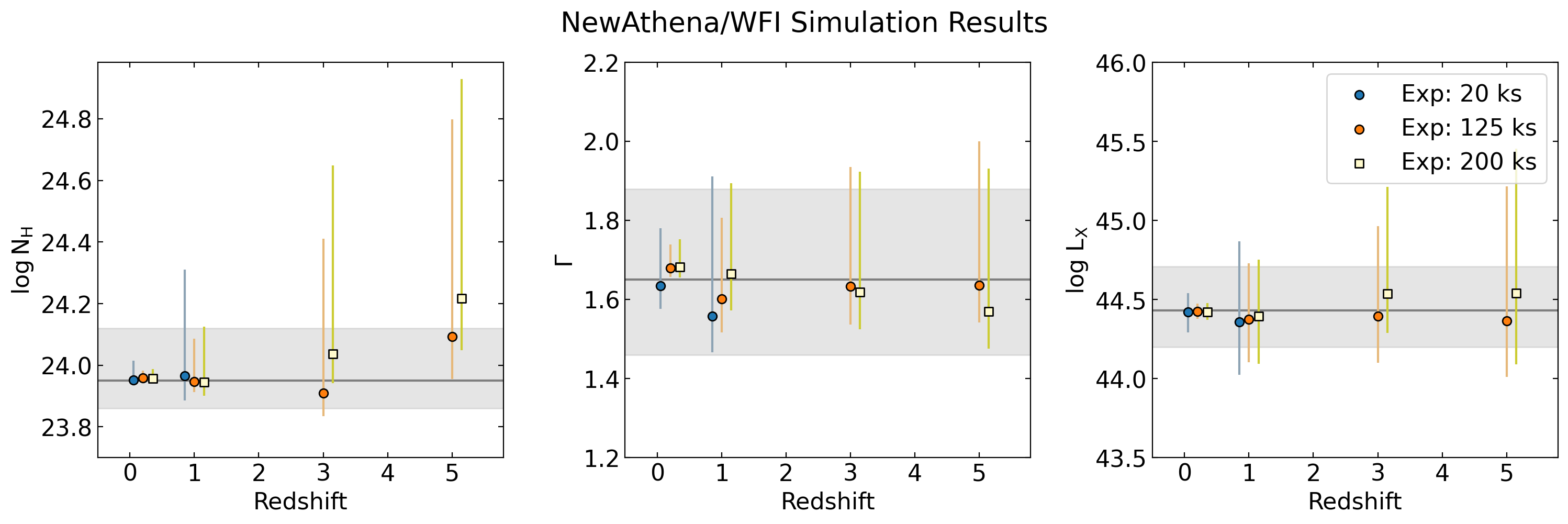}
    \caption{Comparison of column density ($\log N_\mathrm{H}$; left), photon-index ($\Gamma$; center), and intrinsic luminosity ($\log L_{\mathrm{X}}$; right) as a function of redshift for simulated observations with \textit{AXIS} (top panels) and \textit{NewAthena} (bottom panels). Simulations are based on the best-fit model of 2MASX J01290761-6038423 (BAT ID 80, z = 0.203), redshifted to z = 1, 3, and 5, and modeled for a range of exposure times. Single-pointing-like exposures of 20 ks (blue) and 125 ks (orange) were simulated for both missions. Survey-depth exposures include 335 ks (green) and 7 Ms (red) for \textit{AXIS}, and 200 ks (yellow) for \textit{NewAthena}, consistent with its planned deep survey. Circles represent single-exposure simulations, while squares indicate survey-depth cases. Points show the median values from the MCMC posteriors; error bars represent the 5th–95th percentile range. Points are slightly offset in redshift for visual clarity. Horizontal grey lines mark the original best-fit values from the real source, with shaded regions indicating their associated uncertainties. All simulations used field-of-view averaged response files and realistic background levels for each mission. Points are omitted where simulations failed to produce sufficient counts for spectral analysis.\vspace{4mm}}
    \label{fig:axis_athena}
\end{figure*}

While these predictions highlight the detection power of future surveys, detailed spectral analysis requires higher photon statistics than simple source detection. To evaluate the ability of \textit{AXIS} and \textit{NewAthena} to recover key spectral parameters for sources like those in our sample, we performed dedicated simulations based on one of our best-studied AGNs, 2MASX J01290761-6038423 (BAT ID 80), which has the largest number of spectra in our dataset. Its best-fit spectral model at $z = 0.203$ was redshifted to $z=1$, 3, and 5, while preserving the intrinsic luminosity \citep[e.g.,][]{peca24axis,tokayer25}.
We considered both typical single-pointings exposures of 20 ks and 125 ks, as well as planned survey exposures of 335 ks and 7 Ms for \textit{AXIS}, and 200 ks for \textit{NewAthena}. On-axis responses and backgrounds were used for single-pointings, while field averaged files were used for survey cases. 
Simulations were limited to $z \leq 5$, as with these exposures \textit{NewAthena} does not meet the minimum count threshold (see below) at higher redshifts, and \textit{AXIS} is constrained by the low number of high-luminosity sources expected in its planned surveys. At $z = 3$ and 5, we excluded the thermal \textit{apec} component, as it shifts out of the observable band.
Simulated spectra were analyzed over the 0.5--10 keV band using the same methods discussed in Section \ref{sec:analysis}, with 100 realizations per redshift–exposure combination. For spectra with less than 200 counts (down to 70 total counts, following \citealp{peca23}), we grouped at a minimum of 5 counts per bin and applied Cash statistics \citep{cash79}.

Results are shown in Figure \ref{fig:axis_athena}. At the same redshift of the simulated source ($z = 0.203$), both observatories perform well across all exposures. At higher redshifts, \textit{NewAthena} provides better results for single-pointing exposures, while \textit{AXIS} is limited to $z=1$ under the same conditions. For the deeper exposures in its planned surveys, \textit{AXIS} achieves excellent parameter recovery out to $z = 5$, particularly in the 7 Ms case, benefiting from deep exposure, low background, and minimal off-axis degradation. In contrast, \textit{NewAthena}’s 200 ks survey offers only marginal improvement over its 125 ks single-pointing, likely due to background and off-axis effects, but still delivering measurments up to $z=5$.

Overall, this comparison shows the complementary strengths of the two missions: \textit{NewAthena} is well suited for detailed studies of bright sources with short to moderate exposures, while \textit{AXIS} is optimized for deep surveys of faint AGNs due to its low background and stable off-axis performance. Together, these complementary capabilities will expand the redshift and luminosity range over which detailed spectral studies can be conducted. In this context, the best-fit models derived in this work offer valuable templates for testing and interpreting future observations of high-luminosity, obscured AGNs with both current and next-generation X-ray missions.

\section{Summary} \label{sec:summary}
In this work, we analyzed a sample of 21 highly luminous X-ray AGNs ($\log L_{\rm X}/\mathrm{erg\,\, s^{-1}}>44.6$), optically classified as Seyfert 1.9 and 2. We conducted a detailed spectral analysis using \nustar, \xmm, \chandra, and \suzaku\ data, incorporating multi-epoch observations where available. 
This sample probes a region of the luminosity–absorption column density parameter space that is covered almost exclusively by the BASS survey. This high-quality dataset enabled physically motivated modeling to constrain both physical and geometrical parameters, with extensive MCMC analyses used to fully explore the parameter space. In addition, we analyzed the available optical spectra, with particular focus on the detection and interpretation of the \nev\ emission line.

Our results can be summarized as follows: 
\begin{itemize}
    \item We found a weak correlation ($2<\sigma<3$) between the photon index ($\Gamma$) and Eddington ratio ($\lambda_{\rm Edd}$), consistent with prior studies. Our results confirm that the relation extends to highly luminous and obscured AGNs;
    \item Half (6/12) of the sources with black hole mass estimates lie in the absorbing column density ($N_{\rm H}$)-$\lambda_{\rm Edd}$ ``forbidden'' region. This suggests that highly luminous, obscured AGNs are likely to be caught in a transitional phase, where radiation from the central engine is beginning to clear the surrounding obscuring material.
    In support of this scenario, we found that sources within and near this region exhibit a higher incidence of $N_{\rm H}$ variability and outflows compared to those located outside this regime;
    \item The covering factors in our sample are consistent with those of Compton-thin AGNs and are systematically higher than values typically found for Compton-thick sources. This supports a scenario in which Compton-thin material extends to larger covering factors than Compton-thick obscuration;
    \item The measured iron K$\alpha$ equivalent widths (EWs) reflect the observational trend of increasing EW with higher column density, as the continuum becomes more suppressed relative to the line emission. We also observe a flattening of this trend at $\log N_{\rm H}/\mathrm{cm^{-2}} < 23$, consistent with the iron line originating from material in the inner edge of the torus or from the broad line region. Additionally, our results are consistent with the Baldwin effect, with higher-luminosity sources showing lower EWs;
    \item For sources with available multi-epoch spectroscopy, we performed a variability analysis for our sample. We found that 11 out of 13 sources ($85^{+5}_{-15}\%$) exhibit variability either in flux (10/13, $77^{+8}_{-15}\%$) or $N_{\rm H}$ (4/12, $33^{+15}_{-10}\%$). These results are in excellent agreement with recent findings on lower-luminosity ($\log L_{\rm X}/\mathrm{erg\, s^{-1}}\sim43$), Compton-thin AGNs.
    \item For 20/21 sources with adequate optical spectroscopy, we measured the \nev\ emission line and found a detection rate of $85^{+5}_{-11}\%$ (17/20). This confirms that the \nev\ emission line is a strong tracer of AGN activity even in cases of heavy obscuration, in agreement with previous studies;
    \item We stacked the optical spectra in our sample and compared the result with stacked spectra of JWST-selected, narrow-line AGNs at $2 \lesssim z \lesssim 9$, which show X-ray weakness. We found that while \nev\ is strongly detected in our local AGN stack, it is absent in the high-redshift JWST stack. Given that we matched samples in \OIII\ to ensure comparable ionization conditions, this discrepancy suggests that the nature of X-ray weak JWST AGNs at high redshift may be more complex than that of their local analogs, potentially involving different physical conditions, evolutionary stages, or environments in which \nev\ is more difficult to detect.
    \item We performed simulations using \textit{AXIS} and \textit{NewAthena} to simulate how sources similar to those in our sample would appear with next-generation X-ray telescopes. Using our best-fit models as templates, we show that both missions will offer complementary capabilities and will be essential for extending this study to higher redshifts.
\end{itemize}

Our results highlight two key considerations. First, they illustrate how high-quality X-ray observations are necessary for performing detailed spectral modeling, which can be challenging at higher redshifts with current instrumentation. Thus, the models and results presented here can serve as valuable benchmarks for future studies with planned next-generation X-ray facilities, such as \textit{AXIS} and \textit{NewAthena}. Second, local AGN studies that combine detailed X-ray spectroscopy with multi-wavelength data are crucial for interpreting recent JWST discoveries of high-redshift sources whose nature remains ambiguous. In particular, our analysis of the \nev\ emission line demonstrates how local AGNs can serve as useful laboratories, providing valuable insights into the physical conditions and possible obscuration properties of AGNs in the early Universe.

\bigskip
\facilities{\nustar, \xmm, \chandra, \suzaku}

\medskip
\software{
          NuSTARDAS v2.1.4\footnote{\href{https://heasarc.gsfc.nasa.gov/docs/nustar/analysis/}{https://heasarc.gsfc.nasa.gov/docs/nustar/analysis/}},
          SAS v21.0.0 \citep{SAS},
          CIAO v4.16.0 \citep{fruscione06},
          PyXspec v2.1.3 \citep{pyxspec,xspec},
          TOPCAT v4.94.9 \citep{topcat},
          Astropy v7.0.1 \citep{astropy:2013,astropy:2018,astropy:2022};
          Pandas v2.1.4 \citep{pandas1,pandas2},
          SciPy v1.15.2 \citep{scipy},
          Matplotlib v3.10.1 \citep{matplotlib},
          linmix v0.1.0 \citep{kelly07}.
          }


\bigskip
\section*{acknowledgments}
We acknowledge the anonymous referee for the valuable comments that improved the quality of the paper. A.P. acknowledges D. Costanzo for all the support over the years. 
A.P. and M.K. acknowledge support from NASA through ADAP award 80NSSC22K1126 and NuSTAR grants 80NSSC22K1933 and 80NSSC22K1934.
K.O. acknowledges support from the Korea Astronomy and Space Science Institute under the R\&D program (Project No. 2025-1-831-01), supervised by the Korea AeroSpace Administration, and the National Research Foundation of Korea (NRF) grant funded by the Korea government (MSIT) (RS-2025-00553982).
B.T. acknowledges support from the European Research Council (ERC) under the European Union's Horizon 2020 research and innovation program (grant agreement number 950533) and from the Israel Science Foundation (grant number 1849$/$19).
C.R. acknowledges support from Fondecyt Regular grant 1230345, ANID BASAL project FB210003 and the China-Chile joint research fund. 
A.T. acknowledges financial support from the Bando Ricerca Fondamentale INAF 2022 Large Grant ``Toward an holistic view of the Titans: multi-band observations of $z>6$ QSOs powered by greedy supermassive black holes''.
I.MC acknowledges support from ANID programme FONDECYT Postdoctorado 3230653.
M.S. acknowledges financial support from the Italian Ministry for University and Research, through the grant PNRR-M4C2-I1.1-PRIN 2022-PE9-SEAWIND: Super-Eddington Accretion: Wind, INflow and Disk - F53D23001250006 - NextGenerationEU.

\clearpage
\appendix

\section{Notes on individual sources}\label{app:srcs_notes}
The baseline model described in Section \ref{sec:analysis} was adjusted for single sources, where necessary, according to the model selection procedure described in Section \ref{sec:results_xrays_dic}.
A second \textit{apec} model was included to improve the fit at soft energies for LEDA2816387 (BAT ID 199).
For 2MASS J02162672+5125251 (BAT ID 119), LEDA 511869 (BAT ID 714), PKS 1549-79 (BAT ID 787), GALEXMSC J025952.92+245410.8 (BAT ID 1248), and GALEXASC J063634.15+591319.6 (BAT ID 1296), we turned off the \textit{apec} emission, as the secondary power-law component adequately accounted for the soft photons from either \chandra\ and \xmm\ spectra, without the need for an additional component. 
For 2MASS J17422050–5146223 (BAT ID 1515), we turned off both the \textit{apec} and scattering components, as the primary power-law emission alone provides an adequate fit to the soft X-ray spectrum. Indeed, the source is unobscured in the X-rays, with an upper limit on $N_{\rm H}$ of $1.2 \times 10^{21} \, \mathrm{cm}^{-2}$, consistent with the high Galactic column density in that direction.

For RBS 2043 (BAT ID 1204) and 2MASX J09003684+2053402 (BAT ID 1346), which reside at the center of the Phoenix and Z2089 galaxy clusters, respectively, we used two \textit{apec} components to model the cluster thermal emission. Specifically, we used sub-solar abundances ($\sim0.33$ and $\sim0.76$ for the two components) as described by \cite{ueda13} in a detailed analysis of the Phoenix cluster. The temperatures of the components, shown in Table \ref{tab:best_results}, agree with their results. 
We used the same approach for Z2089, as we found a satisfactory fit using the same abundances. Here, the temperatures of the thermal components are in agreement with the thermal profile reported in \cite{giles17}.

For RBS 2043 (BAT ID 1204), we found a larger scattering fraction ($\sim14\%$) than the other sources in our sample. This value remains consistent, within the uncertainties, with the results of \citet{ricci17}. While such a high scattering fraction is in line with findings for high luminosity AGNs in X-ray surveys \citep[e.g.,][]{ricci17,peca23}, we acknowledge that it may be affected by the cluster thermal emission, as both the components cover the same energy range.

In the \xmm\ 2021 observation of 2MASX J01290761-6038423 (BAT ID 80), we used only the PN data, as the MOS cameras lacked good coverage in the detector chips where the source was observed after background filtering. We utilized only the PN camera also for 2MASS J02162672+5125251 (BAT ID 119) and PKS 1549-79 (BAT ID 787), as both MOS cameras were affected by pile-up issues.

The \chandra observation used for BAT ID 1248 is contained in the Chandra Data Collection~\dataset[DOI: 10.25574/cdc.408]{https://doi.org/10.25574/cdc.408}

\begin{figure*}[!tp]
    \centering
    \includegraphics[scale=0.4]{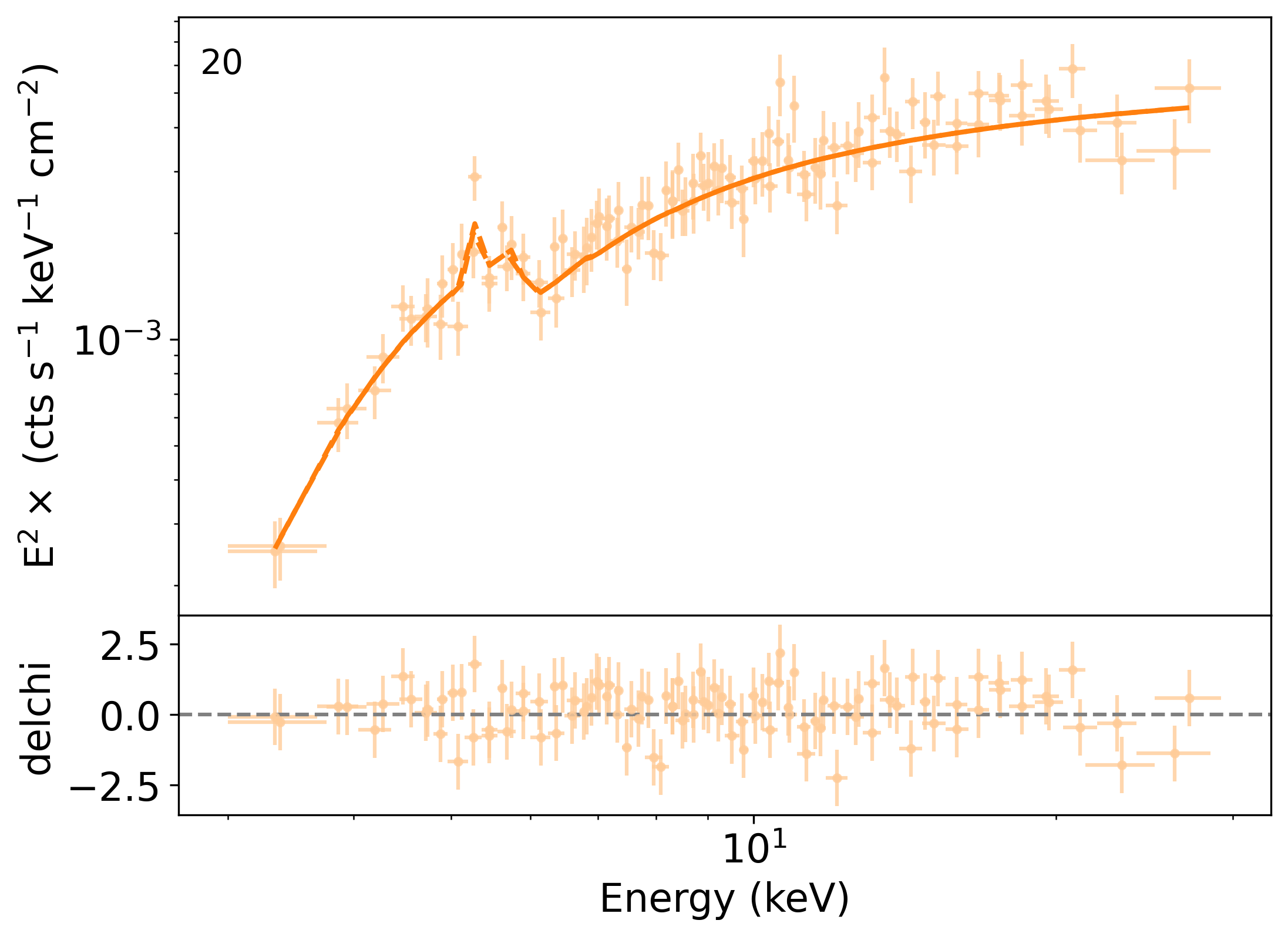}  \hspace{2mm}
    \includegraphics[scale=0.4]{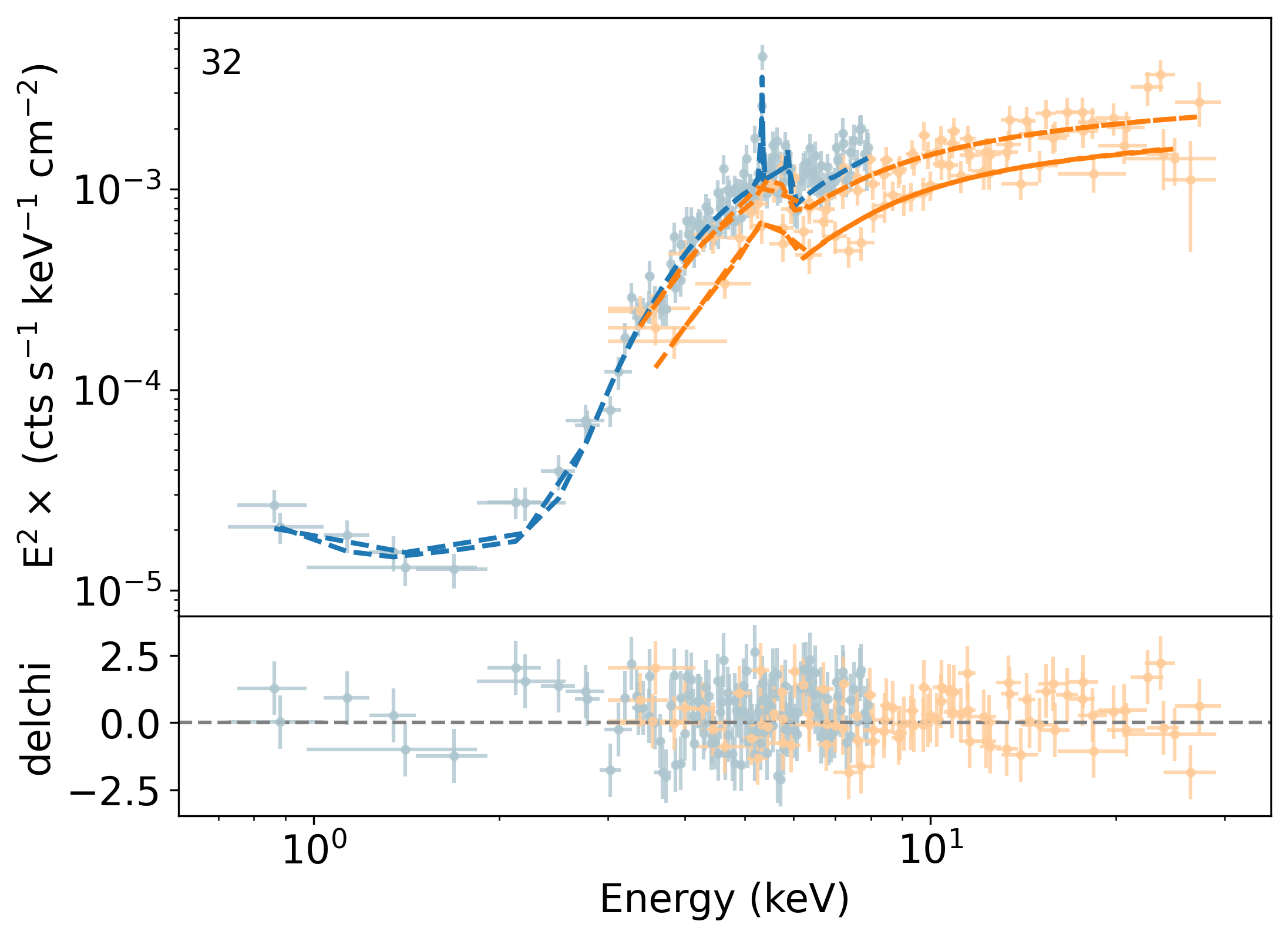}  \hspace{2mm}
    \includegraphics[scale=0.4]{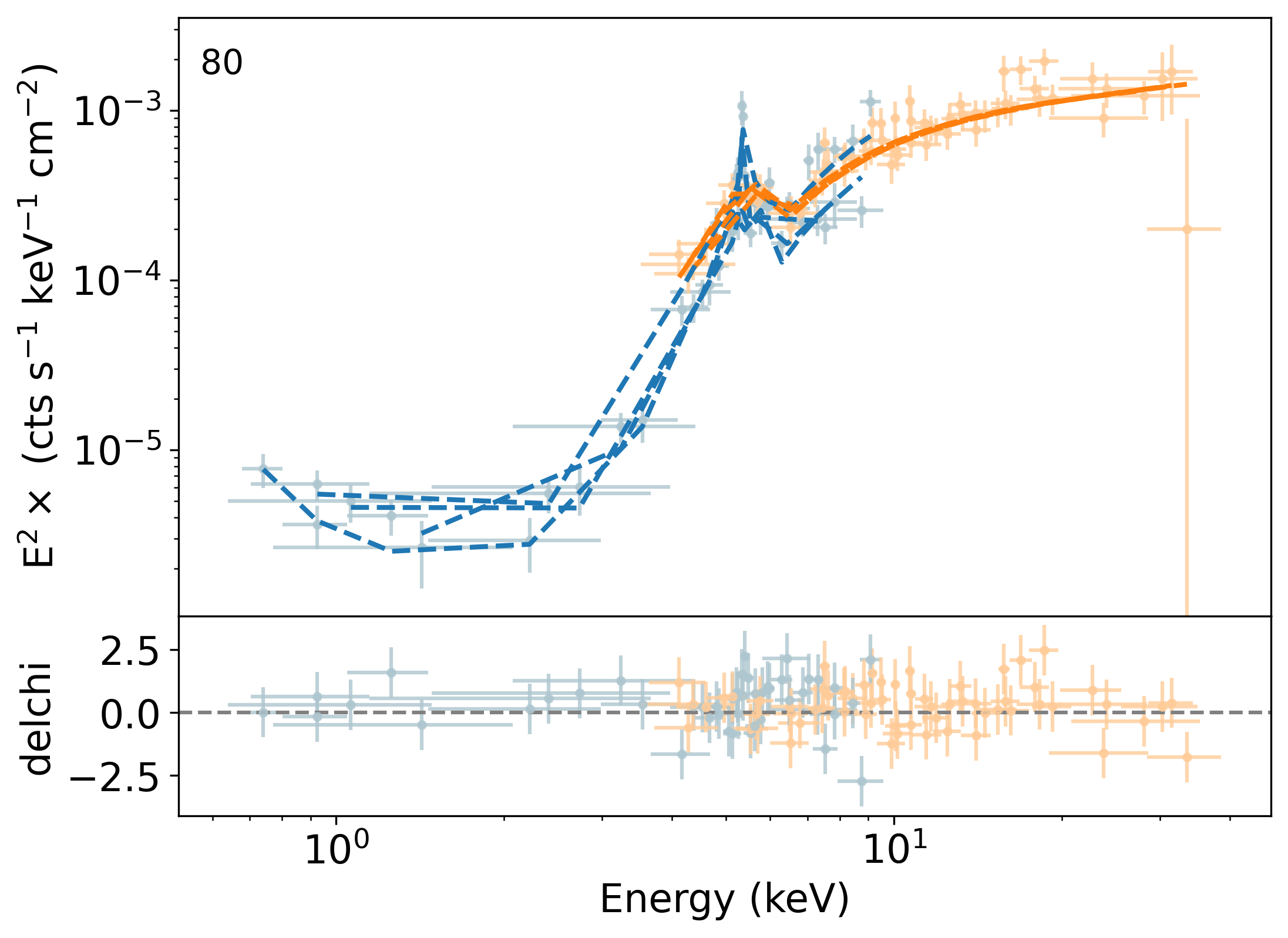}  \hspace{2mm}
    \includegraphics[scale=0.4]{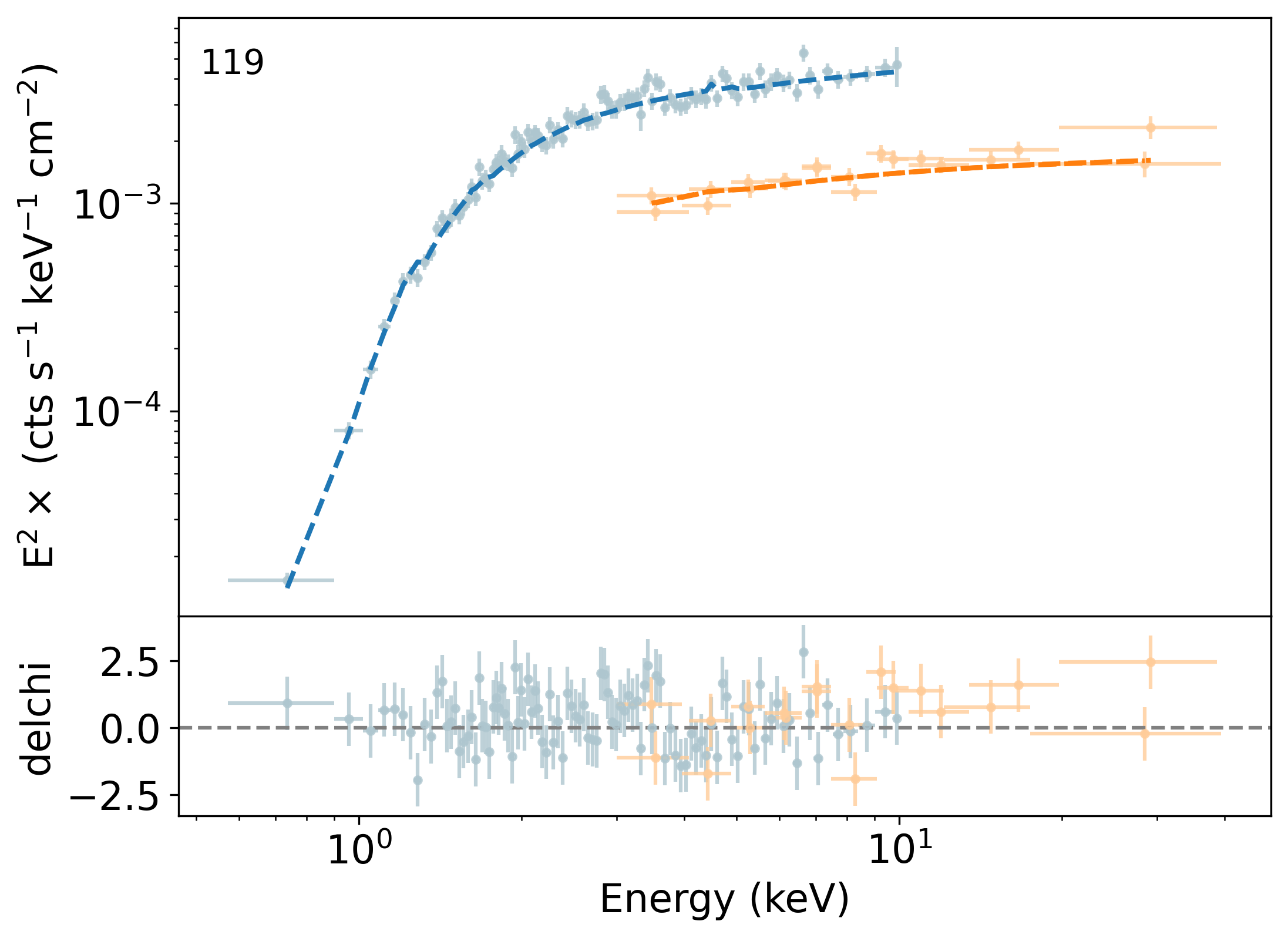}  \hspace{2mm}
    \includegraphics[scale=0.4]{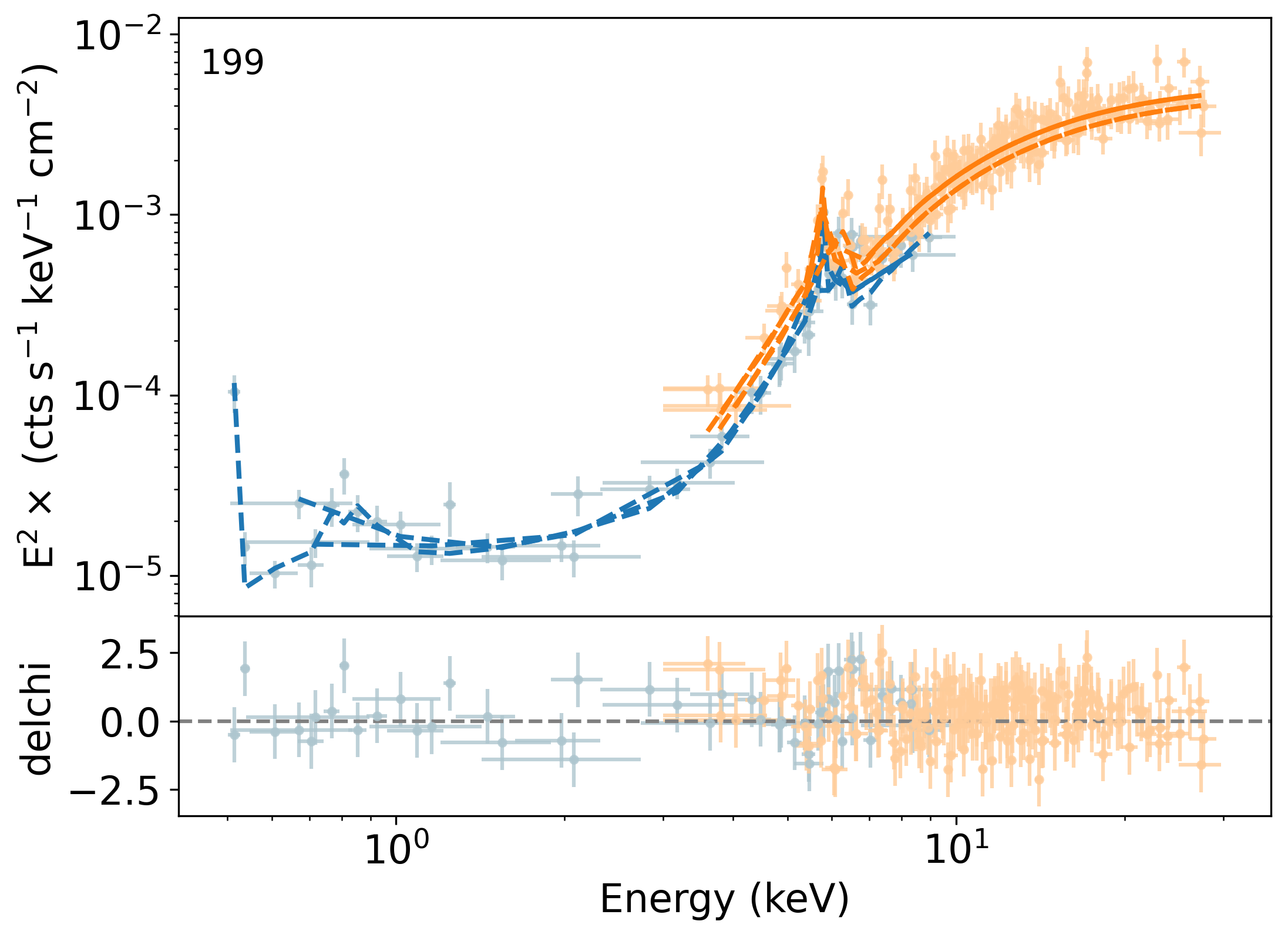}  \hspace{2mm}
    \includegraphics[scale=0.4]{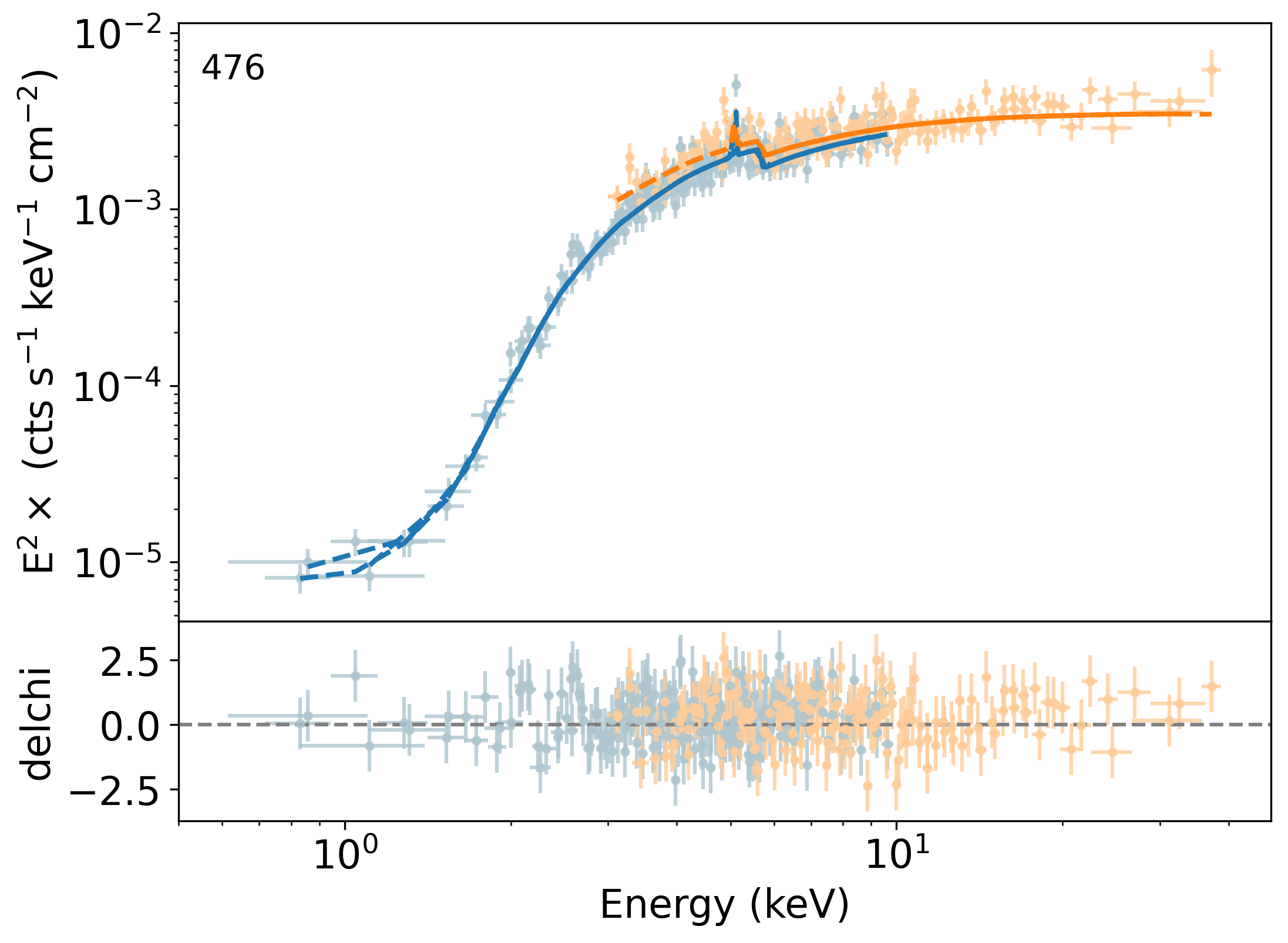}  \hspace{2mm}
    \includegraphics[scale=0.4]{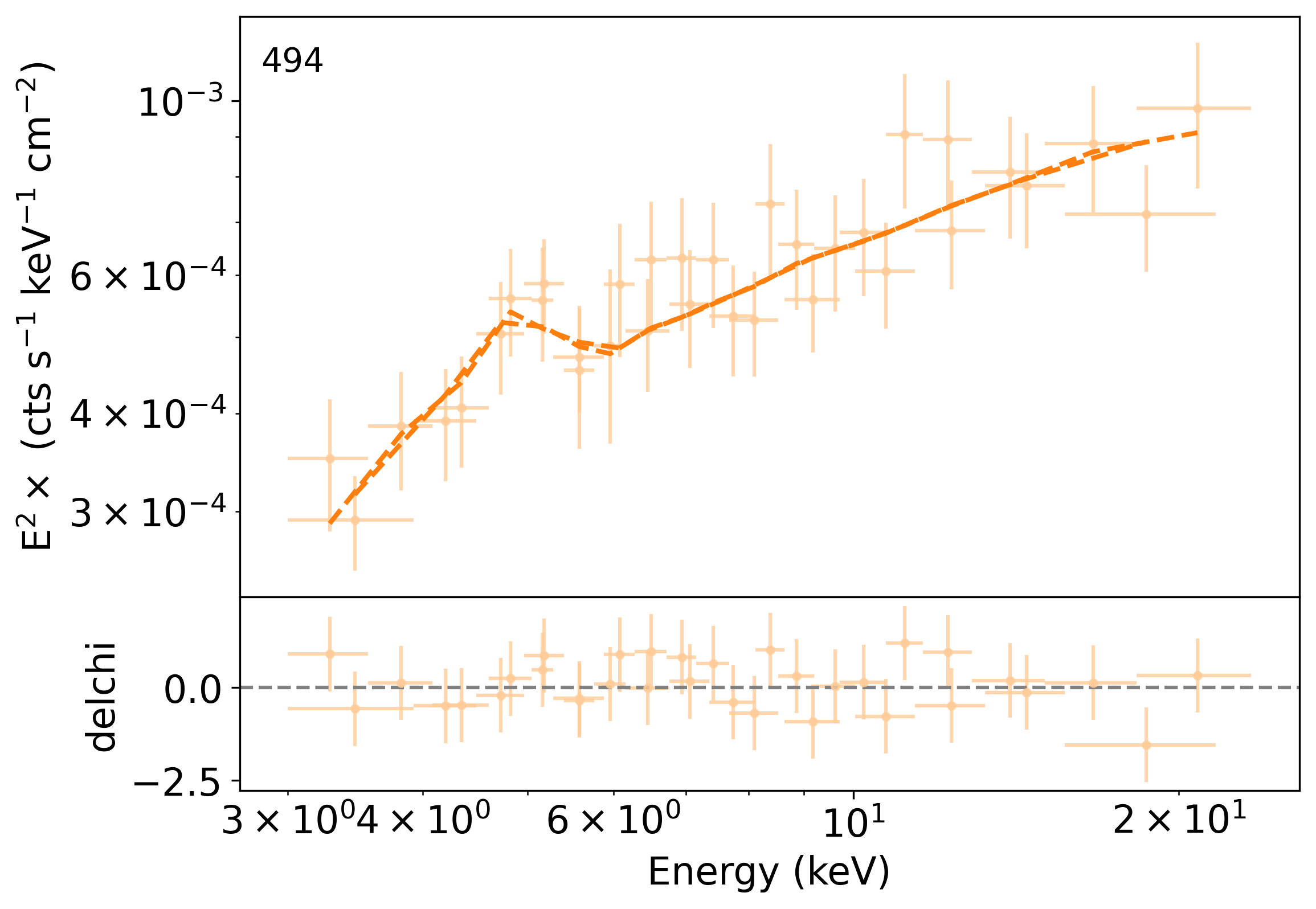}  \hspace{2mm}
    \includegraphics[scale=0.4]{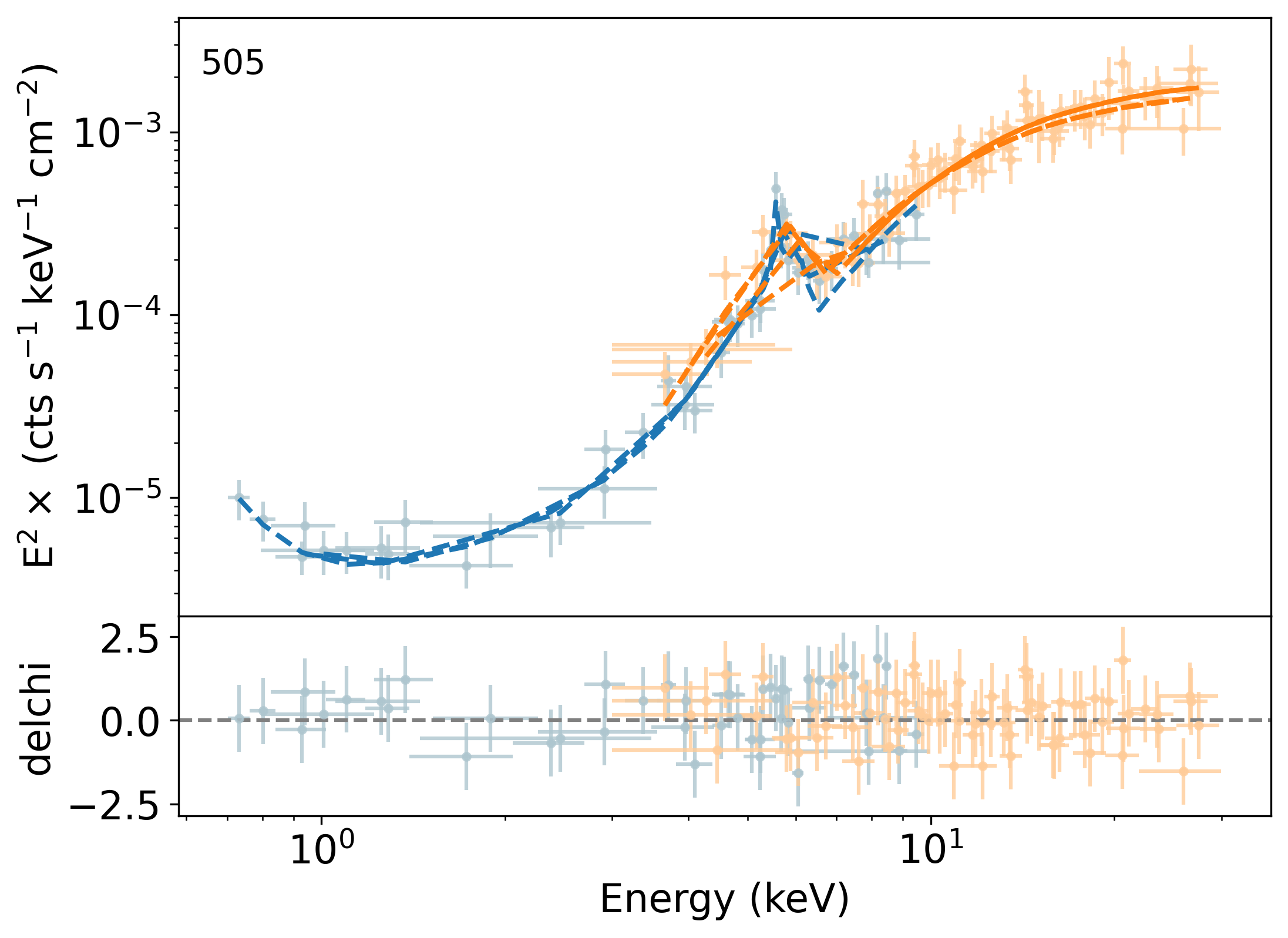}  \hspace{2mm}
    \caption{Best-fit unfolded spectra of the sources analyzed in this work. The corresponding models are shown as dashed lines, and the BAT ID is indicated in the top-left corner of each panel. For clarity, \textit{NuSTAR} spectra are shown in orange, while soft X-ray spectra (\textit{XMM-Newton}, \textit{Chandra}, and \textit{Suzaku}) are shown in blue. Residuals are displayed in the lower subpanels. All spectra have been rebinned for visual clarity.}
    \label{fig:bestfits1}
\end{figure*}

\begin{figure*}[!tp]
    \centering
    \includegraphics[scale=0.4]{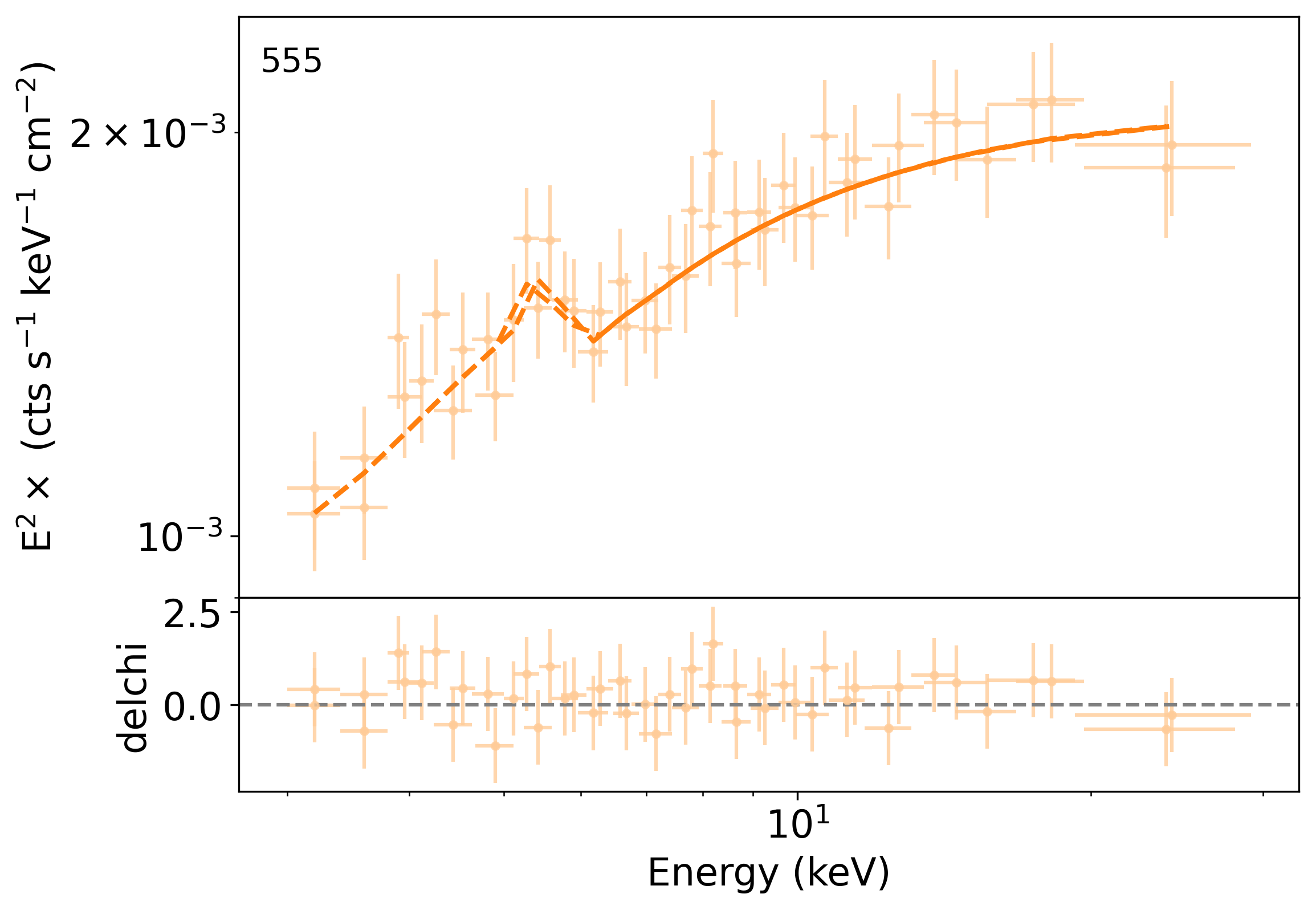}  \hspace{2mm}
    \includegraphics[scale=0.4]{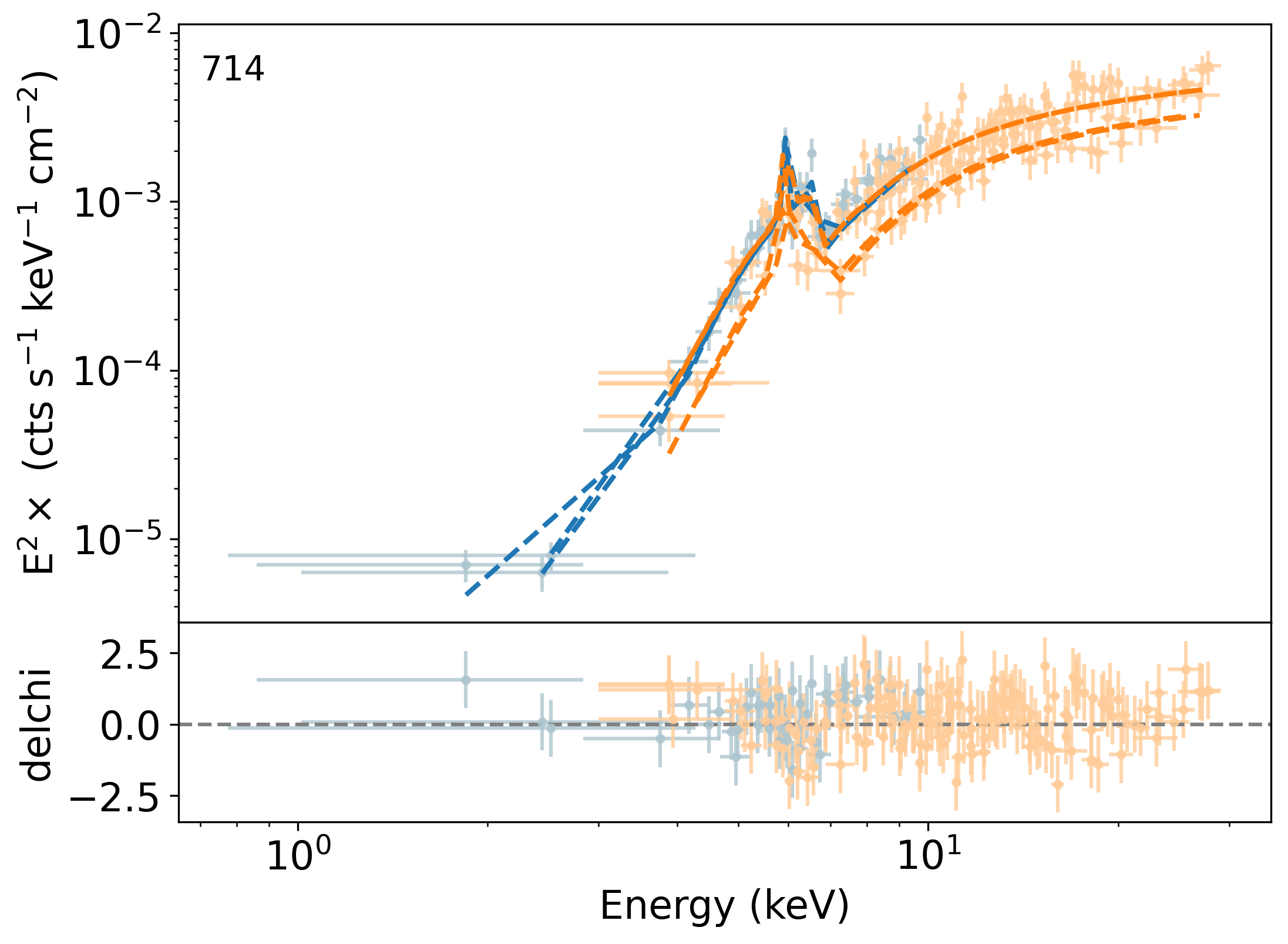}  \hspace{2mm}
    \includegraphics[scale=0.4]{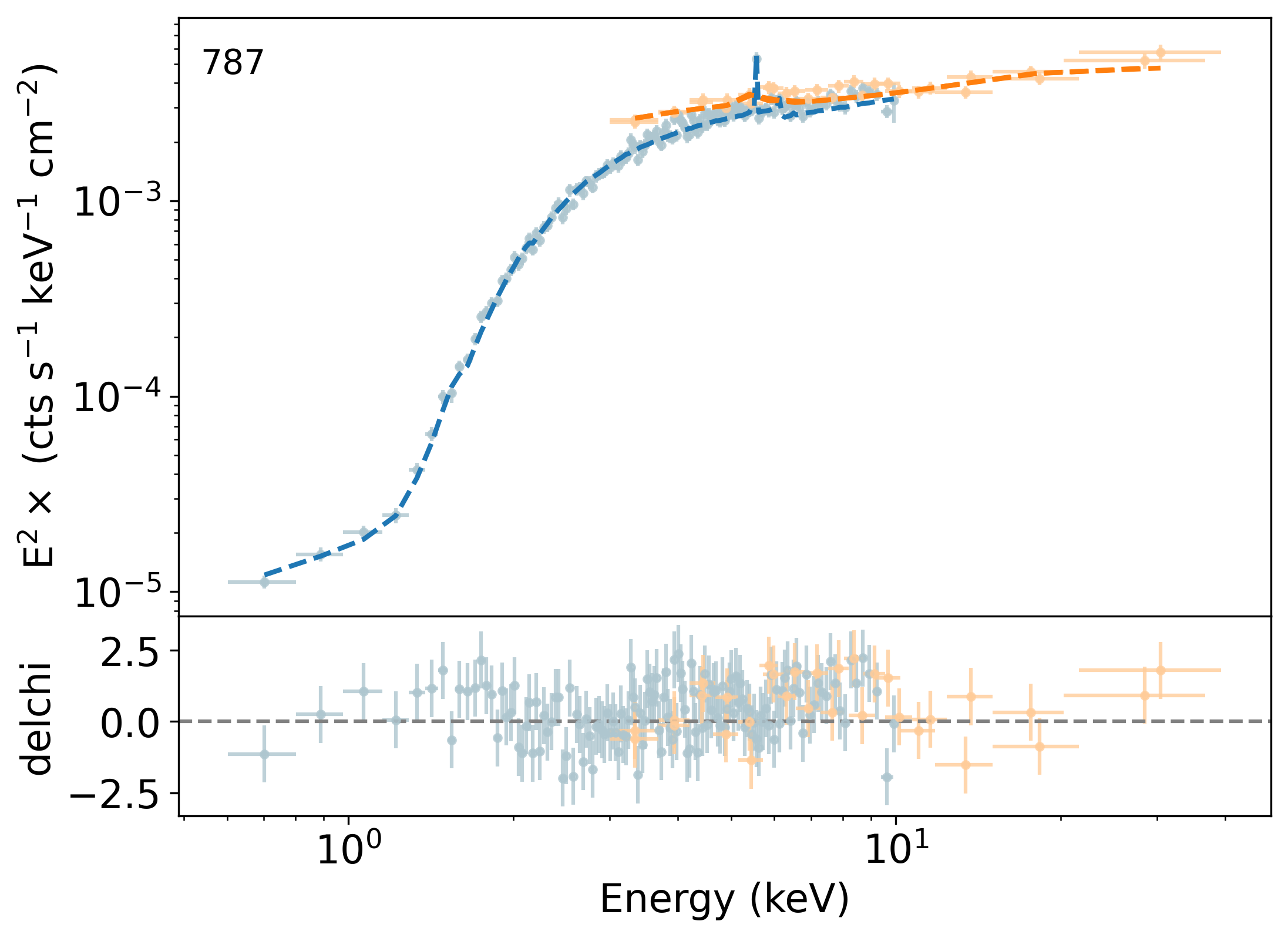}  \hspace{2mm}
    \includegraphics[scale=0.4]{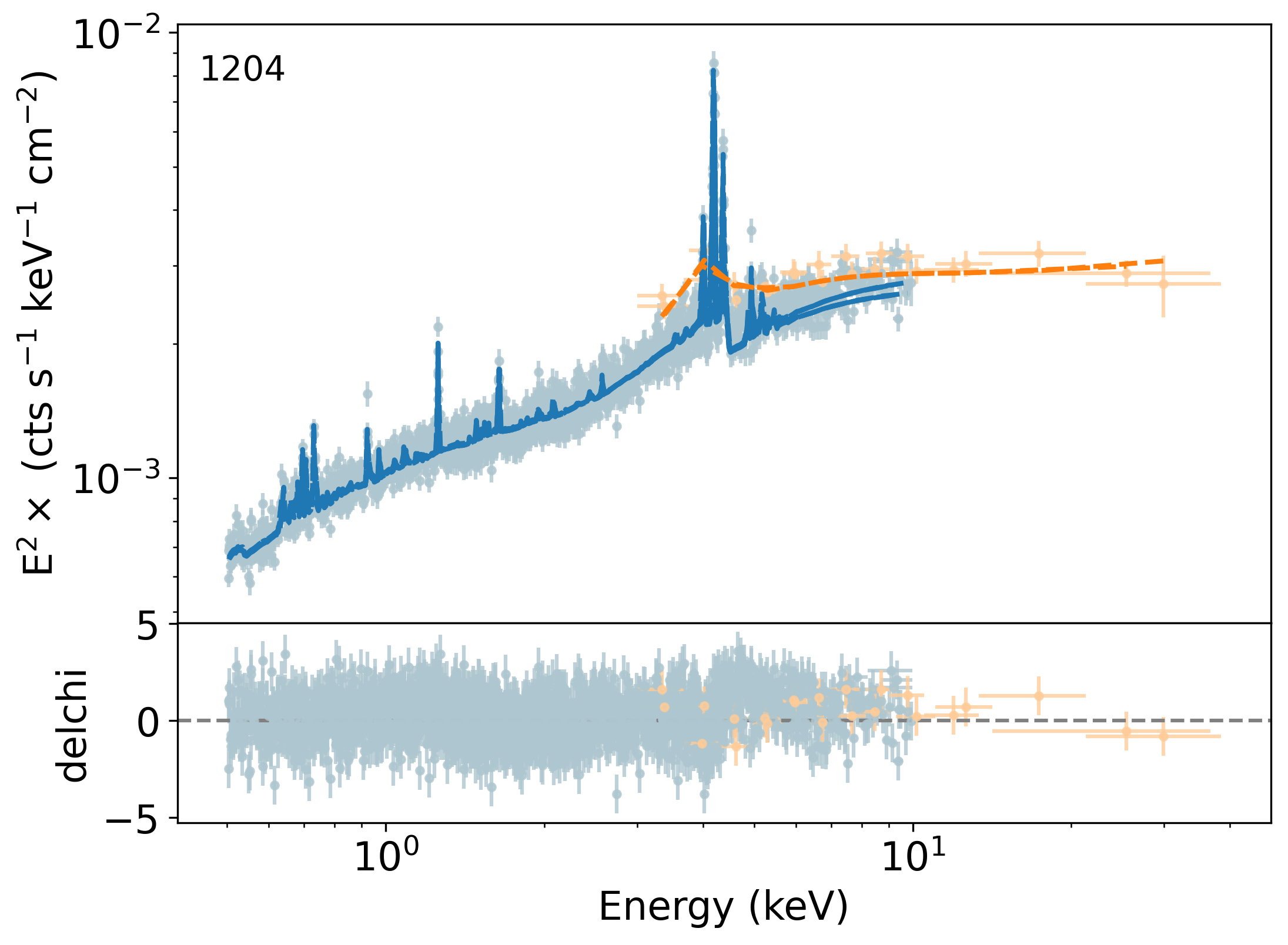}  \hspace{2mm}
    \includegraphics[scale=0.4]{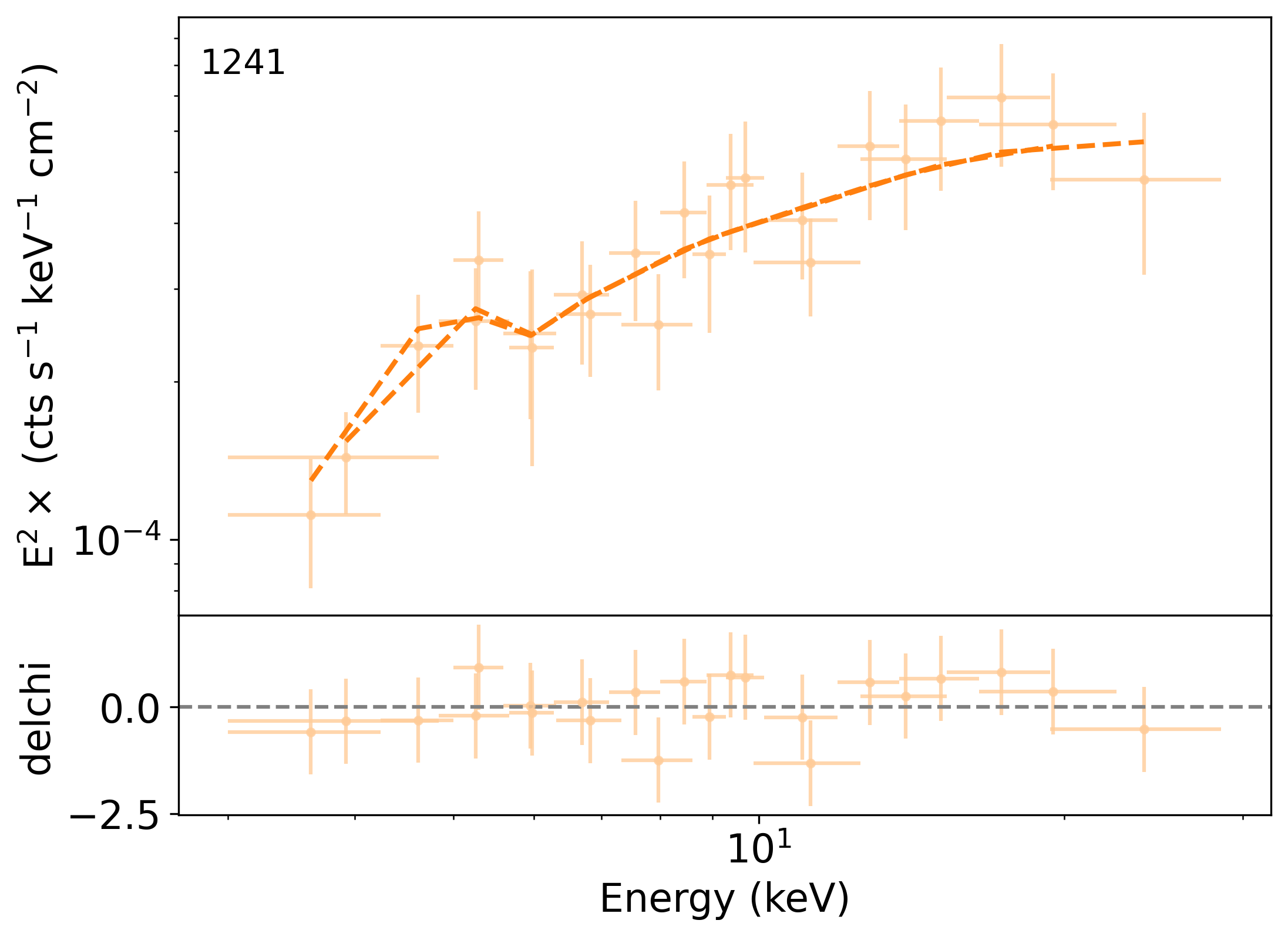}  \hspace{2mm}
    \includegraphics[scale=0.4]{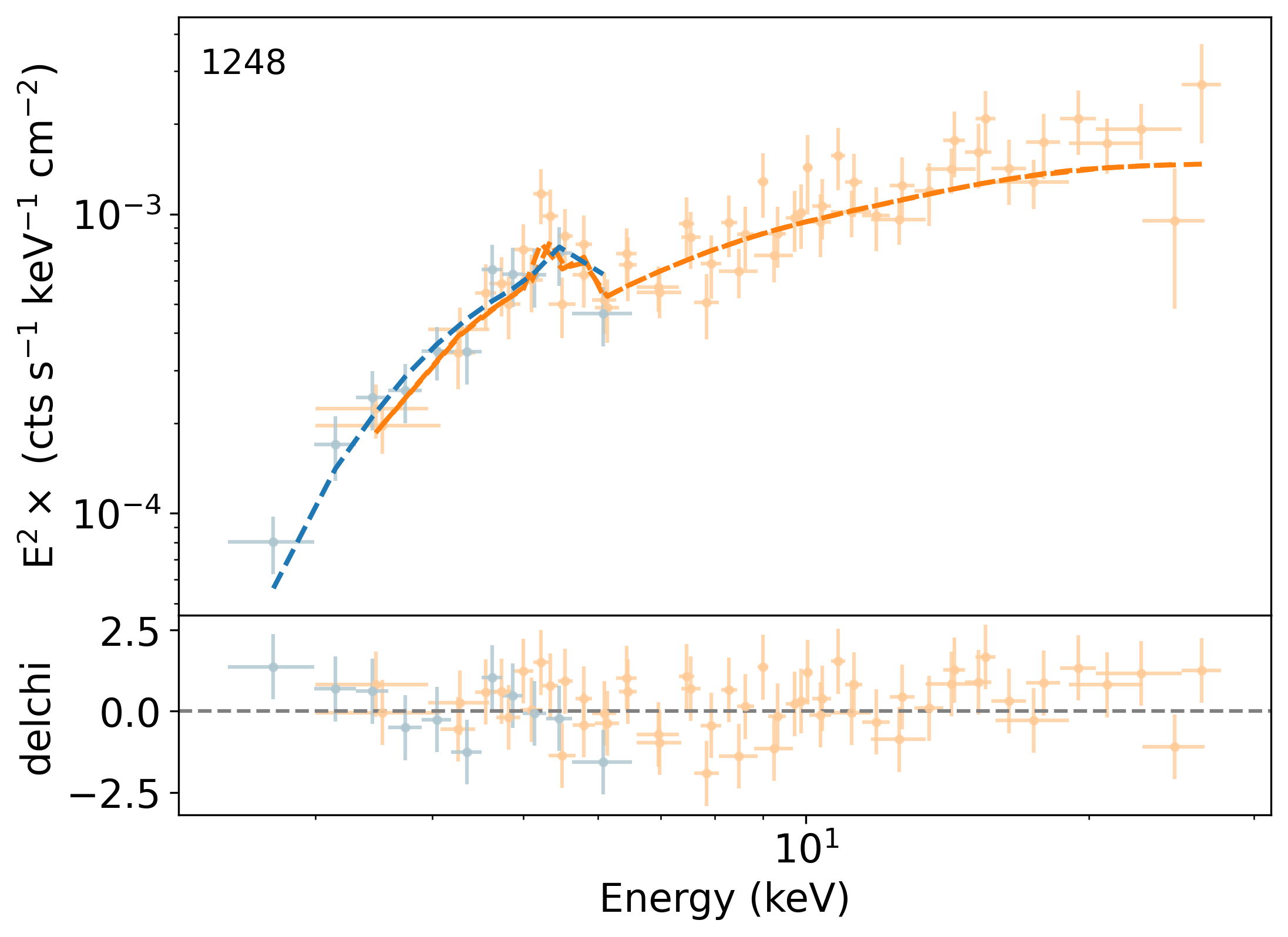}  \hspace{2mm}
    \includegraphics[scale=0.4]{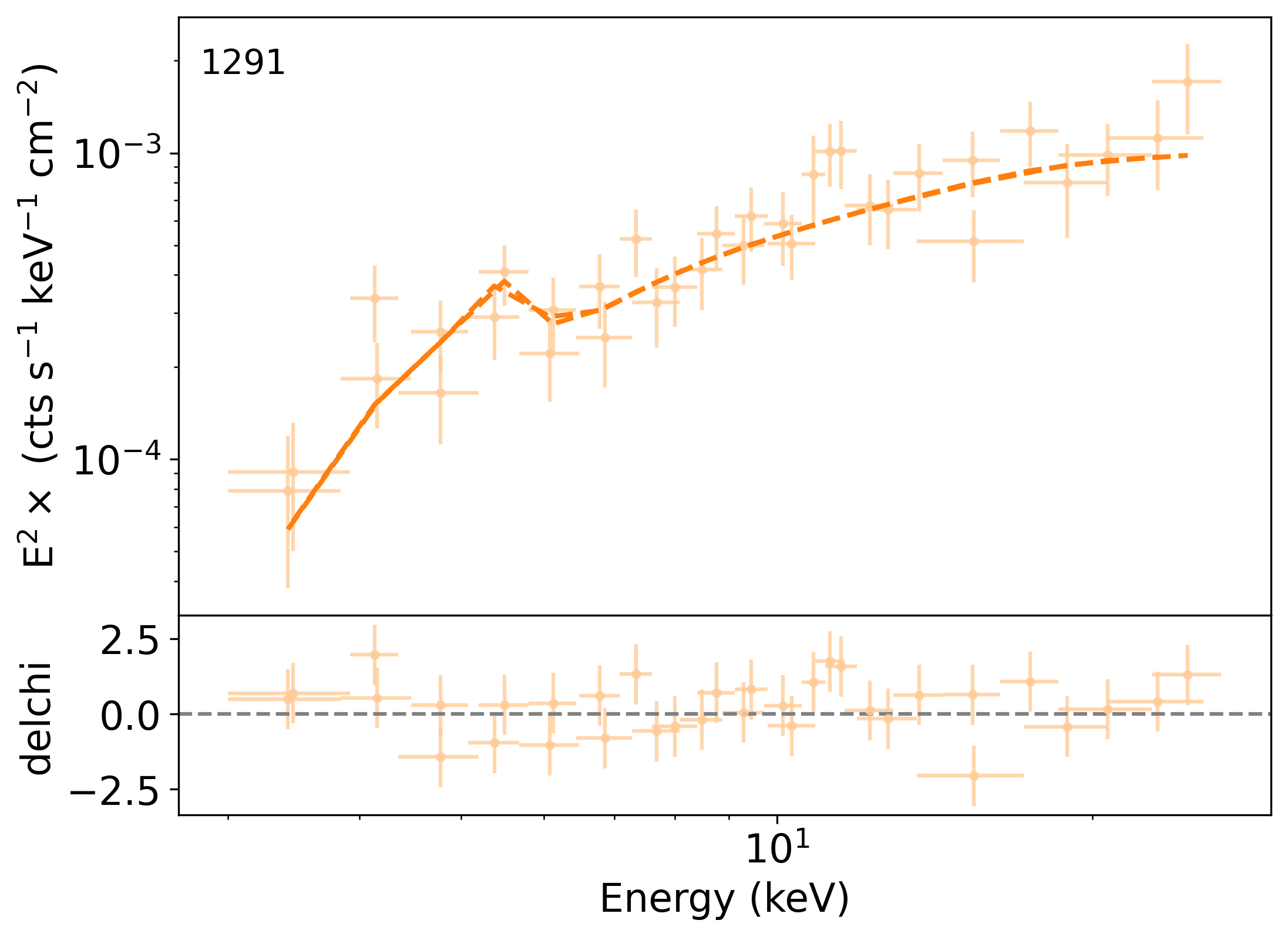}  \hspace{2mm}
    \includegraphics[scale=0.4]{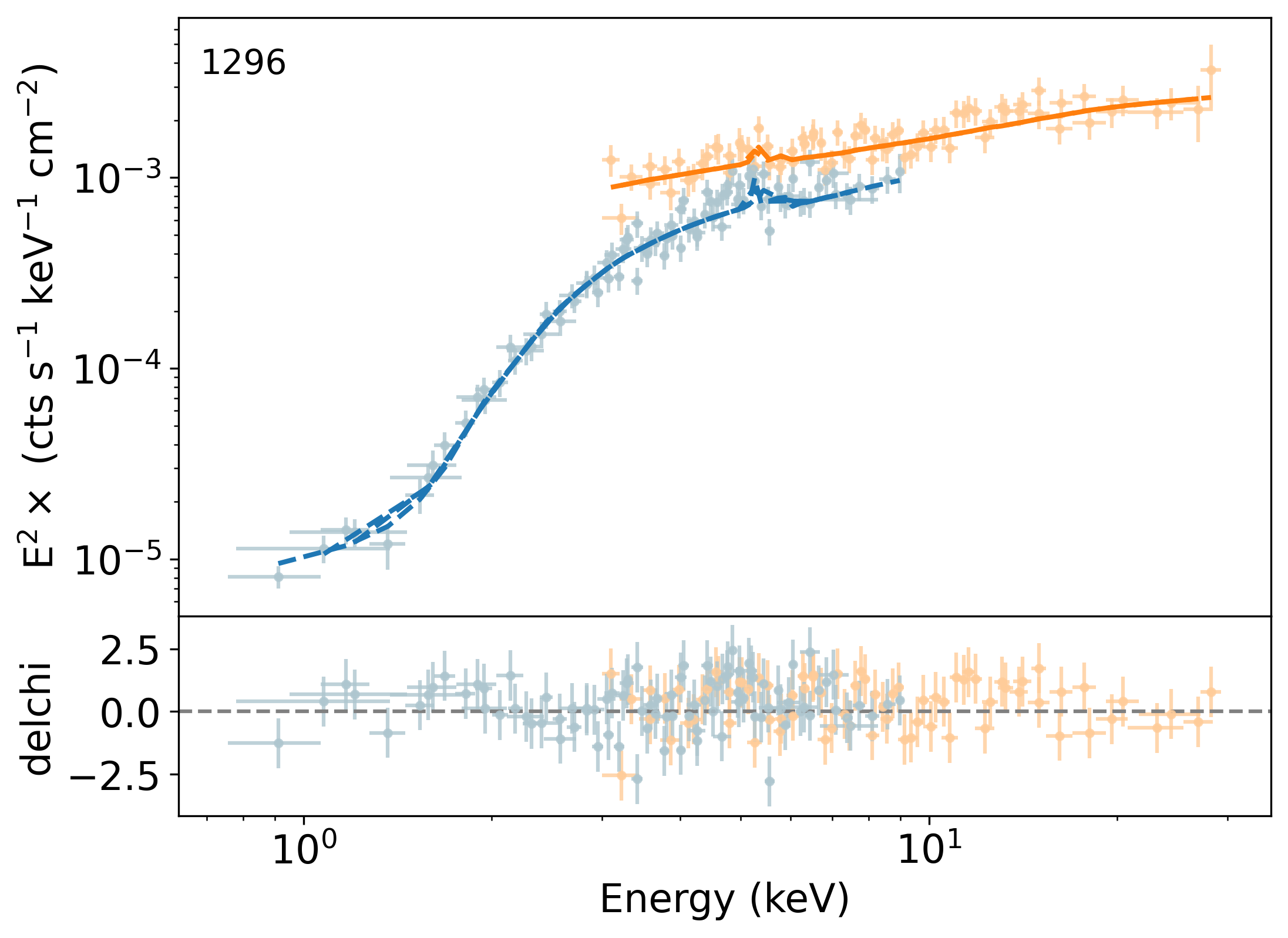}  \hspace{2mm}
    \caption{Same as Figure \ref{fig:bestfits1}.}
    \label{fig:bestfits2}
\end{figure*}

\begin{figure*}[!tp]
    \centering
    \includegraphics[scale=0.4]{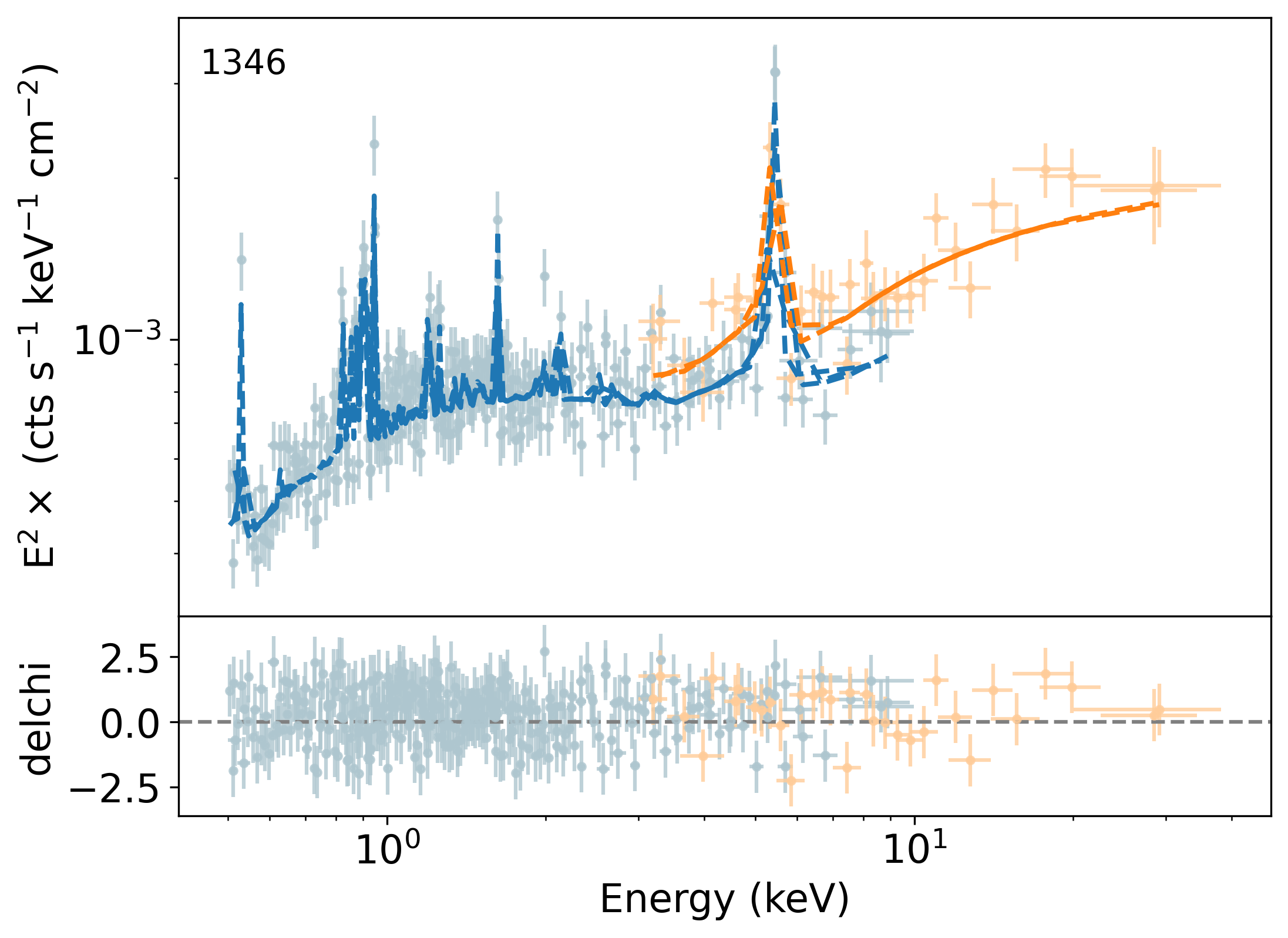}  \hspace{2mm}
    \includegraphics[scale=0.4]{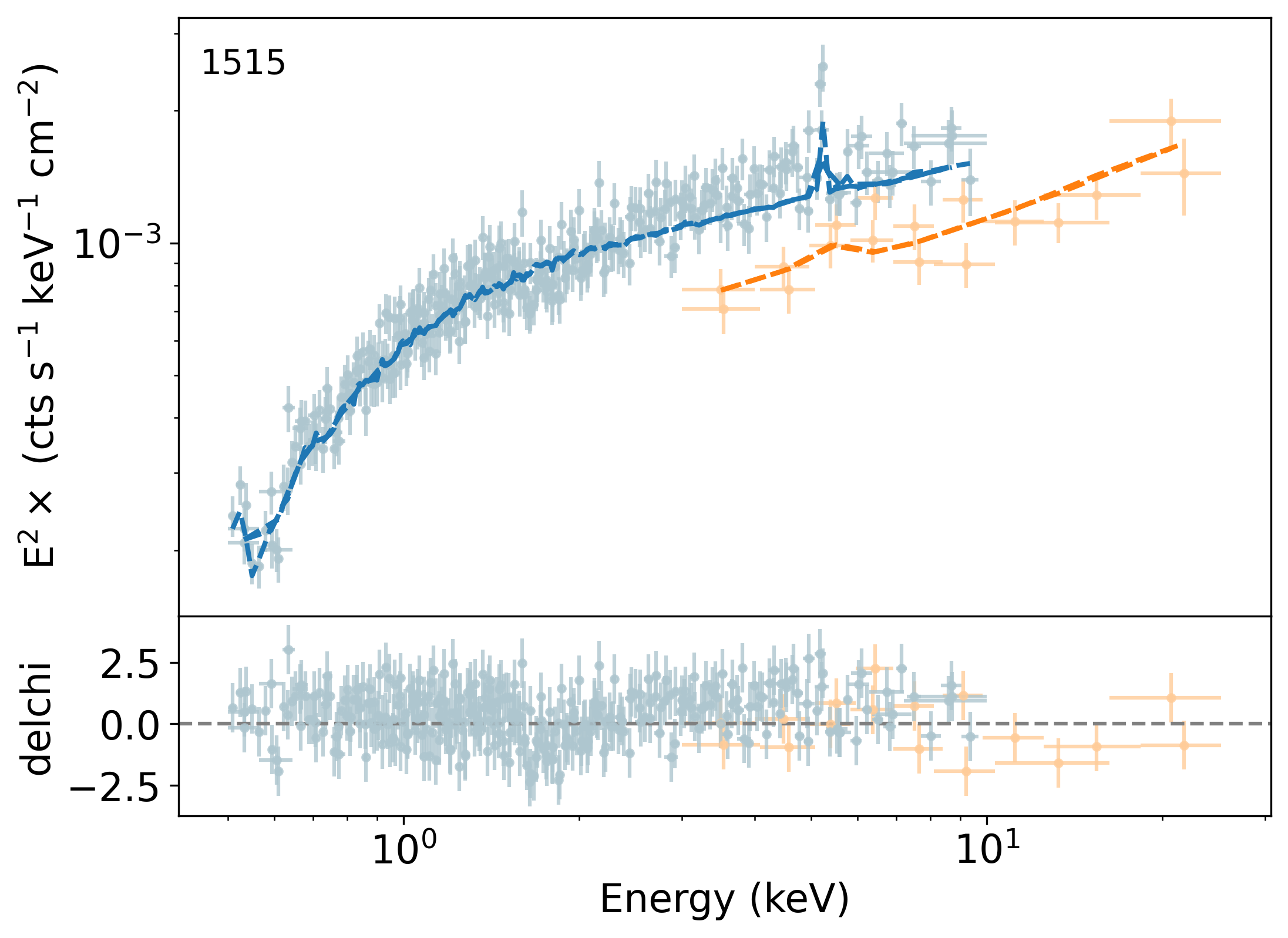}  \hspace{2mm}
    \includegraphics[scale=0.4]{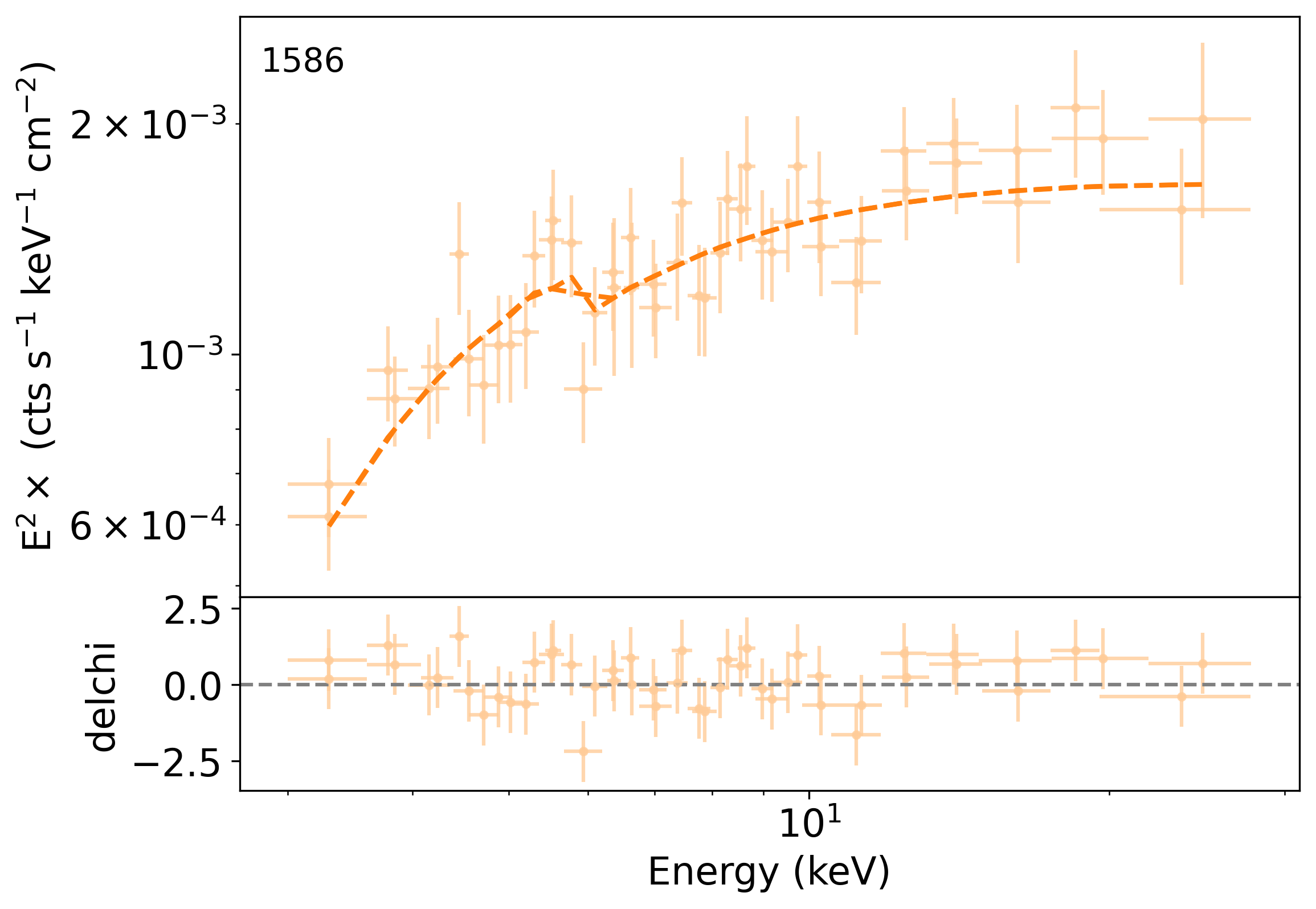}  \hspace{2mm}
    \includegraphics[scale=0.4]{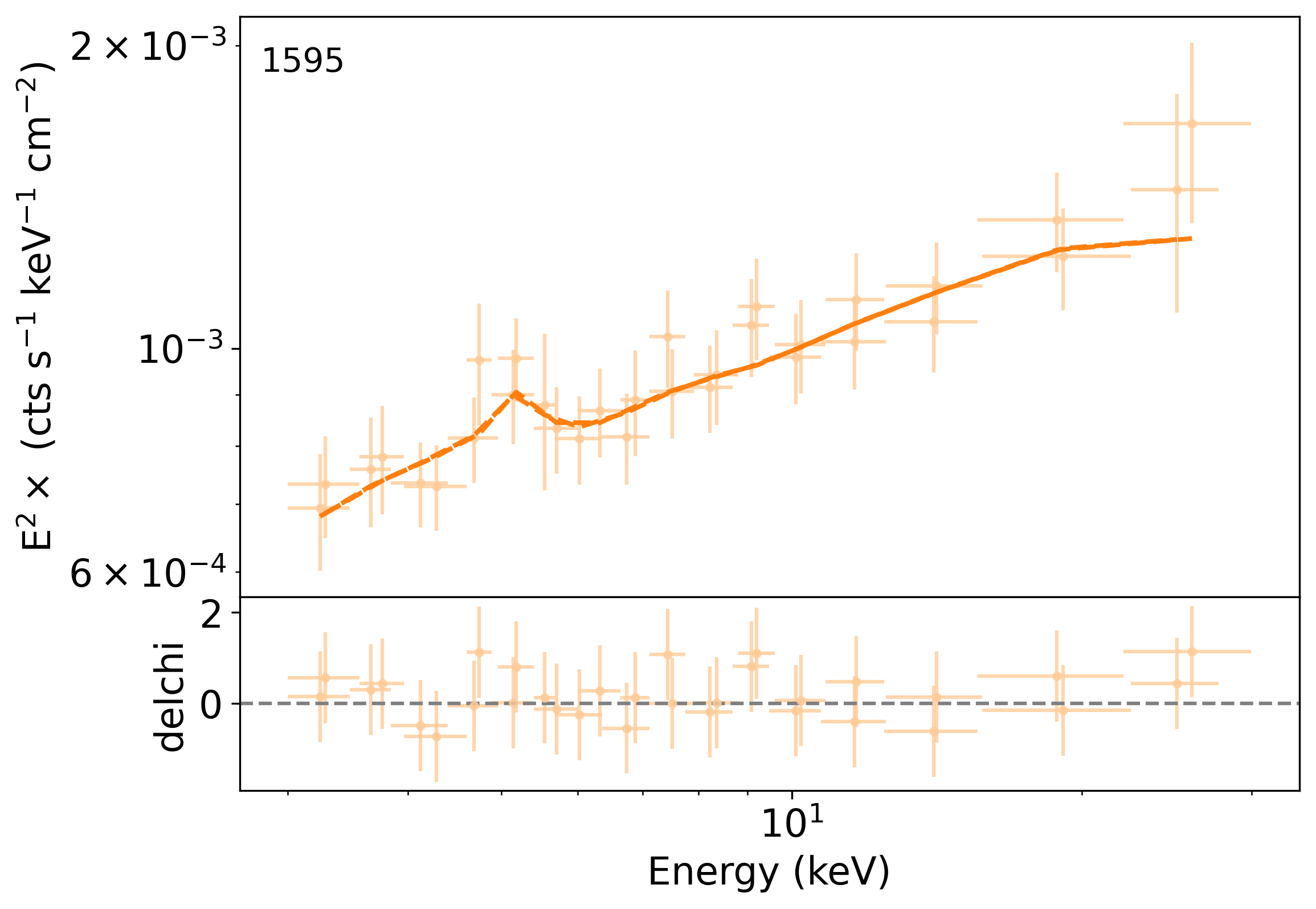}  \hspace{2mm}
    \includegraphics[scale=0.4]{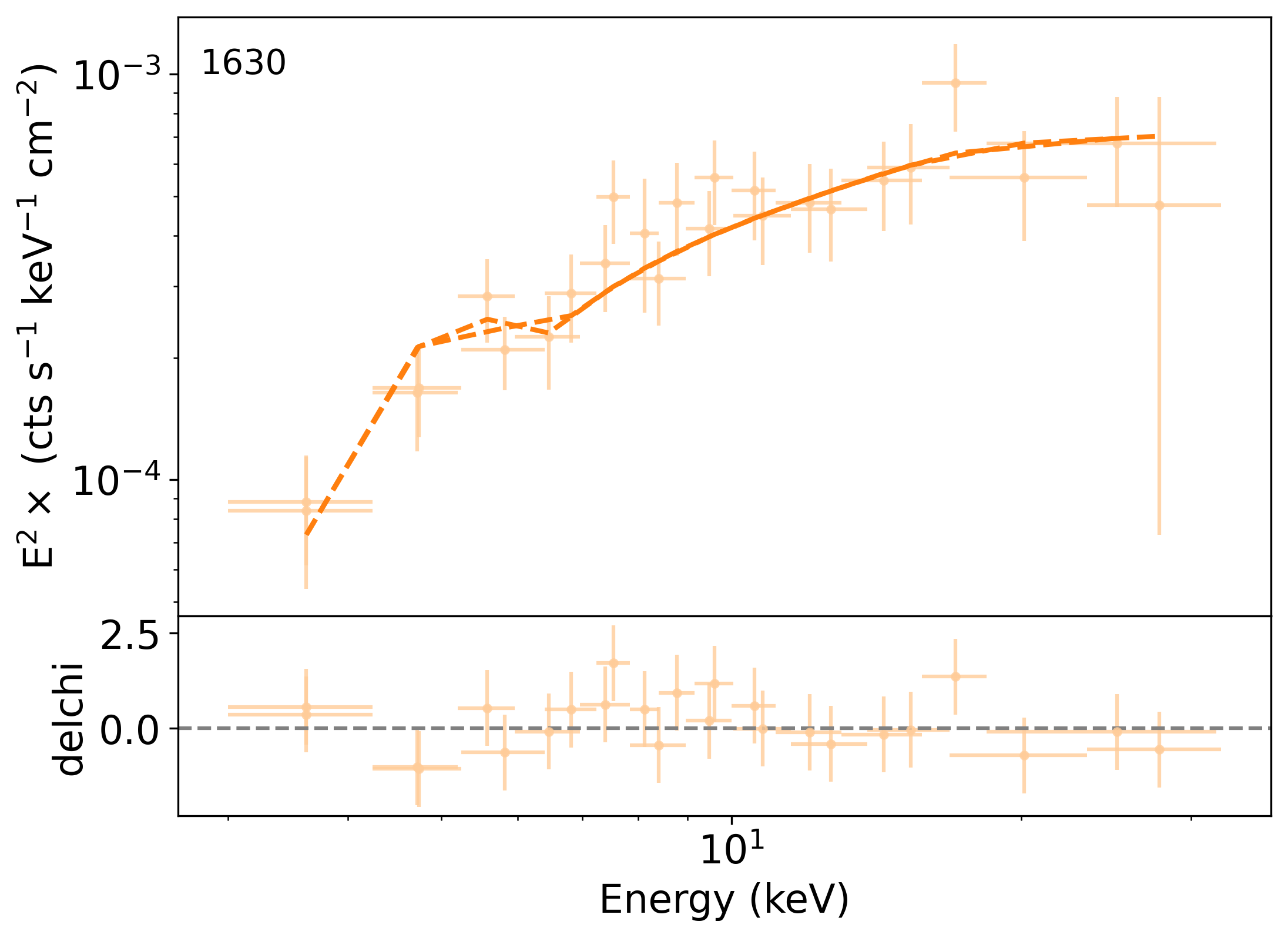}  \hspace{2mm}
    \caption{Same as Figure \ref{fig:bestfits1}.}
    \label{fig:bestfits3}
\end{figure*}

\begin{figure*}[!tp]
    \centering
    \includegraphics[scale=0.57]{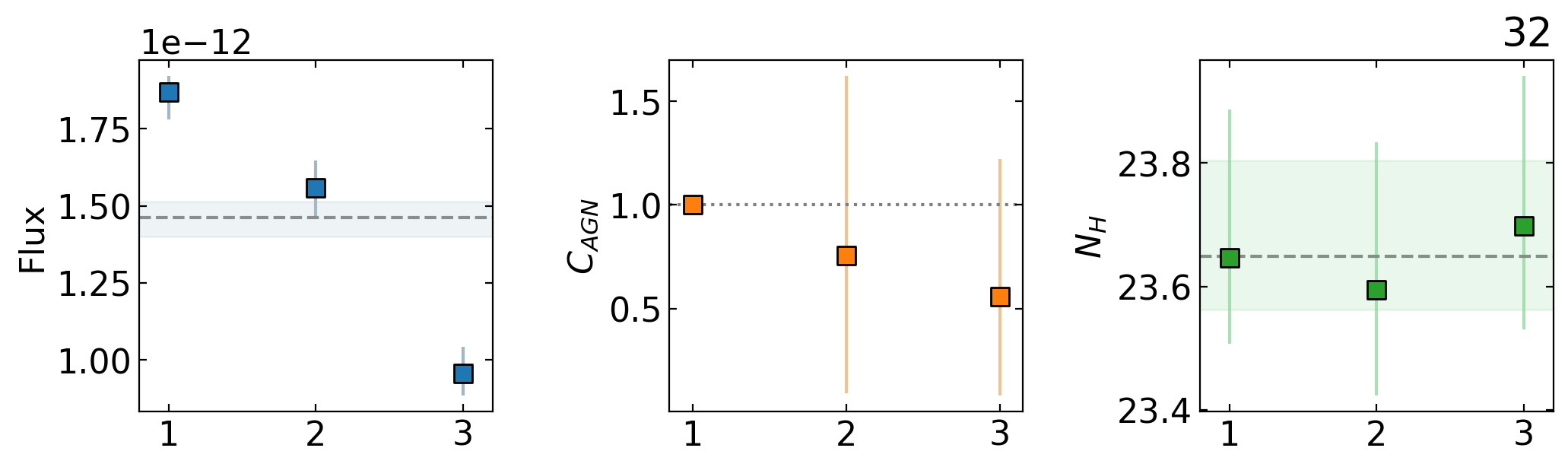}
    \includegraphics[scale=0.57]{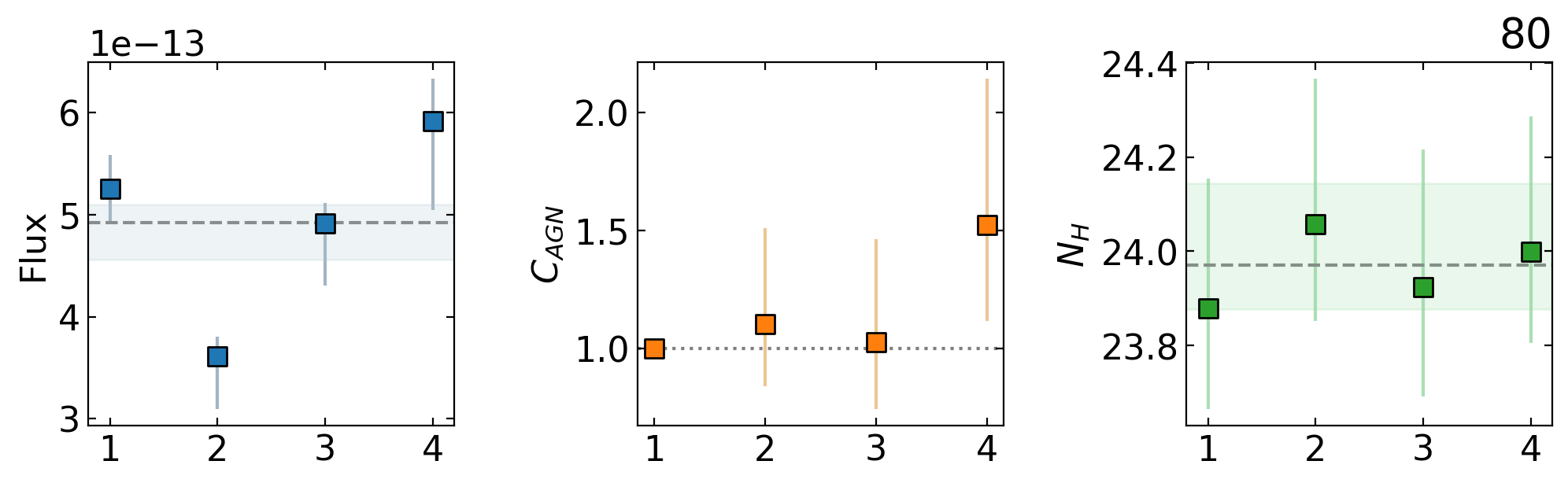}
    \includegraphics[scale=0.57]{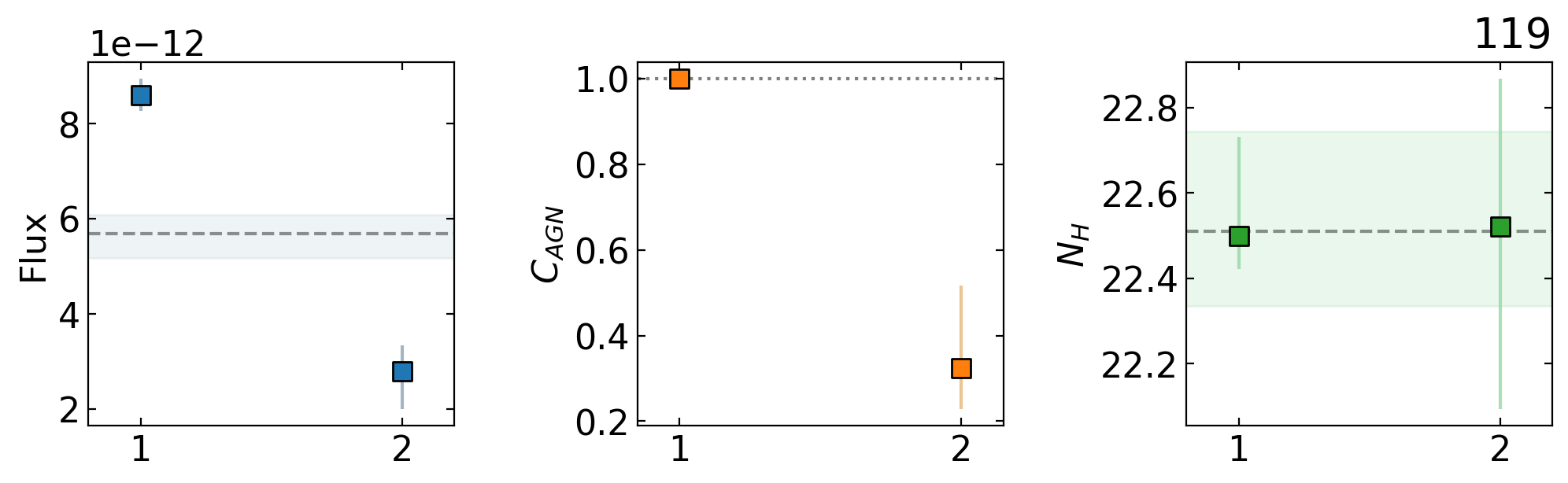}
    \includegraphics[scale=0.57]{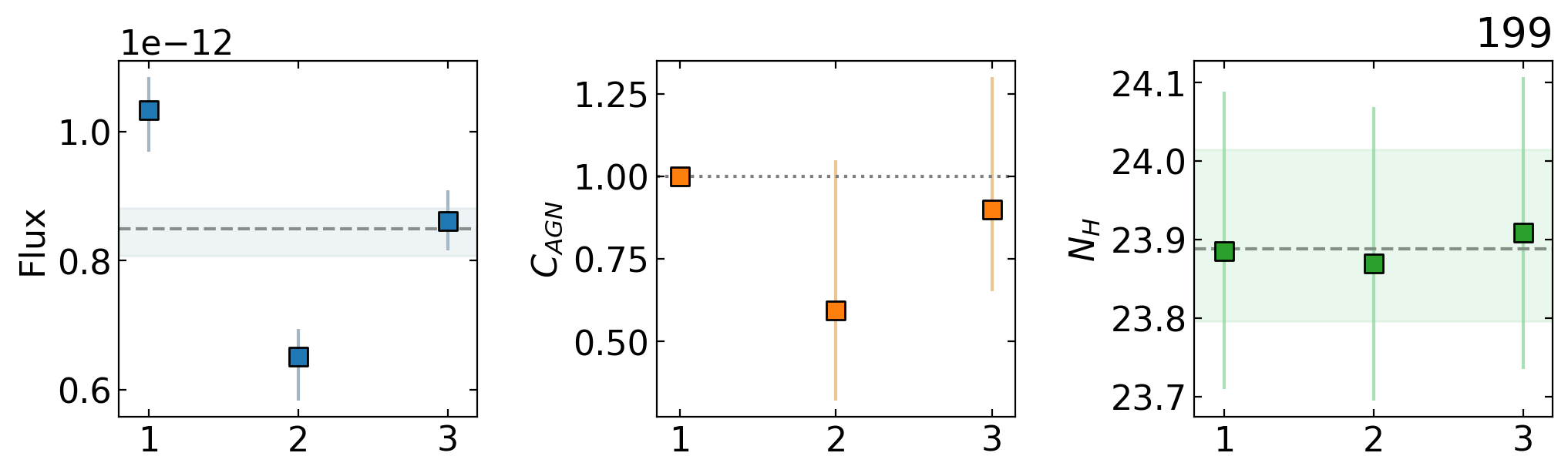}
    \caption{Observed 2--10\,keV best-fit flux in erg/s/cm$^2$ (left panels), the constant $C_{AGN}$ accounting for intrinsic variability (middle panels), and logarithm of line-of-sight $N_{\rm H}$ in cm$^{-2}$ (right panels) for sources with multi-epoch observations. BAT IDs are indicated in the top-right of each panel. Horizontal lines in the flux and $N_{\rm H}$ panels indicate the average values, with shaded regions representing the propagated uncertainties. For $C_{AGN}$, the horizontal line and the first data point correspond to $C_{AGN}=1$, fixed for the first epoch in the simultaneous fitting procedure. Observations are shown in temporal order (as listed in Table \ref{tab:var}), though the x-axis spacing is not to scale.}
    \label{fig:variability1}
\end{figure*}

\begin{figure*}[!tp]
    \centering
    \includegraphics[scale=0.57]{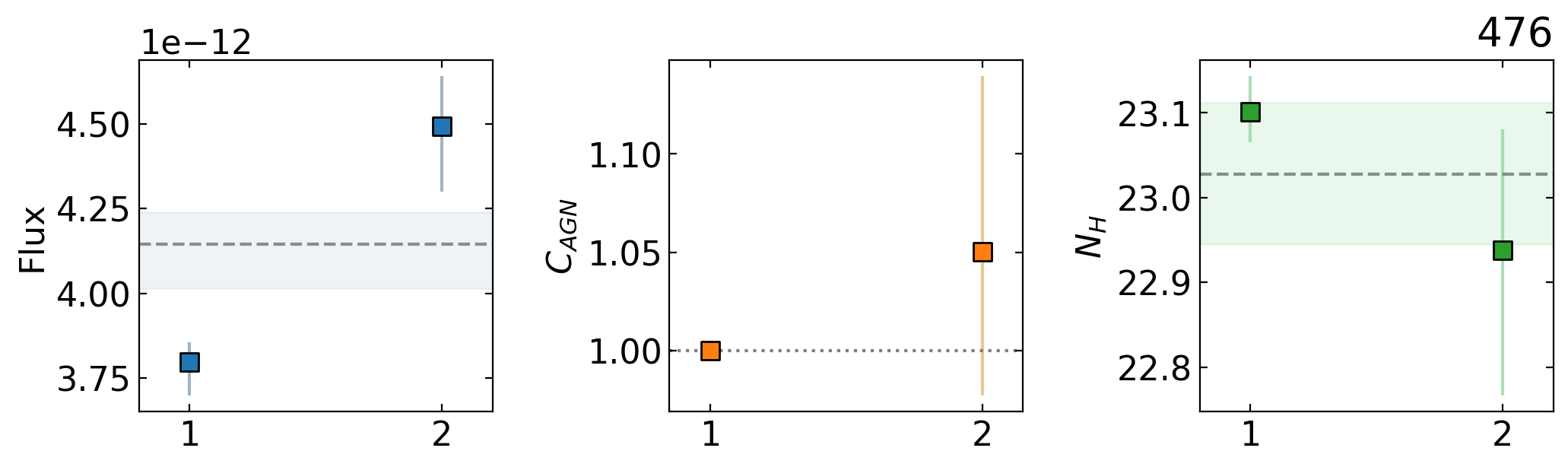}
    \includegraphics[scale=0.57]{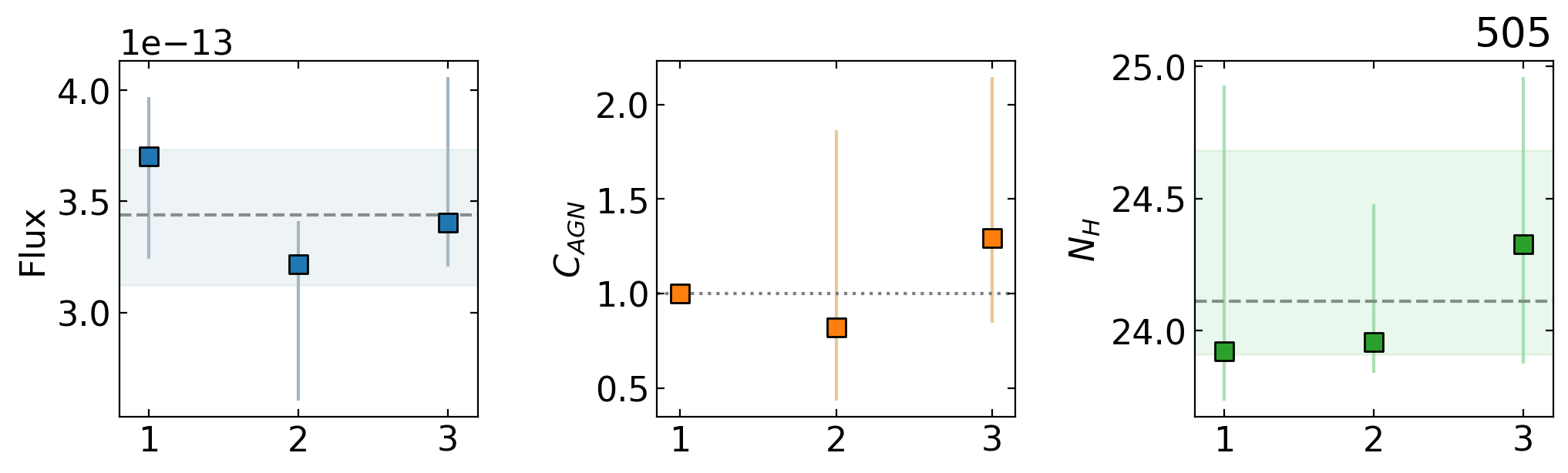}
    \includegraphics[scale=0.57]{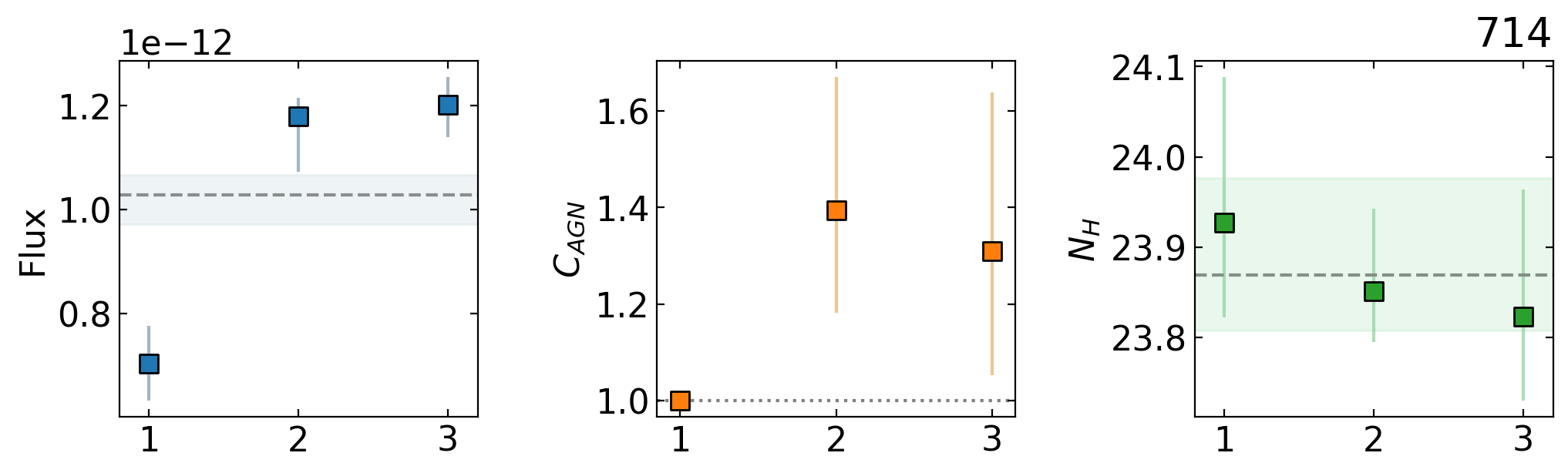}
    \includegraphics[scale=0.57]{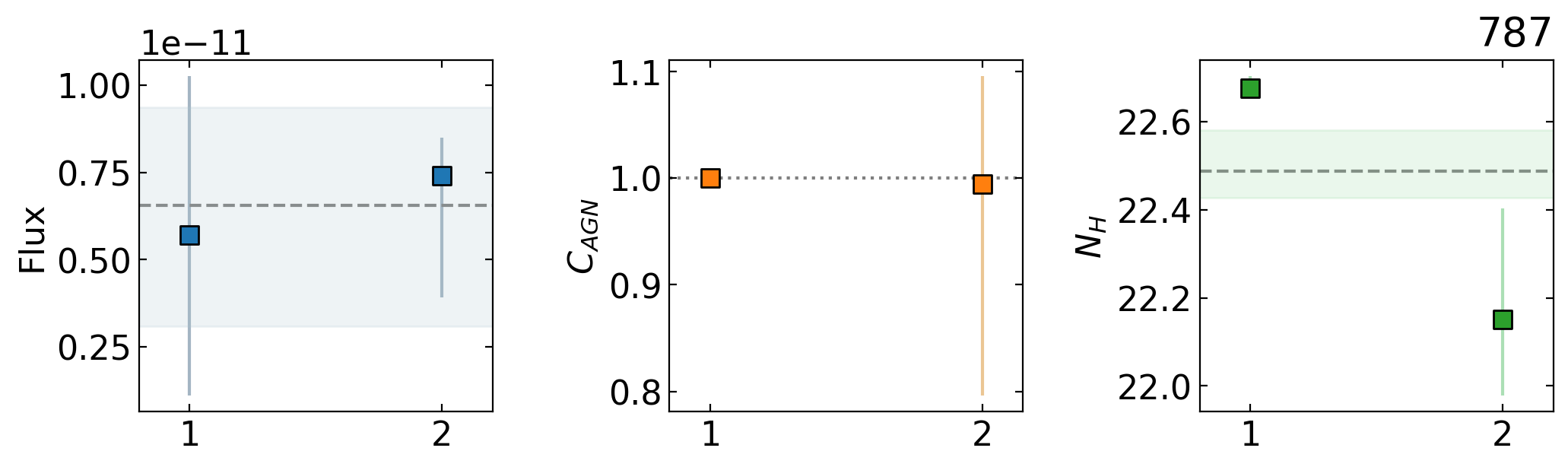}
    \includegraphics[scale=0.57]{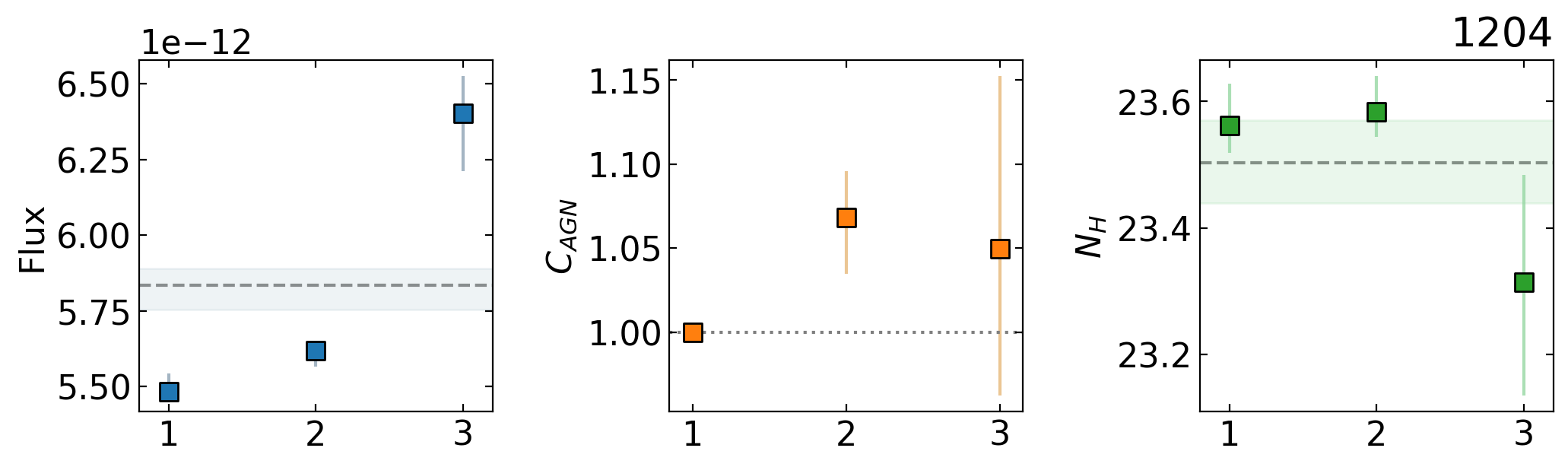}
    \caption{Same as Figure \ref{fig:variability1}.}
    \label{fig:variability2}
\end{figure*}

\begin{figure*}[!tp]
    \centering
    \includegraphics[scale=0.57]{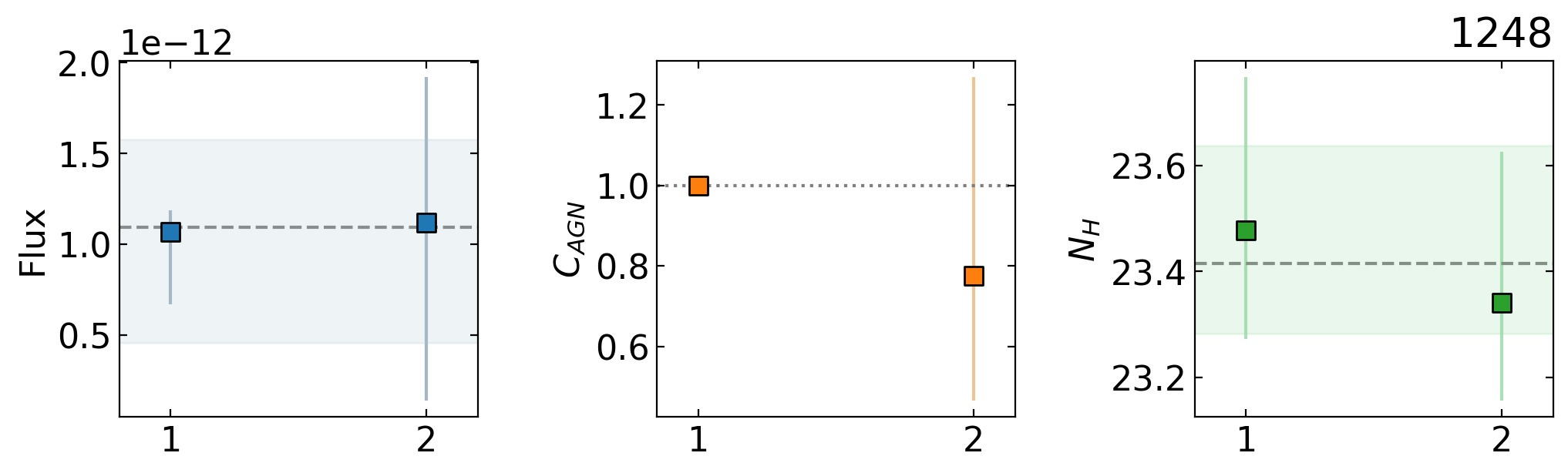}
    \includegraphics[scale=0.57]{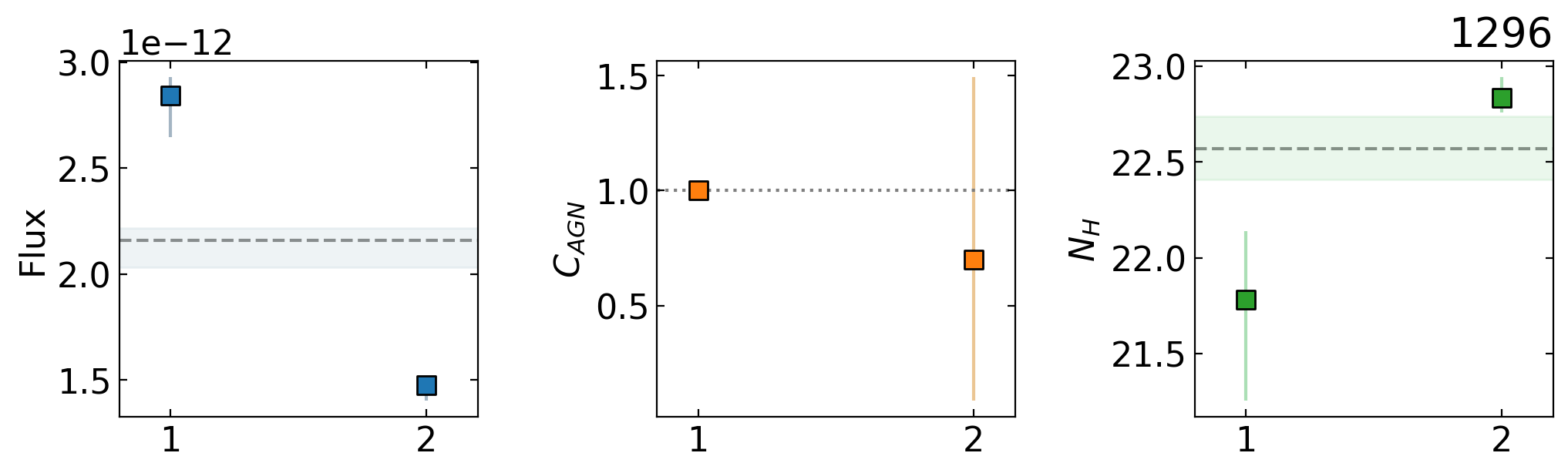}
    \includegraphics[scale=0.57]{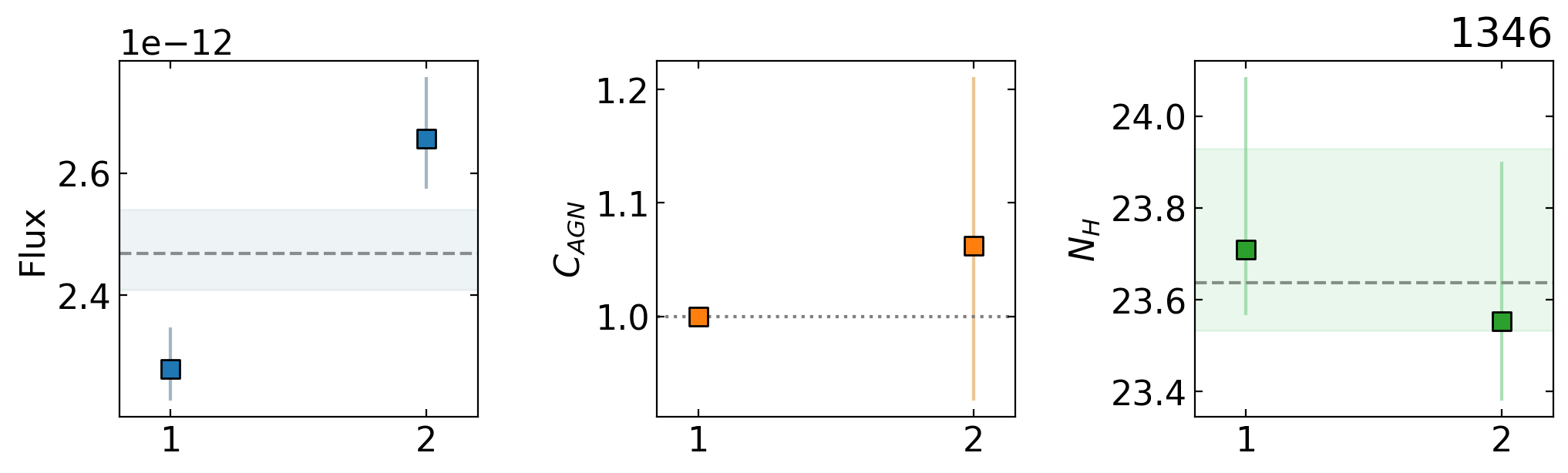}
    \includegraphics[scale=0.57]{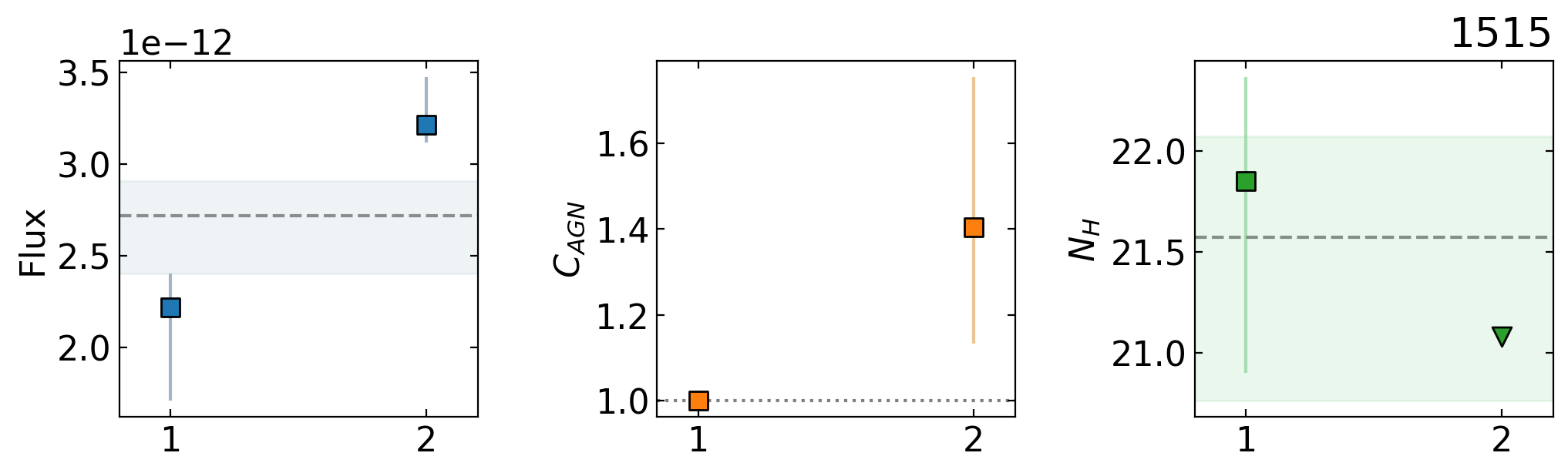}
    \caption{Same as Figure \ref{fig:variability1}. For BAT ID 1515, the second observation yielded only an upper limit, which we indicate with a triangle, as the best-fit returned a value lower than the Galactic column density of $N_{\rm H} = 1.2 \times 10^{21}$ cm$^{-2}$ \citep{kalberla05}.}
    \label{fig:variability3}
\end{figure*}

The best-fit unfolded spectra, models, and residuals are presented in Figures~\ref{fig:bestfits1}, \ref{fig:bestfits2}, and \ref{fig:bestfits3}. For sources with multi-epoch observations, the temporal evolution of $C_{\rm AGN}$, $N_{\rm H}$, and the observed 2--10\,keV flux (see Section~\ref{sec:variability}) is shown in Figures~\ref{fig:variability1}, \ref{fig:variability2}, and \ref{fig:variability3}.
To maintain graphical clarity and flow in the text, all these figures are placed at the end of the paper.

\begin{figure*}[!tp]
    \centering
    \includegraphics[scale=0.69]{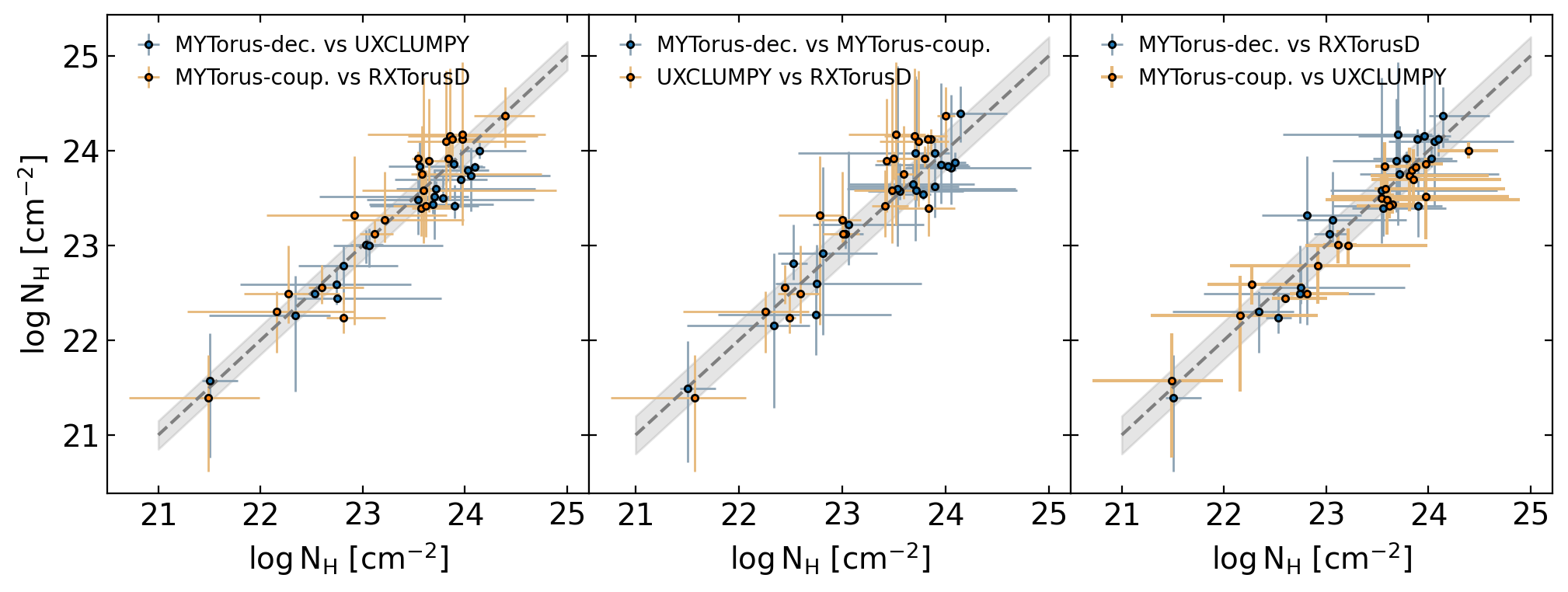}
    \caption{Comparison of $N_{\mathrm{H}}$ values derived using different torus models for the sources in our sample. Each panel shows pairwise comparisons between models, with the 1:1 relation indicated by a dashed grey line and a shaded region representing $\pm0.2$ dex deviation.
	\textit{Top panel:} MYTorus-decoupled vs. UXCLUMPY (blue) and MYTorus-coupled vs. RXTorusD (orange).
	\textit{Middle panel:} MYTorus-decoupled vs. MYTorus-coupled (blue) and UXCLUMPY vs. RXTorusD (orange).
	\textit{Bottom panel:} MYTorus-decoupled vs. RXTorusD (blue) and MYTorus-coupled vs. UXCLUMPY (orange). Across all comparisons, the $N_{\mathrm{H}}$ values are consistent within the uncertainties, with no significant offsets.}
    \label{fig:model_comp}
\end{figure*}

We also show in this Section that the $N_{\mathrm{H}}$ values derived from different torus models presented in Section \ref{sec:analysis} are consistent within the uncertainties. This is shown in Figure~\ref{fig:model_comp}, which compares the fitted column densities for individual sources across models. All values agree within 0.2 dex when considering the uncertainties, with no significant systematic offsets found. This indicates that the inferred obscuration levels are consistent and not significantly dependent on, or biased by, the choice of torus model.

\section{Comparision with BASS DR1 and DR2}\label{app:ricci_comp}

\begin{figure*}[!tp]
    \centering
    \includegraphics[scale=0.55]{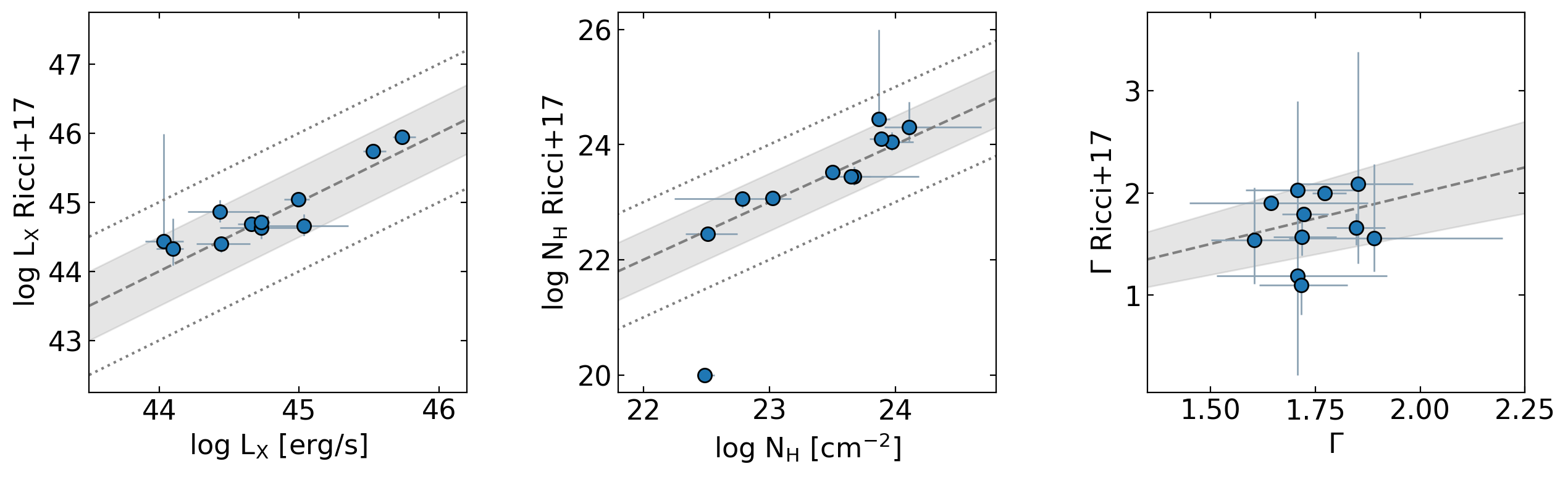}
    \caption{Comparison of $L_{\rm X}$, $N_{\rm H}$, and $\Gamma$ (left, middle, and right panels, respectively) with \cite{ricci17} for the 11 sources that are in both samples. The grey dashed lines represent the 1:1 relations. The shaded grey areas in the left and middle panels indicate the 0.5 dex deviation, and the grey dotted lines show the 1 dex deviation. For the right panel, the grey shaded area shows the 20\% deviation from the 1:1 relation. There is an overall agreement for these parameters (details in text), with differences primarily due to the inclusion of \nustar\ data, different modeling, and variability.}
    \label{fig:comp_ricci}
\end{figure*}

\begin{figure}[!tp]
    \centering
    \includegraphics[scale=0.75]{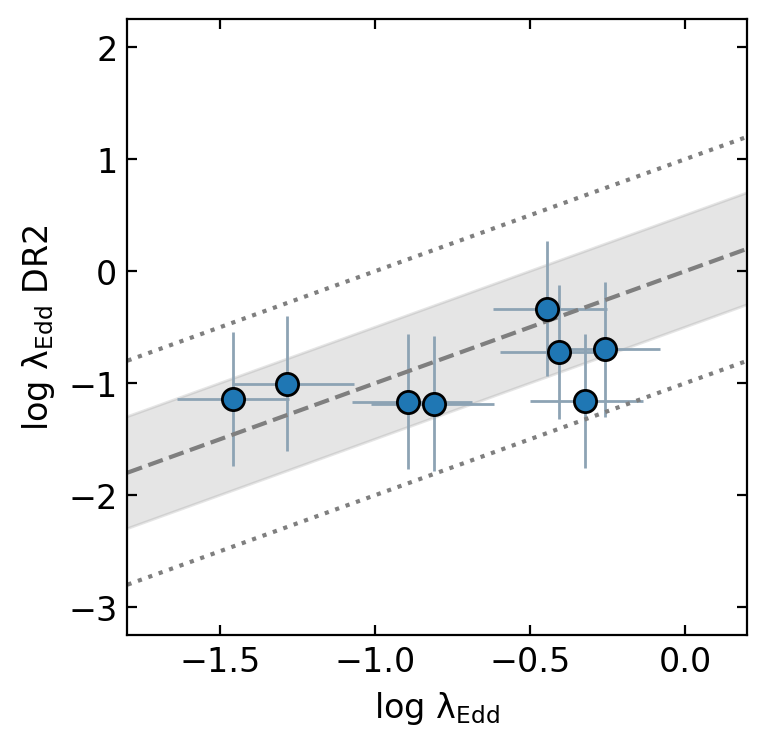}
    \caption{Comparison of $\lambda_{\rm Edd}$ with an updated version of the BASS DR2 catalog \cite{koss22_dr2catalog} (see text). $\lambda_{\rm Edd}$ has been estimated for 10 sources in this work. The grey dashed line represents the 1:1 relation, the shaded grey area indicates a 0.5 dex deviation, and the grey dotted lines show a 1 dex deviation. There is an overall agreement for these parameters (details in text), with differences primarily due to the inclusion of \nustar\ data, different modeling, and variability.}
    \label{fig:comp_dr2edd}
\end{figure}

In this appendix, we compare the results from our spectral analysis with those of the BASS DR1 and DR2, which are based on the \textit{Swift}/BAT 70-month catalog. Specifically, we compare $N_{\rm H}$, $L_{\rm X}$, and $\Gamma$ with \cite{ricci17}, and $\lambda_{\rm Edd}$ with \cite{koss22_dr2catalog}, as shown in Figure \ref{fig:comp_ricci} and \ref{fig:comp_dr2edd}, respectively. The primary difference between this work and \cite{ricci17} is the inclusion of \nustar\ observations, which were unavailable for the earlier analysis. Additionally, there are differences in spectral modeling: while \cite{ricci17} employed a compilation of simple to complex models, they did not use the self-consistent AGN torus models adopted here.

Figure \ref{fig:comp_ricci} shows the comparison for the 11 sources analyzed both in this work and \cite{ricci17}. The remaining sources are part of the BASS 105-month catalog only, and one source (BAT ID  494) has not been fitted in \cite{ricci17}. Overall, our results are consistent with those of \cite{ricci17}, yet with some scatter. Specifically, for $L_{\rm X}$, all sources are within 0.5 dex, while for $N_{\rm H}$ all but two sources are within 0.5 dex, with 10/11 within one dex.
A notable outlier in the $N_{\rm H}$ measurements is PKS 1549-79 (BAT ID 787), which \cite{ricci17} identified as unobscured. However, our analysis classifies it as obscured. Our $N_{\rm H}$ measurement for this source is consistent with \cite{tombesi14}, who analyzed the \xmm\ data. We note that \cite{ricci17} modeled this source based on its optical classification as a blazar, whereas more recent SED analyses incorporating optical and radio data now classify it as a Seyfert galaxy (\citealp{koss22_dr2catalog} and references therein). This, together with hints of variability (see Section \ref{sec:variability}) and the inclusion of \nustar\ data, might explain the difference in the results.
For $\Gamma$, 9 out of 11 sources lie within a 20\% scatter, and one of the two remaining sources is consistent with this range when uncertainties are considered. The only clear outlier is ESP 39607 (BAT ID 32), which shows a very flat photon index of $\Gamma \sim 1.1$ in \cite{ricci17}. We note that the results from \cite{peca_ufi}, which focus specifically on this source, align well with the value obtained in this work.

To place our sample in the context of previous BASS population studies, we compare with the results of \cite{ricci17} by selecting sources similar to those in our sample. Specifically, selecting Seyfert 1.9 and 2 AGNs with intrinsic (absorption-corrected) 2–10 keV luminosities above $10^{44}$ erg s$^{-1}$, the observed obscured fraction (i.e., sources with $N_{\mathrm{H}} > 10^{22}$ cm$^{-2}$) in \cite{ricci17} is $85^{+4}_{-7}\%$. In our sample, the corresponding obscured fraction is $95^{+2}_{-9}\%$, i.e., in agreement within the uncertainties.

Figure \ref{fig:comp_dr2edd} shows a comparison of $\lambda_{\rm Edd}$ for eight sources for which this quantity was available both in our analysis and in BASS DR2.
All the sources are within one dex, with 7/8 within 0.5 dex. The remaining source is consistent with this range when accounting for uncertainties.
Differences in $\lambda_{\rm Edd}$ primary stem from updated black hole masses and new X-ray luminosities derived in this work. Overall, our results are in good agreement with prior estimates in BASS data releases.

\section{Optical/Infrared AGN diagnostics}\label{app:opt_ir_class}

\begin{figure}[!tp]
    \centering
    \includegraphics[scale=0.55]{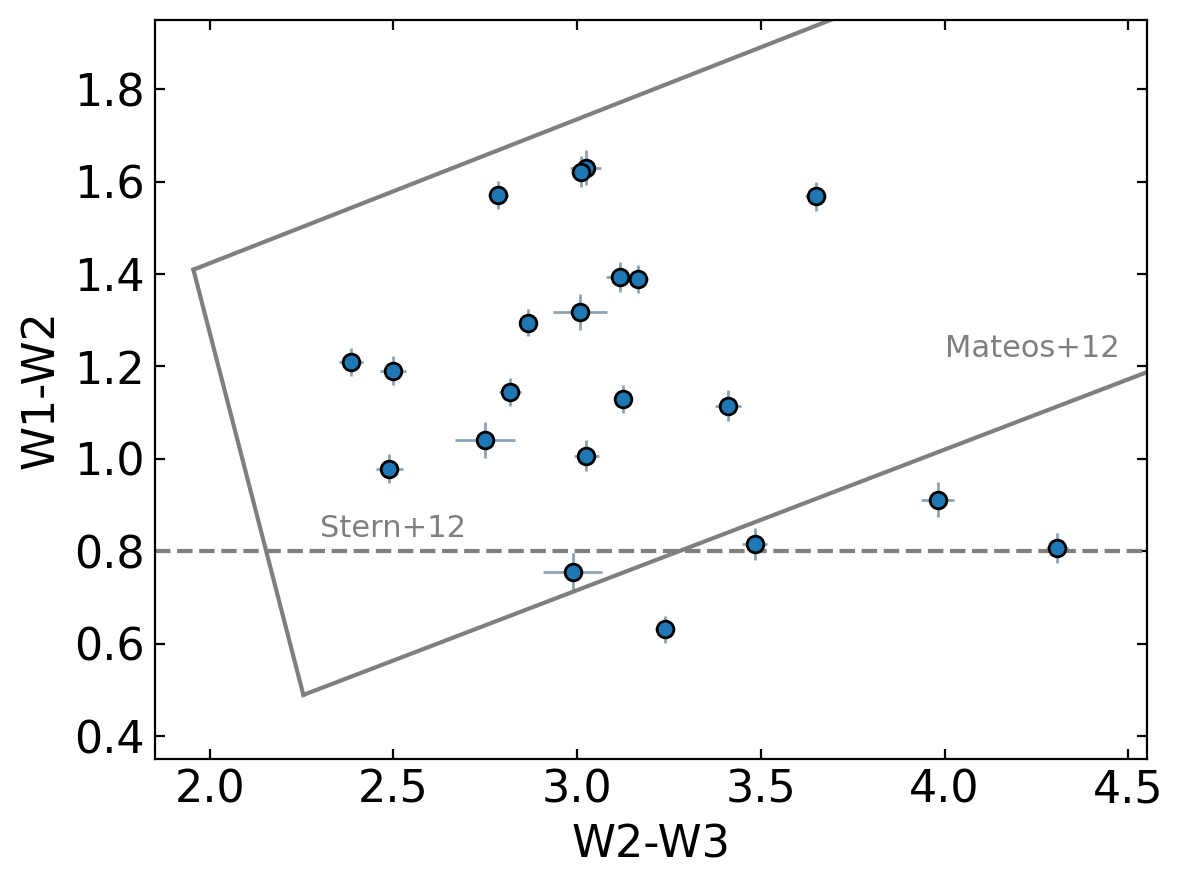}
    \caption{Color-color classification using \textit{WISE} \textit{W1}, \textit{W2}, and \textit{W3} color cuts. Sources within the region outlined by the solid grey lines \citep{mateos12} or above the dashed grey line \citep{stern12} are classified as AGN. One source falls outside both criteria but lies very close to the boundaries; see also \cite{ichikawa17}.}
    \label{fig:wise_selection}
\end{figure}

This section presents the infrared and optical classifications for our sample.
Figure \ref{fig:wise_selection} shows AGN selection criteria from \cite{stern12} and \cite{mateos12} based on \textit{WISE} \citep{wright10,mainzer11} mid-infrared color diagnostics. We matched our sample to the AllWISE catalog\footnote{\href{https://wise2.ipac.caltech.edu/docs/release/allwise/}{https://wise2.ipac.caltech.edu/docs/release/allwise/}} using a 1\arcsec matching radius, finding counterparts for all the sources. Vega magnitudes are used.
All but one source satisfy the AGN selection criteria from either \citet{stern12} or \citet{mateos12}. The single outlier (BAT ID 714) lies just outside both boundaries; however, \citet{ichikawa17} showed that WISE-based classifications can depend on luminosity, suggesting this source can still be safely classified as AGN.

\begin{figure}[!tp]
    \centering
    \includegraphics[scale=0.84]{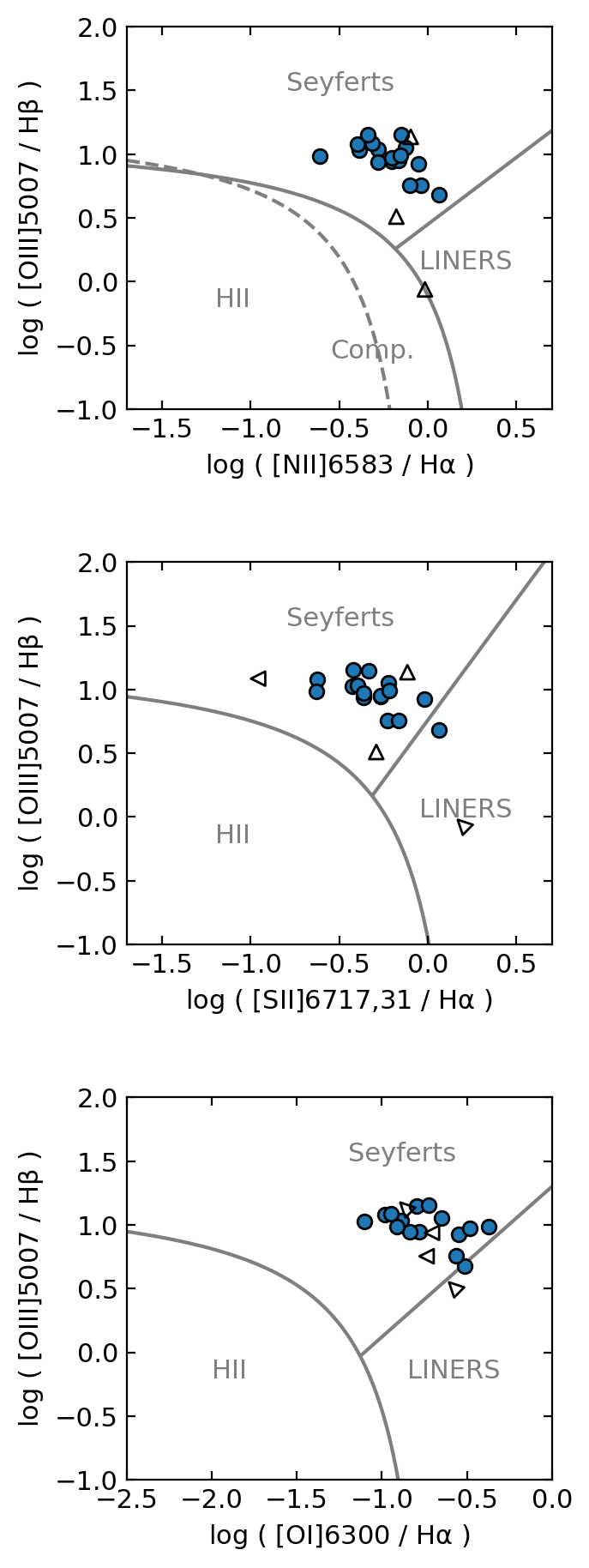}
    \caption{Optical classification from emission line diagnostic diagrams \citep{baldwin81, kewlwy01, kauffmann03, kewley06, schawinski07}. Upper limits are represented by empty markers oriented toward the limit. The uncertainties on emission line measurements are dominated by a systematic error of $<$0.1 dex \citep{koss22_dr2catalog}. All the sources without upper limits are classified as Seyfert in at least one diagnostic diagram.}
    \label{fig:bpt_selection}
\end{figure}

Figure \ref{fig:bpt_selection} shows the standard optical emission-line diagnostic diagrams (BPT; \citealp{baldwin81, veilleux87}). Emission line measurements for BASS DR2 sources are taken from \citet{oh22}, while those for the remaining sources are derived from our new optical spectra (see Section \ref{sec:opt_data}) using the same methodology. The line diagnostics could be performed for 20 out of 21 sources, with the remaining source (BAT ID 119) lacking sufficient SNR for reliable line measurements. Specifically, we obtained measurements for 19 sources in the \NIIHa\, and \SIIHa\, versus \OIIIHb\ diagrams, and for 18 sources in the \OIHa\, versus \OIIIHb\ diagram.  As shown in the figure, our sources consistently fall within the Seyfert region across all three diagnostics.
All sources described in this work are, therefore, classified as AGNs based on their infrared properties and exhibit Seyfert-like characteristics in the optical diagnostics.

\clearpage

\newpage


    

\newpage\clearpage
\bibliography{sample63}{}
\bibliographystyle{aasjournal}



\end{document}